\documentclass[aoas,generic,preprint]{imsart}
\pdfoutput=1
\usepackage[english]{babel} 
\usepackage[T1]{fontenc}
\usepackage{amsfonts,amsmath,amssymb}
\usepackage{natbib} 
\usepackage{graphicx}
\usepackage{calligra}
\usepackage{latexsym}
\usepackage{a4wide}
\usepackage{apalike}
\usepackage{url}
\usepackage[utf8]{inputenc}
\usepackage{fancyref}
\usepackage{color}
\usepackage{multirow, tabularx}
\usepackage{bbm}
\usepackage{mathpazo}
\RequirePackage[colorlinks,citecolor=blue,urlcolor=blue]{hyperref}

\arxiv{0000.0000}
\setattribute{journal}{name}{}

\newcommand{\Exp}{\mathbb{E}} 

\newcommand{\X}{\mathcal{\mathbf{X}}}
\DeclareMathOperator{\design}{\boldsymbol{X}} 
\newcommand{\Thet}{{\Theta}} 

\newcommand{\lp}{\left(}
\newcommand{\rp}{\right)} 

\DeclareMathOperator{\CSupp}{A}
\newcommand{\Fish}{\Gamma} 

\newcommand{\Sel}{\Pi} 
\DeclareMathOperator{\TS}{\mathcal{T}} 
\newcommand{\indeK}{k} 
\newcommand{\indeL}{l} 
\newcommand{\inde}{\lp\indeK,\indeL\rp} 
\DeclareMathAlphabet{\mathcalligra}{T1}{calligra}{g}{n}
\newcommand{\indg}{\mathit{g}} 
\newcommand{\indb}{b} 
\newcommand{\inds}{j} 

\newcommand{\Tp}{\mathbf{T}}
\newcommand{\Tr}{\operatorname{Tr}}

\newcommand{\Rs}{R^2~=~(RS,RE)}
\newcommand{\Rp}{R^2~=~(RS,RP)}

\renewcommand{\inds}{i}

\newcommand{\vertiii}[1]{{\left\vert\kern-0.25ex\left\vert\kern-0.25ex\left\vert #1 
    \right\vert\kern-0.25ex\right\vert\kern-0.25ex\right\vert}}

\usepackage{float}
\floatstyle{ruled}
\newfloat{alg}{thb}{lop}
\floatname{alg}{Algorithm}

\usepackage{amsthm}
\usepackage{theoremref}

\theoremstyle{plain}
\newtheorem*{defn*}{}

\theoremstyle{plain}

\theoremstyle{remark}

\renewcommand{\cite}{\citep}
\newcommand{\citenb}[1]{\citet{#1}} 

\usepackage[lofdepth,lotdepth]{subfig}
\usepackage{epstopdf}
\graphicspath{{figures/}}

\begin{document}
\begin{frontmatter}

\title{Two Sample Inference for Populations of Graphical Models with
Applications to Functional Connectivity}
\runtitle{Inference for Populations of Graphical Models}
\author{Manjari Narayan\thanksref{1}\ead[label=a1]{manjari@rice.edu}} ,
\author{Genevera I. Allen\thanksref{1,2,3}}
 \and  
 \author{Steffie N. Tomson\thanksref{4}}
\runauthor{Narayan et al.}
\affiliation{Dept. of Electrical Engineering\thanksmark{1} \& Dept. of  Statistics, Rice University\thanksmark{2}, \\ 
Dept. of Pediatrics-Neurology, Baylor College of Medicine \& Jan
and Dan Duncan Neurological Research Institute, Texas Children's Hospital\thanksmark{3}, 
Dept. of Psychiatry, UCLA\thanksmark{4} }







\begin{abstract}
\ Gaussian Graphical Models (GGM) are popularly used in neuroimaging
studies based on fMRI, EEG or MEG to estimate functional connectivity,
or relationships between remote brain regions. In multi-subject
studies, scientists seek to identify the functional brain connections
that are different between two groups of subjects, i.e. connections
present in a diseased group but absent in controls or vice versa. This
amounts to conducting two-sample large scale inference over network
edges post graphical model selection, a novel problem we call
Population Post Selection Inference.  Current approaches to this
problem include estimating a network for each subject, and then
assuming the subject networks are fixed, conducting two-sample
inference for each edge. These approaches, however, fail to account
for the variability associated with estimating each subject's graph,
thus resulting in high numbers of false positives and low statistical
power.  By using resampling and random penalization to estimate the
post selection variability together with proper random effects test
statistics, we develop a new procedure we call $R^{3}$ that solves
these problems.  Through simulation studies we show that $R^{3}$
offers major improvements over current approaches in terms of error
control and statistical power.  We apply our
method to identify functional connections present or absent in
autistic subjects using the ABIDE multi-subject fMRI study. 
\end{abstract}

\end{frontmatter}

\section{Introduction}\label{section:intro}

Functional connectivity seeks to find statistical dependencies between
neural activity in different parts of the brain \cite{Smith:2012nx,smith2013functional}.  It has been studied at a systems or whole brain level using noninvasive imaging
techniques such as functional MRI (fMRI), EEG and MEG, or using more
invasive techniques at the micro level such as electrophysiology or
ECoG.  Whole brain functional connectivity has become especially
popular to study in resting state fMRI where subjects lie in the scanner
passively at rest \cite{Smith:2012nx}. For this data, connectivity is typically
modeled as a network with functionally or anatomically derived brain
regions as nodes and connections as 
undirected edges \cite{Bullmore:2009vn}.  In this paper, we are particularly
interested in
studying multi-subject functional connectivity for resting-state fMRI
data; the statistical challenges we outline and methods we develop,
however, are applicable to functional connectivity in many
neuroimaging modalities.

Many have sought to use functional connectivity and more broadly
connectomics to better understand neurological conditions and
diseases.  Specifically, we seek to address the question - How are
functional connections different in a group of diseased subjects than in
healthy controls? - by conducting inference across a population of
brain networks.  This question has been well studied in the neuroimaging 
literature; see \cite{milham2012open,Craddock:2013uq} for detailed reviews.   Indeed, neuroscientists have used these techniques to find connectivity
biomarkers for diseases such as Alzheimer's and clinical depression \cite{tam2014connectome,Tao:2013kx}.
However, we will show that these widely
used methods suffer from major statistical flaws that can
result in high error rates.  Understanding and solving these flaws
presents us with a new type of statistical problem, something that we
will term 
{\em Population Post Selection Inference (popPSI)}, that has been
previously unaddressed in the statistics literature.  
Thus in this paper, we have three major objectives: (1) To introduce this
completely new problem that arises in population functional
connectivity to the statistical community; (2) To discuss the open
statistical challenges that arise with this problem and diagnosis
problems associated with the currently used methods in neuroimaging; and (3)
To introduce a new statistical method that partially solves these
problems, leading to much improved performance in 
terms of statistical power and error control.

\subsection{Current Standard in Neuroimaging} \label{intro:std}

Before proceeding to define our problem, we pause to
understand current approaches in the neuroimaging literature to
conducting inference across a population of brain networks.  The
current standard as described in \cite{Zalesky:2010wb,Bullmore:2009vn,Zalesky:2012uq,Meda:2012lq,Palaniyappan:2013fk} follows three main steps
after a pre-processing step to format the data:
\begin{enumerate}
\item[Step 0.] Parcellate data for each subject. 
\item[Step 1.] Estimate a brain network for each subject.
\item[Step 2.] Aggregate graph metrics for each subject.
\item[Step 3.] Conduct two-sample inference on the graph metrics
across subjects.   
\end{enumerate}
Henceforth, we will refer to this approach as the standard method.  We
discuss each of these steps in further detail.

Resting-state functional MRI (fMRI) data is acquired as
three-dimensional volumes ($\approx 10,000 - 100,000$ voxels) over
time ($\approx 50 - 500$ time points captured every 2-3 seconds) as
each subject lies in the scanner at rest.   Studying functional brain
networks at the voxel level is not desirable as most connections would
be due to close spatial proximity and hence subject to physiological confounds \cite{Craddock:2013uq,turk2013functional}. Thus, voxel level connections are
difficult to interpret.  As a result, most study brain
networks where each node is an anatomical or functionally
derived brain region.  After standard fMRI pre-processing which
includes registering each subject's volume to a common
template \cite{beckmann2003general}, each subject's brain scan is parcellated by mapping
voxels to anatomical regions (e.g. AAL, Talaraich, or Harvard-Oxford
atlas \cite{fischl2004automatically}), or functionally derived regions \cite{power2011functional}.  The time
series of the voxels are then averaged within each region, yielding a
matrix, $\X_{p \times T}$, for $p$ brain regions ($\approx 90 - 500$)
and $T$ time points for each subject.

Given the parcellated fMRI data for each subject, Step 1 estimates a
brain network connecting the $p$ brain regions for each subject.
While there are many statistical models that have been used to
estimate brain networks (see \cite{Craddock:2013uq,simpson2013analyzing} for a thorough review), by far the most common is to use thresholded correlation matrices \cite{Zalesky:2010wb,Bullmore:2009vn,Zalesky:2012uq,Palaniyappan:2013fk}.
However, thresholded partial correlations have also been employed as
in a recent paper \cite{Tao:2013kx}. 

In Step 2, neuroimagers take the subject networks as fixed and study
topological properties of these networks using techniques adapted from physics and
computer science.  These so-called ``graph metrics'' summarize certain
properties of the networks such as degree, node centrality,
participation coefficient, modularity, and efficiency among many
others; see \cite{Sporns:2011ys} and the software \cite{rubinov2010complex} for a complete list of commonly used topological metrics in neuroimaging.

Finally in Step 3,
neuroimagers compare values of the graph metrics from Step 2 across
the population of subjects using large-scale statistical inference.
For two group populations (e.g. controls vs. diseased), this typically
entails using standard two-sample test statistics such as a two-sample
$t$-test for continuous graph metrics or a two-sample test for
proportions for binary graph metrics.  As many have noted the
benefits of non-parametric procedures, most use permutation null
distributions instead of asymptotic theoretical nulls \cite{Zalesky:2010wb,simpson2013permutation}.
Finally, as many of the graph metrics result in a statistic for each
node of the network (i.e. the degree of each network node),
neuroimagers correct for multiplicity, typically by controlling the
false discovery rate (FDR) \cite{Zalesky:2010wb}.  To summarize, the final inference step to test for differences in a single graph metric across two
subject groups consists of three sub-steps - two-sample test
statistics, permutation nulls, corrections for multiple testing \cite{Zalesky:2010wb}.

\subsubsection{Our Problem}

\renewcommand{\indg}{g}

The above outline of population inference for functional connectivity
is broad and is used for testing many types of graph metrics
and with many types of statistical network models.  In this paper, we
wish to study this problem carefully and hence focus on a very
specific statistical problem: Using Markov Networks and specifically
Gaussian Graphical Models (GGMs) as the model for subject-level
networks, we seek to test for the differential presence of a single
network edge in one group of subjects.  Thus, we assume that the
observed multi-subject fMRI data, $\mathcal{\mathbf{X}}_{n \times
p \times T}$ for $n$ subjects, $p$ brain regions, and $T$ whitened
time points, arises from the following model:
\begin{align}
\label{our_model}
\textrm{Subject-Level:} & \ \ x_{j}^{(i)} \overset{iid}{\sim} N(
0, (\Thet^{(i)})^{-1}) \ \ \forall j = 1, \ldots T.   \nonumber  \\ 
\textrm{Group-Level:} & \ \ Y_{k,l}^{(i)}
= \mathbb{I}( \theta^{(i)}_{k,l} \neq 0 ) \sim
Bern( \pi^{\indg}_{k,l} ) \ \ \forall
i \in \mathcal{G}_{\indg} \ \& \ \forall  1 \leq k < l \leq p.
\end{align}
Here, $\Thet_{i}$ is the $p \times p$ sparse inverse covariance matrix
for subject $i$ with $\theta_{k,l}$ denoting the $k,l^{th}$ matrix
element, $\indg$ denotes group membership, and
$\pi^{\indg}_{k,l}$ denotes the group level probability of an
edge at $(k,l)$.  We assume that each subject follows a separate
Gaussian graphical model (GGM), but that the support of each edge in
the graph structure follows some group-level probability.  Note that
this permits each subject to have a potentially different brain
network, an important attribute as we expect each subject to have a
slightly different brain network.  
Given this population of GGMs, we seek to test for 
differential edge support between two groups of subjects, by testing
the following hypothesis for each edge, $(k,l)$:
\begin{align}
\label{edge_prob_test}
\mathcal{H}_{0}: \pi^{A}_{(k,l)} = \pi^{B}_{(k,l)} \ \
vs. \ \ \mathcal{H}_{1}: \pi^{A}_{(k,l)} \neq \pi^{B}_{(k,l)}.
\end{align}
This corresponds to asking whether a single functional connection in
the brain network is more present in one group of 
subjects than the other.  For example with autistic subjects (which we
will study further in our case study in Section~\ref{section:abide}), we may hypothesize that
autistic subjects will have fewer edges than healthy controls between
the fusiform gyrus which is responsible for facial cognition and other
regions such as the occipital lobe associated with social cognition.   
Our ultimate goal is to develop an inferential procedure for tesing \eqref{edge_prob_test} 
based on the model \eqref{our_model} that has high statistical power and controls or limits the false positives, 
either for testing a single edge or the false discovery rate (FDR) for testing many edges.

While the inference problem we study is a special case of the
general framework employed in neuroimaging, it is nonetheless a new
problem that has not been specifically addressed by the neuroimaging
community.  Several, however, have used Markov Networks and GGMs to study
functional connectivity \cite{Huang:2010fk,Smith:2011uq,Ryali:2012fk}; many others have used closely related partial correlations to model connectivity 
\cite{marrelec2006partial}. These models offer
several advantages for connectivity as they capture  more direct 
functional connections compared to correlation-based networks,
correspond to a coherent statistical model, and have been shown to be
more robust to physiological constructs such as head motion \cite{yan2013comprehensive}.
Also, while most conduct inference on graph metrics, several have
proposed to test individual edges \cite{Zalesky:2010wb,varoquaux2013learning,Lee:2014fk}; moreover, several specific
functional connections have been associated with clinical
conditions \cite{Bullmore:2012kx,Tao:2013kx}.  Finally, we note that testing edges in Markov
Networks is more powerful that of testing correlations as differential
connections can be pinpointed to precise brain regions because of
the conditional dependence relationships.

Our problem is also new from a statistical perspective, but related to
several other problems in the statistical literature.  For example,
some have noted that testing for functional connections in a
population is akin to testing for zero entries in the covariance or
precision matrix \cite{ren2013asymptotic}.  Others have proposed to test for
differences between the elements of two covariances \cite{cai2013two,zhao2014direct}.  When
applied to functional connectivity, however, these inference
procedures make the key assumption that all subjects share the same
network model, an assumption that we do not make.  Also, some have
proposed methods to find network differences based on perturbations to random networks \cite{balachandran2013inference} or testing procedures for the stochastic block model \cite{vogelstein2013graph}.  Importantly, these classes of methods assume a model
that generates the networks and not one that generates the observed
subject-level data directly.  Finally, many have sought to
characterize differences in subject networks through estimation via
multiple GGMs \cite{Guo:2011vn,Danaher:2011vq} instead of through direct inference as we propose.

\subsection{Population Post Selection Inference} \label{intro:popPSI}

Our model, \eqref{our_model}, is a two-level model, and there is a large body of
statistical literature on estimation and inference for multi-level and
random effects models; see \cite{Searle:2009jh} for an overview.  Unfortunately, we will not be able to directly employ any of these classical estimates
and inference procedures for our problem.  To estimate the
subject-level parameters, $\Thet$, corresponding to subject-level
brain networks, we will need to use sparse graph selection
techniques.  This is necessary as first, we are testing the sparse
support of $\Thet$; additionally, we expect functional connectivity
to be a sparse network; and finally, often the number of brain regions
considered, $p$, is larger than the number of resting-state time
points, $T$, thus necessitating regularized maximum likelihood
estimation.  By using a selection procedure to estimate the
subject-level parameters, however, our parameter estimates no longer
follow known distributions, negating the possibility of employing
classical random effects methods.

Inference for multi-subject functional connectivity then gives rise to a
completely new class of challenging multi-level statistical inference
problems.  We term this new class of problems {\em Population Post Selection 
Inference (popPSI)} and define these as follows: PopPSI problems are
two-level problems in which  
a variable selection procedure is used for parameter estimation at the
subject-level and inference is to be conducted on parameters at the
group level.   
In our case, the variable selection problem at level one corresponds
to using graph selection to estimate the brain networks for each
subject, and the inference problem at level two corresponds to testing
for differential edge support between two groups of subjects.  
Indeed, any multi-subject inference problems for functional
connectivity can be seen as popPSI
problems.

We employ the name Population PSI to denote the close connection to Post
Selection Inference (PSI) \cite{berk2012valid}.  A growing literature on PSI has
focused on inference, including p-values and confidence intervals, for the
coefficients of linear regression after selection via lasso-type
estimators \cite{wasserman2009high,zhang2011confidence,van2013asymptotically,javanmard2013confidence} others have discussed PSI for graph
selection \cite{wasserman2013estimating}.  This current PSI literature, however, has focused
on conducting inference directly on the selected parameters in a
single-level model.  Our Population PSI problem, on the other hand,
seeks to aggregate selected parameters across subjects and conduct
inference between subject groups at the population level.  This then,
presents a new class of statistical problems that poses many new
challenges.

In this paper, we focus on better understanding this new popPSI
problem, especially our specific inference problem outlined
in \eqref{edge_prob_test}, and propose a novel methodological approach that offers
dramatic improvements over the current standard in neuroimaging.  In
Section~\ref{section:challenge}, we seek to understand the performance of the standard method in neuroimaging for our inference problem \eqref{edge_prob_test}; namely, we show that the standard method has very high error rates with
low statistical power.  Investigating the standard method, we outline
two challenges characteristic of our popPSI problem that are
unaddressed by the standard method: two levels of network variability,
Section~\ref{subsec:challenge1}, and biases resulting from graph
selection errors, 
Section~\ref{subsec:challenge2}. In 
Section~\ref{section:methods}, we propose a novel method named $R^{3}$ that uses
resampling, random effects, and random penalization to address the
first challenge and partially address the second challenge raised
previously.  Our new $R^{3}$ method integrates the three steps of the
standard approach into one procedure and by doing so offers
substantial gains in statistical power and error control.  We
investigate our method, variations of our
approach, and the standard method in extensive simulation studies in
Section~\ref{section:sims}.  In Section~\ref{section:abide}, we apply
our method to the ABIDE multi-subject fMRI study to find functional
connections that are associated with autism. 
We conclude with a discussion in Section~\ref{section:disc}. 
Also we note that while there are a plethora of open theoretical 
questions that arise with new popPSI problems and our specific
problem, in this paper, we focus on building an intuition behind the
challenges associated with these problems and propose a 
methodological solution; we save theoretical investigations for
future work.

\section{Challenges of Population Post Selection Inference}
\label{section:challenge} 

Our new Population PSI problem introduced in
Section~\ref{section:intro} will pose 
many challenges both methodologically and theoretically. We identify
two challenges that are broadly characteristic of popPSI problems when
conducting inference on multiple unobserved networks in high
dimensions. In order to understand these challenges, we carefully
examine the standard approach and highlight its shortcomings in the
context of our particular inference problem, \eqref{edge_prob_test}.

\subsection{Investigating the Standard Approach}\label{subsec:standard}

We begin by motivating the need for alternatives to the standard
approach, outlined in Section~\ref{intro:std}, by studying this in the
context of our model \eqref{our_model} and inference
problem \eqref{edge_prob_test}.  Recall that the standard 
approach begins by estimating a graph structure independently for each
subject; for our 
problem, this entails selection and estimation for Gaussian graphical
models for which there are many well known procedures~\cite{Friedman:2008ys,meinshausen2006high}.  (We discuss
these further for our particular problem in
Section~\ref{section:methods}).  Next, the standard method aggregates
graph metrics for each subject which for our problem are the simple
binary edge presence or absence indicators for each edge in the
network.  Finally, the standard approach conducts inference across
subjects on the graph metrics; for our problem, this means testing for
differences in the edge support across the two groups of subjects.
For this, we can use a two-sample difference of proportions
test for each edge $(k,l)$:  $T
= \frac{\hat{\pi}^A - \hat{\pi}^B}{\sqrt{s^2_A + s^2_B}}$, where
$\hat{\pi}^{A}$ is the observed proportion of edge $(k,l)$ in subject
group $A$, and $s^2_g = \frac{1}{n_g} \hat{\pi}^g(1-\hat{\pi}^g)$ is
the usual estimate of 
the sample binomial variance. As previously mentioned, most use
permutation testing to obtain p-values and correct for multiplicity by
controlling the FDR; we do the same noting that as our test statistics
are highly dependent due to the network structure, we use the
Benjamini-Yekutieli procedure for dependent tests~\citep{Benjamini:2001hl}.

\begin{figure}[!htp]
\centering{
    \subfloat[Small World]{\label{fig:1a:small}	
          	\includegraphics[width=.35\linewidth,height=.3\linewidth]{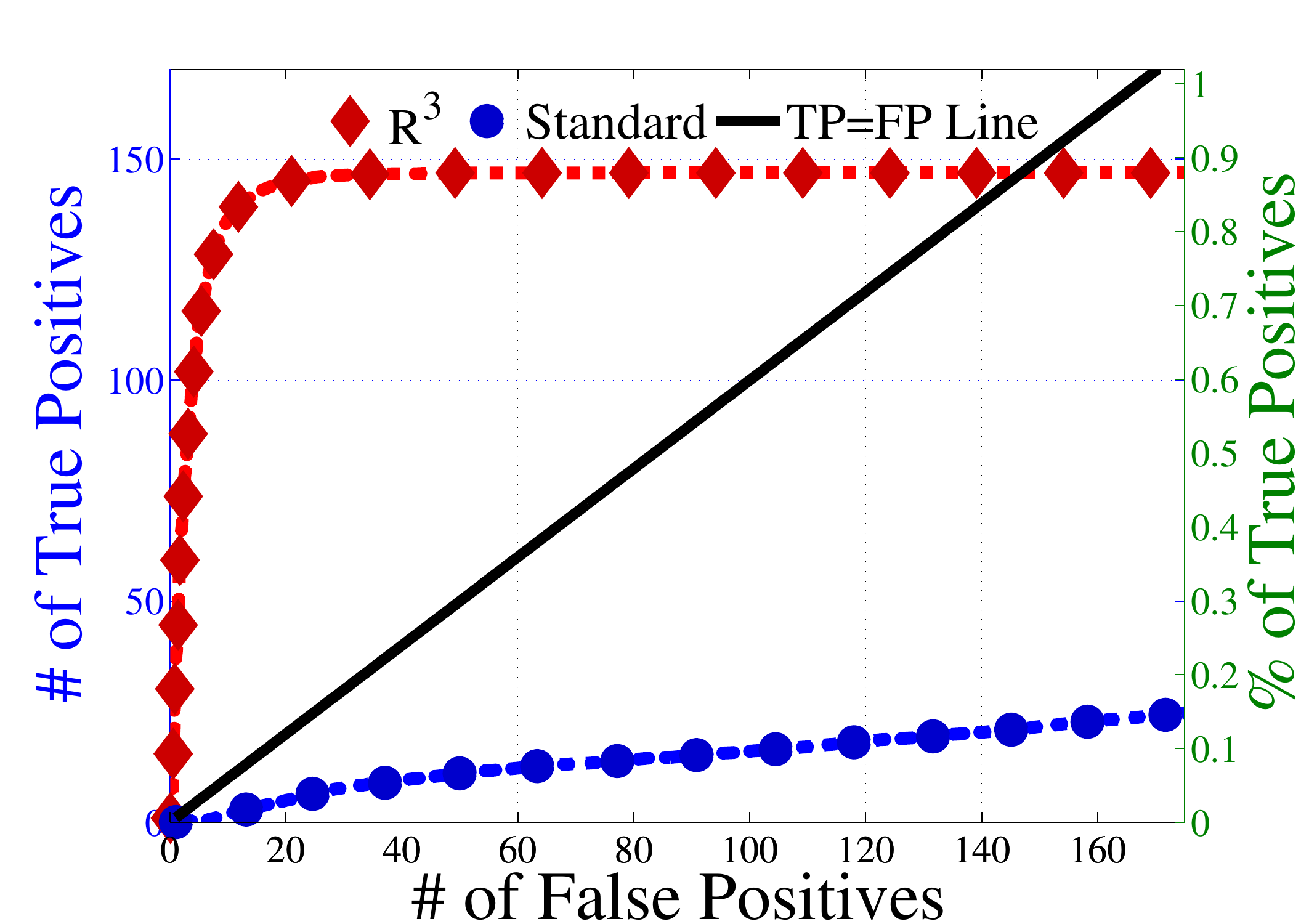}} 
	\subfloat[Standard]{\label{fig:1a:small}	
          	\includegraphics[width=.3\linewidth]{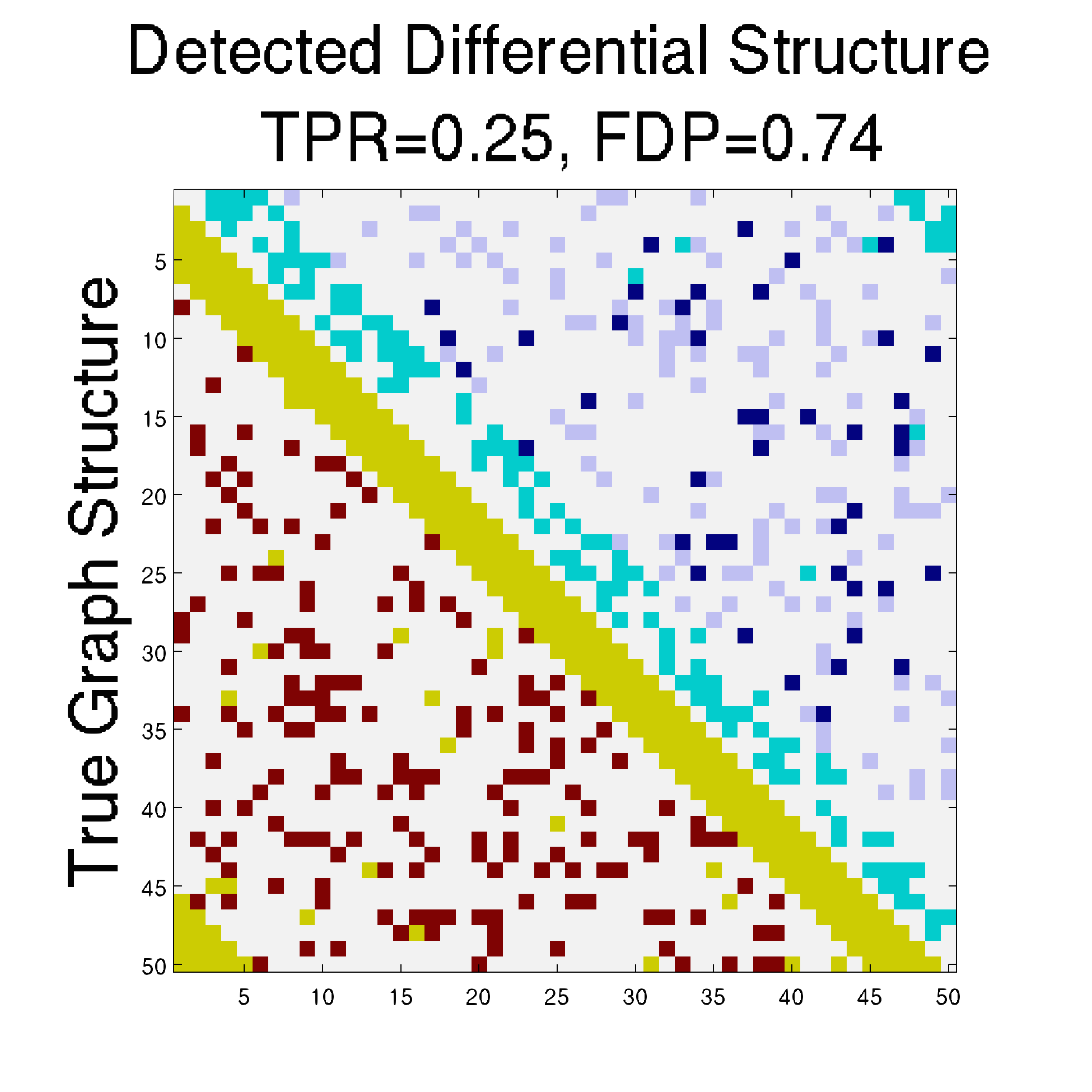}}
    \subfloat[$R^3$]{\label{fig:1a:small}	
          	\includegraphics[width=.3\linewidth]{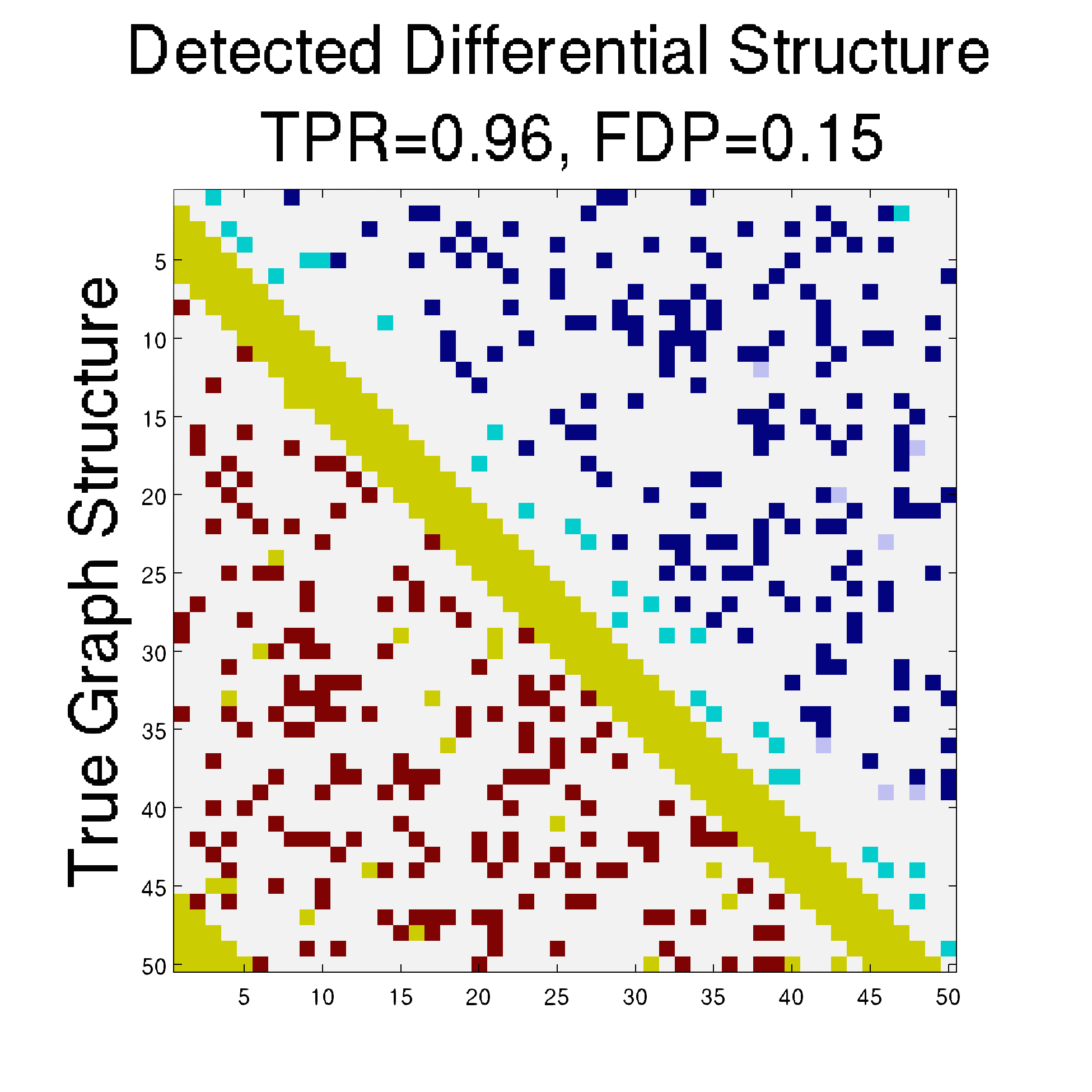}}\\
  \includegraphics[width=.55\linewidth,height=.12\linewidth]{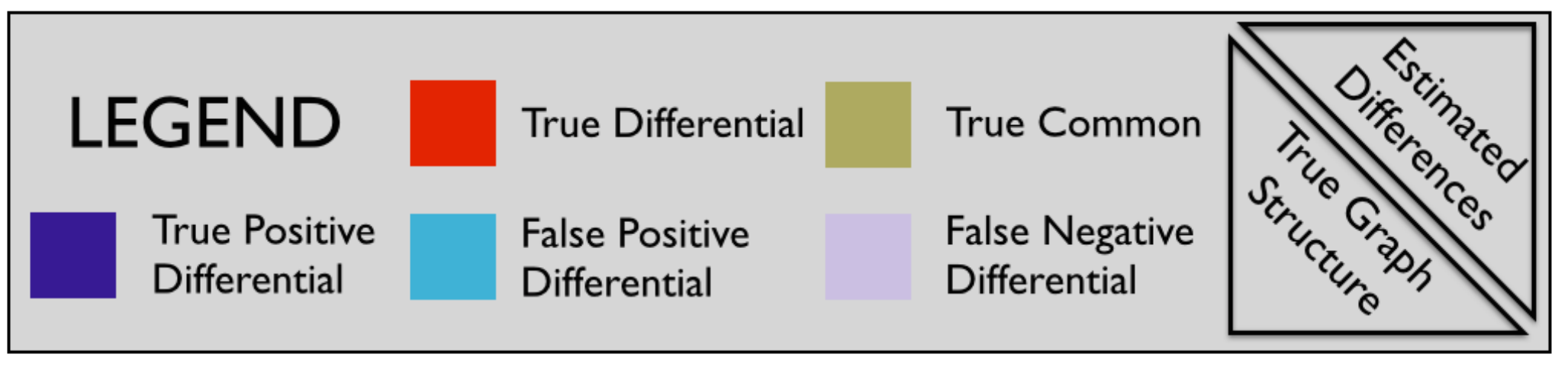}}
\caption{\label{fig:motivation} Motivating simulation study comparing
          	the standard approach to our proposed procedure,
          	$R^3$, in terms of (a) ROC curves for sequentially
          	rejected tests, and confusion adjacency
          	matrices along with observed true positive rate (TPR)
          	and false discovery proportion (FDP) for (b) the
          	standard method and (c) our approach.  The standard
          	method performs poorly in terms of both error control
          	and statistical power.}
\end{figure}

To understand the performance of this standard method, we present a
small preview of our simulation study discussed later in
Section~\ref{section:sims}.  Briefly, we assume that each subject graph
in group $A$ follows a small-world structure on $p=50$ nodes; in
group $B$, there are in addition 150 differential edges, meaning that
$\pi^{A}_{(k,l)} = 1$ and $\pi^{B}_{(k,l)} = 1$ for all differential
edges, $(k,l)$.  We generate data according to this model with $T=400$
time points and $n_{A} = n_{B} = 20$ subjects in each
group. Figure~\ref{fig:motivation} illustrates the results of this
standard approach as well as our new procedure, $R^{3}$, which we will
introduce later in Section~\ref{section:methods}.  Part (a) gives
ROC curves for the number of false positives verses true positives as
each sequential test is rejected; parts (b) and (c)
give the adjacency confusion matrix illustrating where the true and false
positive as well as false negative edges are detected in the graph structure.  

Our motivating simulation shows that the standard approach performs
terribly in terms of both error control and statistical power.  While
the magnitude of the poor performance of this approach may seem
astonishing, the poor performance should come as no surprise: The inferential
procedure (e.g. test statistics) of the standard approach assume a
one-level model that would be appropriate when the subject graphs are
fixed and known or directly observed quantities.  When these subject
networks are unobserved, however, and must be estimated from finite
data, these one-level test statistics are incorrect for our two-level
problem.  Specifically for
two-level problems, the variance of parameters estimated by
incorrectly assuming a one-level models is 
underestimated.  For our problem, the extra source of variability
arises from the graph selection procedure; we discuss challenges
associated with this subsequently in Section~\ref{subsec:challenge1}.
Incorrect variance estimates, however, are not the only problem with
the standard approach: A more subtle problem arises from the fact that
the proclivities of graph  
selection procedures for the Gaussian graphical model lead to
biased estimates of the edge proportions, $\hat{\pi}^{g}$.  As
discussed in Section~\ref{subsec:challenge2} and seen in Fig~\ref{fig:motivation}, graph selection false
positives and false negatives do not occur at random throughout the
network structure, leading to biased group level estimates.

Also, it is important to note that the standard approach corresponding
to a one-level problem would be appropriate if we were able to {\em
perfectly} estimate the network structure for each subject as this is
then the same as assuming the subject networks were directly
observed.  For typical fMRI data, however, this is unlikely to ever
happen due to (i) the limited sample size, $T$, relative to $p$ and
(ii) the highly connected 
network structures typical of brain networks \cite{Bullmore:2009vn};
these are known to violate irrepresentable and incoherence
conditions that are necessary for perfect graph
selection \cite{meinshausen2006high,ravikumar2011high,cai2011constrained}.

\subsection{Challenge I: Two Levels of Network Variability}
\label{subsec:challenge1}
\begin{figure}[!htp]
	\subfloat[Autism Subject 1]{
	\includegraphics[width=.3\linewidth]{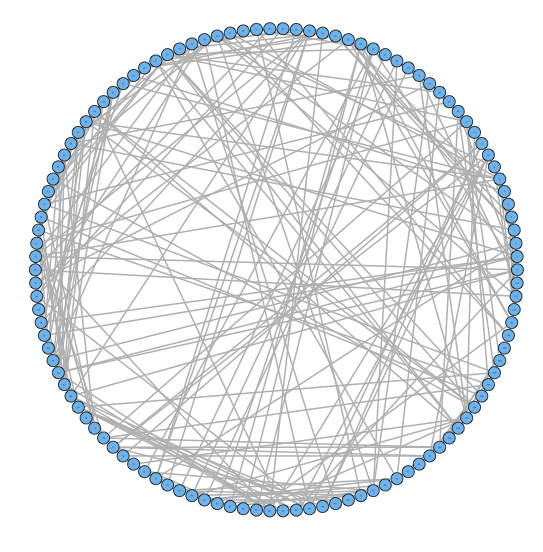}}
	\subfloat[Autism Subject 2]{
	\includegraphics[width=.3\linewidth]{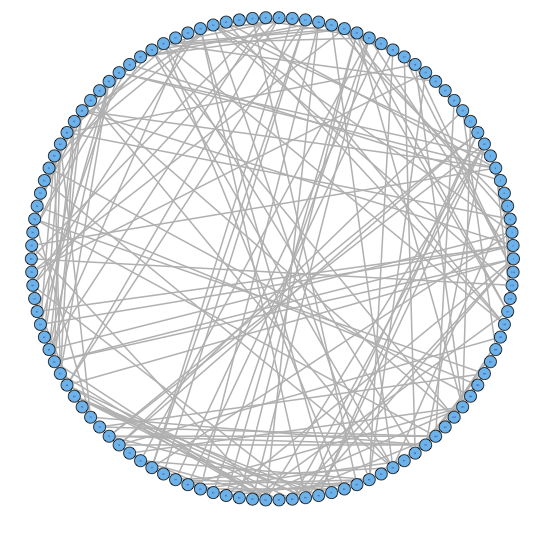}}
	\subfloat[Autism Subject 3]{
	\includegraphics[width=.3\linewidth]{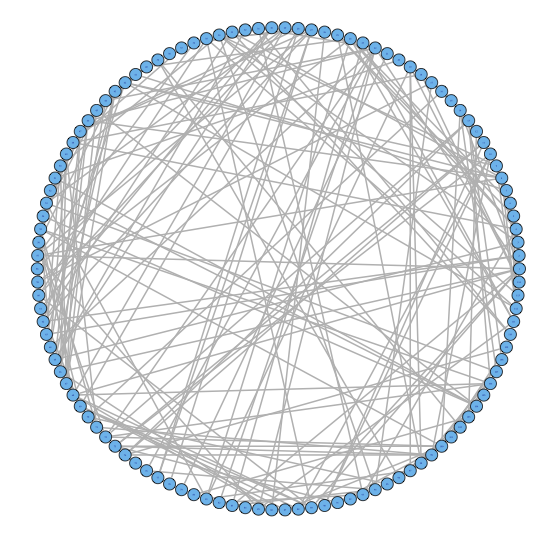}} \\
	\subfloat[Control Subject 1]{
	\includegraphics[width=.3\linewidth]{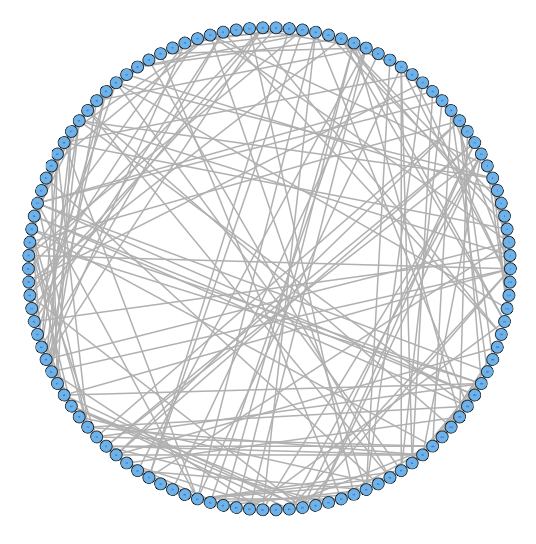}}
	\subfloat[Control Subject 2]{
	\includegraphics[width=.3\linewidth]{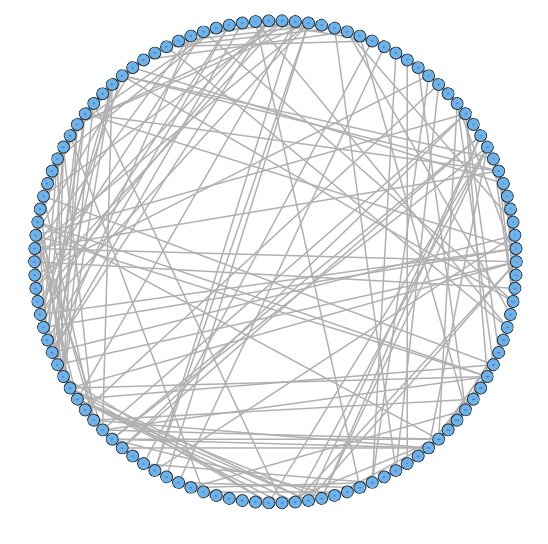}}
	\subfloat[Control Subject 3]{
	\includegraphics[width=.3\linewidth]{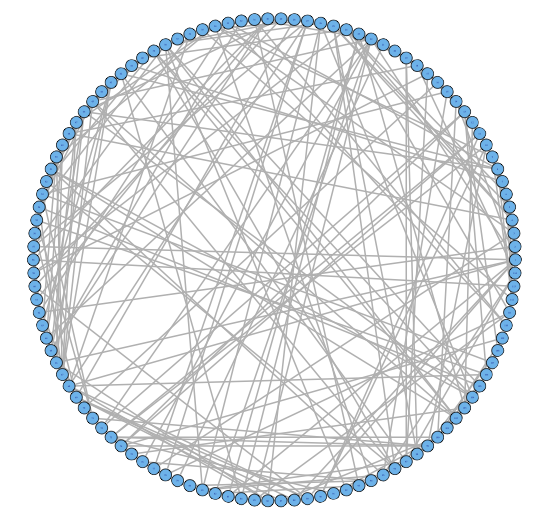}}\\
	\subfloat[Control 4, Resample 1]{
	\includegraphics[width=.3\textwidth]{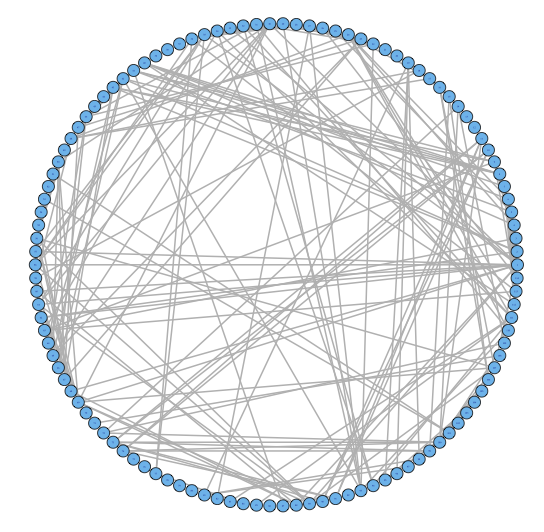}}
	\subfloat[Control 4, Resample 2]{
	\includegraphics[width=.3\textwidth]{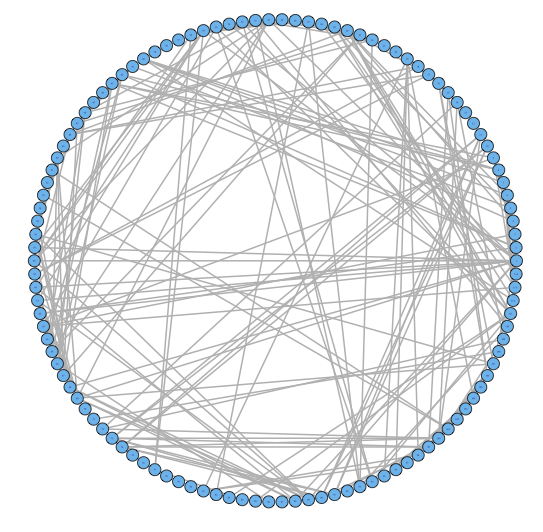}}
	\subfloat[Control 4, Resample 3]{
	\includegraphics[width=.3\textwidth]{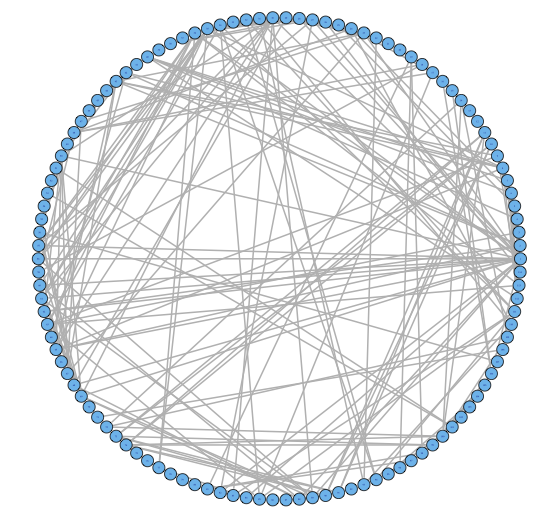}}	
\caption{Motivating example of both inter- and intra-subject network
  variability in estimated functional brain networks.  Gaussian
  graphical models were used to estimate networks from the 
  UCLA ABIDE fMRI data set \cite{ABIDE:2013fk} that we work with further in
  Section~\ref{section:abide} for three autistic subjects (top), three
  control subjects (middle), and three bootstrap resampled data sets from a
  control subject (bottom).  To conduct population inference across two groups,
  we must account for both the network variability between subjects
  (top and middle panels) as well as the variability associated with
  network estimation within a 
  single subject (bottom panel). This also motivates the applicability
  of our two-level model, \eqref{our_model}, for population network inference.
\label{fig:variability}}
\end{figure}

For our two-level problem \eqref{our_model}, we must account for two
sources of network variability when conducting population inference:
(i) variability between subjects within a group and (ii) variability
of the network selection procedure within a single subject.  To see
this, let us study a real multi-subject fMRI example.  In
Figure~\ref{fig:variability}, we show estimated functional brain networks for
subjects from the UCLA fMRI ABIDE data set \cite{ABIDE:2013fk}.  We describe the
details of this data set, our pre-processing, and brain parcellation
later in the Case Study in Section~\ref{section:abide}.  In the top and middle
panels, we estimate brain networks for each subject using graph
selection methods for 
Gaussian graphical models (see Section~\ref{subsec:graph_est} for details) and plot
these as circle graphs to easily visualize network differences.  It is
clear that there are not only differences between autistic
subjects and control subjects, but there is also large heterogeneity
across subjects within each group.  This is well-known in the
neuroimaging literature~\citep{milham2012adhd,nielsen2013multisite},
and makes finding statistically significant differences between subject groups much
more challenging.

Less well studied in neuroimaging, is the second source of variability
which arises from {\em estimating} networks for each subject instead
of directly observing the networks.  In the bottom panel of
Figure~\ref{fig:variability}, we re-estimate brain networks for a single control
subject with bootstrap resampled data.  It is clear that there is major
intra-subject variability arising from our graph selection procedure.
Indeed in neuroimaging, test--re-test studies which conduct brain
imaging on the same subject in repeated sessions have shown high
variability in the subject's estimated brain networks~\citep{wang2011graph}.
This also motivates the necessity of using a two-level model like
\eqref{our_model} for population network inference as opposed to the
one-level model and test statistics of the standard procedure.

Now, let us consider the consequences of these two levels of network
variability for our specific model and edge testing problem.  Studying
the variability via the post selection distribution of the estimated networks,
$\hat{\Thet}^{(i)}$, is a major challenge
that has not yet been tackled in the statistics literature.  Thus,
a direct approach to conducting population inference for the model
\eqref{our_model} is 
beyond the scope of this paper and something that is saved for future
work.  Instead, we opt to break this problem into a series of simpler
ones in an approach that is
more closely aligned with the standard procedure:  We 
consider a separate two-level model for each edge in the network that
can capture the two sources of network variability.

To model the two sources of network variability for each edge, we turn
to the commonly used Beta-Binomial model~\citep{Searle:2009jh}.  
As presented earlier, let
$Y_{k,l}^{(i)} = \mathbb{I}( \theta^{(i)}_{k,l} \neq 0 )$ denote the
edge support statistic for the $(k,l)^{th}$ edge associated with 
the $i^{th}$ subject graphical parameter (precision matrix)
$\Thet^{(i)}$.  Since each estimated network is a random variable,
$Y_{k,l}^{(i)}$ is a random variable whose variability is related to
the selection variability of our estimated network for subject $i$.
Let $\mu^{(i)}_{(k,l)} = \mathbb{P}(\theta^{(i)}_{(k,l)} \neq 0)$ be a
new parameter denoting 
this subject-level probability of observing an edge at $(k,l)$ in the
$i^{th}$ subject; we model the selection variability in the $i^{th}$
subject as  $Y_{k,l}^{(i)} \sim \text{Bern}(\mu^{(i)}_{(k,l)})$.  But, the
edge selection probabilities for each 
subject are themselves random variables related to the group-level
probabilities for each edge.  A common model for such probabilities is
the beta distribution; thus, we let $\mu^{(i)}_{(k,l)} \sim
\text{Beta}( \alpha^{g}_{(k,l)}, \beta^{g}_{(k,l)})$.  Typically, a
reparameterization of this model is used where $\pi^{g}_{(k,l)} =
\alpha^{g}_{(k,l)}/ (\alpha^{g}_{(k,l)} + \beta^{g}_{(k,l)})$ 
denotes the mean, 
$\Exp(\mu^{(i)}_{(k,l)}) = \pi^{g}$, of the Beta distribution, and
where $\rho^{g}_{(k,l)} = 1/(\alpha^{g}_{(k,l)} + \beta^{g}_{(k,l)}+1)$ is related to
the variance of the Beta distribution
given by $\text{Var}(\mu^{(i)}_{(k,l)}) =
\rho^{g}_{(k,l)}\pi^{g}_{(k,l)}(1-\pi^{g}_{(k,l)})$ \citep{Searle:2009jh}. 
Suppose we also observe $m$ iid observations from this model and let
$Z^{(i)}_{(k,l)} = \sum_{j=1}^{m} Y_{j,(k,l)}^{(i)}$.  Then, we arrive at
the familiar form of the Beta-Binomial model:
\begin{align} \label{eqn:beta_binomial}
Z^{(i)}_{(k,l)}| \mu^{(i)}_{(k,l)} & \overset{iid}{\sim} \text{Bin}(\mu^{(i)}_{(k,l)},m), \quad \ 
\mu^{(i)}_{(k,l)} \overset{iid}{\sim} \text{Beta}(\pi^{(g)}_{(k,l)},\rho^{(g)}_{(k,l)}).
\end{align}

This Beta-Binomial model, which is often used to model over-dispersed
binary data, is ideal for modeling both the intra-subject selection
variability and the inter-subject group-level variabilities of each
edge.  To see this, consider the unconditional variance of $Z^{(i)}_{(k,l)}$
which incorporates two levels of variability as follows (for
convenience, we suppress the edge indices): 
\begin{align}\label{eqn:unconditional_var}
\text{Var}(Z^{(i)}) & = \sum_{j} \text{Var}(Y^{(i)}_{j}) + \sum_{j<j'} \text{Cov}(Y^{(i)}_{j},Y^{(i)}_{j'}) \nonumber \\  
& = \ m\pi^{g}(1-\pi^{g}) + \underbrace{m(m-1)\rho^{g}\pi^{g}(1-\pi^{g})}_{\textrm{Additional Binomial Variation}}.
\end{align}
Hence, the first term represents variability across subjects in group
$g$ and the second term represents the variability associated with the
selection procedure within subject $i$, a quantity that we assume to be constant across subjects $i$ in each group $g$.  Consider now what happens
if our true model follows this two-level Beta-Binomial model, but as
with the standard approach, we
use a one-level Binomial model and associated
two-sample test statistic.  The variance is thus
underestimated and the test statistic
is overoptimistic.  Then, when inference is conducted for the population
mean $\pi^{g}$, using the incorrect Binomial model leads to
inflated Type I error rates; this behavior has been well-documented
\citep{weil1970selection,liang1994use}. Hence, failure 
to use the correct two-level model which accounts for the two levels
of network variability partially explains the high
error rates of the standard procedure observed in
Figure~\ref{fig:motivation}.

Notice in \eqref{eqn:beta_binomial} that we have defined our
Beta-Binomial model for 
the edge selection probabilities assuming that we have multiple iid
observations from this model.  For real fMRI data, we typically only
have one scanning session per subject and hence only one estimate of
the functional connectivity network, $\Thet^{(i)}$, per subject.  Then with
only one observation, $Y^{(i)}_{(k,l)}$, for each subject,
the Beta-Binomial model for each edge reduces to a Beta-Bernoulli
model.  In this model, the correlation parameter, $\rho^{g}$, is
unidentifiable and the intra-subject variability associated with graph
selection cannot be estimated.  Thus, estimating the two-levels of
network variability from data with only one observation is a
challenge; in Section~\ref{subsec:RS}, we discuss how we address this by using
resampling techniques to estimate the second source of network
variability.

\subsection{Challenge II: Graph Selection Errors}
\label{subsec:challenge2}

In the previous section, we deconstructed our problem into a
two-level model for each edge to simplify modeling the two sources of
variability.  The models for each edge, however, are clearly
not independent as we are modeling the network support for a
population of Gaussian graphical models.  Here, we discuss how
dependencies in the population network structure can lead to graph selection
errors that bias the estimates of our group-level edge parameters.
These in turn lead to false positives and false negatives
when conducting inference at the group level.

Note that as previously discussed, we are working under the regime
where we cannot obtain perfect estimates of the network support, as
this is the most realistic scenario for real fMRI data.  Thus, it is
constructive to understand the conditions under which perfect network
recovery or graph selection consistency is achievable so that we can
understand the consequences when these conditions are violated.  
\citet{meinshausen2006high} 
first introduced an irrepresentable condition for neighborhood 
selection-based estimation of GGMs that closely follows from
irrepresentable or incoherence conditions for the lasso regression problem \citenb{Zhao:2006uq}. Later, \citenb{ravikumar2011high} characterized a log-determinant based irrepresentable
condition corresponding to
estimating GGMs via penalized maximum 
likelihood, or the so-called graphical lasso method \cite{rothman2008sparse,Friedman:2008ys}.  This
condition places restrictions on the Fisher information matrix,
$\Fish = \Theta^{-1} \otimes \Theta^{-1}$; that is, if we let $S$ denote the
network support and let $S^{C}$ denote the non-support, then the
condition requires that $\||\Fish_{S^{c},S}^{T}\lp \Fish_{S,S}
\rp^{-1} |\|_{\infty} \leq 1 - \eta$, for some $0 < \eta < 1$. 
In addition to irrepresentability conditions, the eigenvalues of the
restricted Fisher information $\lp\Fish_{S,S}\rp^{-1}$ as well as
covariance matrix $\lp\Theta^{-1}\rp_{S}$ need to be bounded away from
zero, and the entries of the precision matrix $\Theta^{\inds}_{(k,l)}$
need to satisfy signal strength conditions in order to prevent false
exclusions of edges in each subject \cite{ravikumar2011high}.  Both
neighborhood and log-determinant irrepresentable conditions limit the
amount of correlation within true edges and between true edges and
non-edges; this, then places severe restrictions on 
the model space and types of network structures where graph selection
consistency is achievable.  As illustrated in \citet{meinshausen2008note},
certain simple network structures nearly guarantee irrepresentable
conditions are violated in the population version, and consequently in
finite samples.   For example,
estimators have a high probability of incorrectly selecting an edge
connecting two nodes that share similar node-neighborhoods. 
Now, let us return to our problem of
conducting group level inference in situations where we know that the
irrepresentable-type conditions are violated.  Differentially present 
edges in one group of subjects can change the network structure in a
manner that graph selection errors are more likely to occur in one
group.  Thus, these 
group-level estimates will be biased.
Following our procedure, biased group-level edge probability estimates
will then bias test statistics and lead to a higher probability of
false positives or false negatives for group-level  inference.

\begin{figure}[!t]
\centering{
\subfloat[Group-Level False Negative at (1,2).]{ 
\includegraphics[width=.48\textwidth]{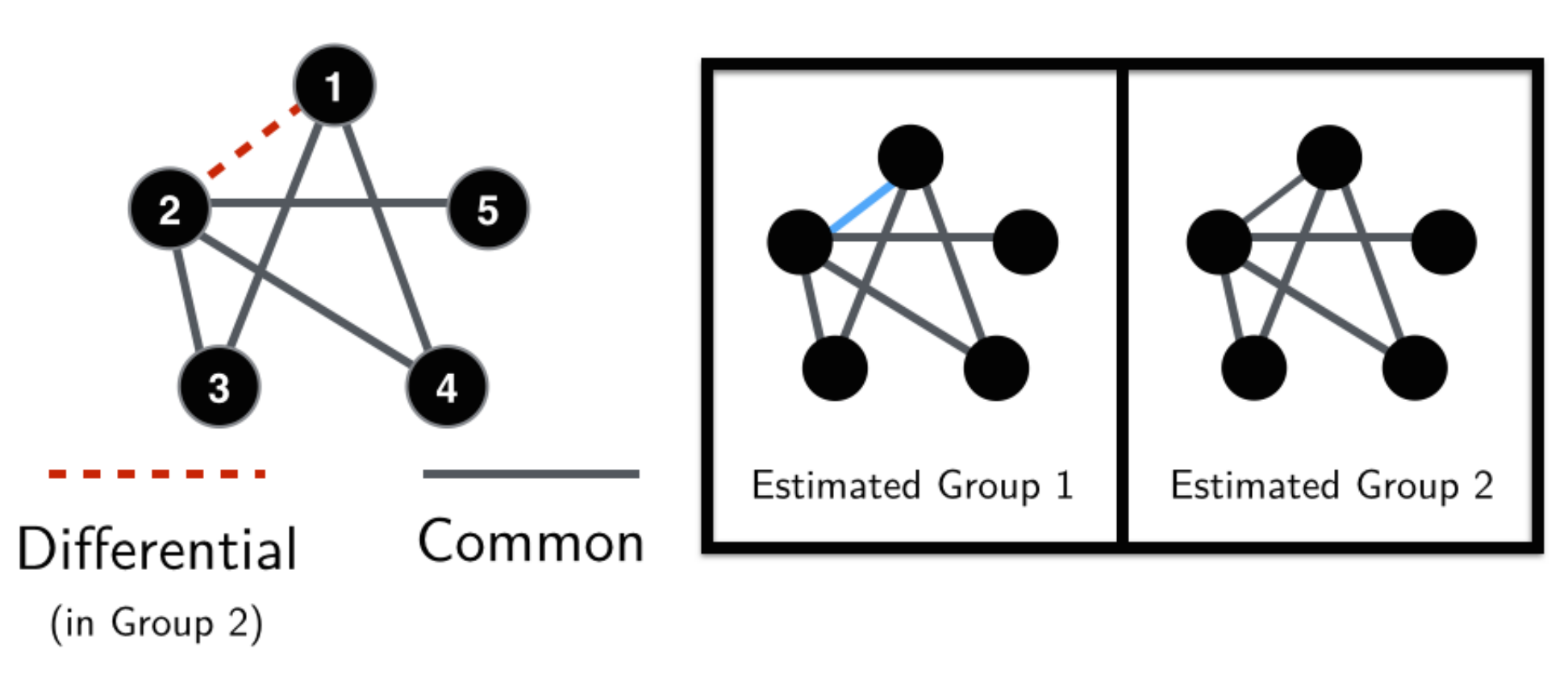}}
\subfloat[Group-Level False Positive at (4,5).]{
\includegraphics[width=.48\textwidth]{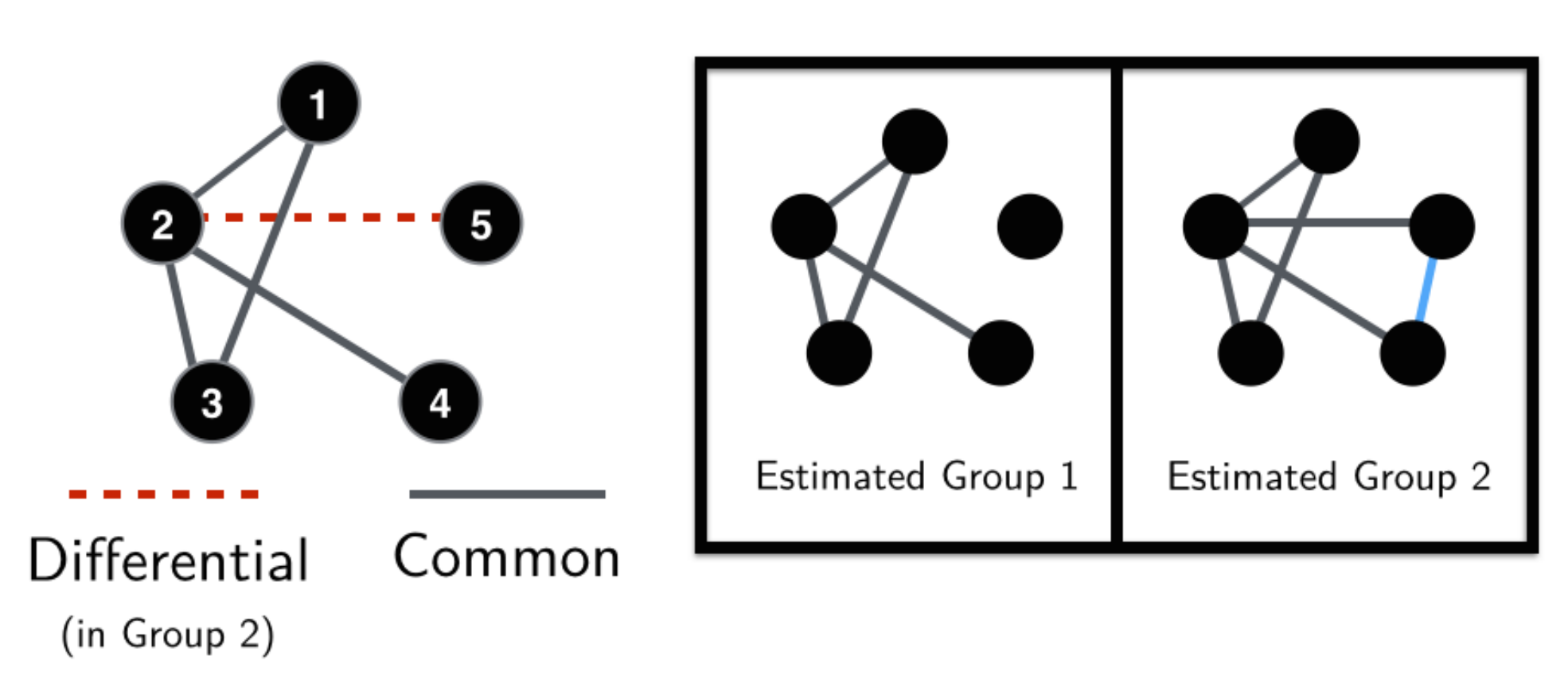}}
\caption{ Illustration of group level biases stemming from graph
selection errors that would result in false negatives (a) and false
positives (b) for group inference. In each figure, true
graphs for a simple 5-node network are given on the left and estimated
graphs on the right.  In (a), an edge is likely to be selected at
(1,2) in group one, resulting in a bias that would yield a false
negative at the group-level.  In (b), an edge is likely to be selected at
(4,5) in group two, resulting in a bias that would yield a false
positive at the group-level.  
\protect\label{fig:toy}}}
\end{figure}

To better understand this, we offer a small illustration in
Figure~\ref{fig:toy}.  For simplicity, we
assume that that the 
group-level probabilities for each edge in \eqref{our_model} are $\{0,1\}$,
meaning that we assume all subjects within a group share the same
network structure.  First in the left panel or Figure~\ref{fig:toy}
(a), we assume that all subjects in the population share the common
edges in black, but that subjects in group two have a differentially
present edge connecting (1,2).  Since nodes 1 and 2 share common
node-neighborhoods, an edge between (1,2) is selected with high
probability in both group 1 
and group 2 subjects.  The group 1 estimate of edge probability (1,2)
will then be biased and lead to a false negative 
when conducting inference across the groups.  Similarly in the left
panel of Figure~\ref{fig:toy} (b), all subjects in group two have an
additional differential edge connecting (2,5).  Unlike in group 1,
when (2,5) are 
connected in group 2, nodes 4 and 5 are also highly correlated due to
common node neighborhoods. 
Thus graph selection in group 2 will estimate an edge at (4,5) with
high-probability, whereas graph
selection will be more likely to estimate the correct network in group
1.  This results in a biased estimate for edge (4,5) in group two,
leading to a false positive at (4,5) when conducting inference at
the group level. 
Thus even for simple graph structures, the location of differentially
present edges in the network structure can lead to graph selection
errors that bias group-level estimates and lead to false positives
and false negatives for group-level 
inference.  With more complicated network structures, this problem
will be further exacerbated.

In general, group-level biases in the edge probability estimates will
occur when graph selection consistency does not hold for each
subject.  It is then difficult to control the overall error rates of
any inferential 
procedure at the group level. Analogous to standard irrepresentability
conditions, we conjecture that there exists irrepresentability-like
conditions for our problem \eqref{edge_prob_test}, that limit 
the correlation between differential and non-differential edges of the
graph. That is, differentially present edges cannot be too correlated 
with common edges (as illustrated in Figure~\ref{fig:toy} (a)) and
differentially present edges cannot be too correlated with non-edges
(as illustrated in Figure~\ref{fig:toy} (b)). While proving such
conditions is beyond 
the scope of this paper, we explore these empirically in
Section~\ref{section:sims}.  Note that as we expect large violations
of irrepresentable-like conditions with real fMRI data, it may be
unrealistic to expect that this problem can be fully solved and
error rates properly controlled.  However, we would expect that any
method that weakens irrepresentability conditions for graph estimation
at the subject level will ameliorate biases in group-level edge
estimates and lead to an inferential procedure with both
higher statistical power and a lower false positive rate.

\section{The \protect $R^{3}$ Method}\label{section:methods}
 
We develop a novel procedure to conduct two-sample inference for our problem \eqref{edge_prob_test}, 
namely, testing for the differential presence / absence of edges across a
population of networks.  Our approach integrates the network
estimation and inference problems to address the two popPSI challenges
outlined in Section~\ref{section:challenge}.  To achieve this, we employ three
key ingredients - resampling, random penalization, and random
effects; hence we call our procedure $R^{3}$.

In this section, we briefly discuss each of the components of the
$R^{3}$ procedure 
separately before putting them all together in Section~\ref{subsec:R3}. 
As discussed in Section~\ref{subsec:challenge1}, we use two level
models at the edge level to account for estimation variability as well
as between 
subject variability of networks. However, we only observe one network
per subject. In the absence of multiple networks per subject, we use
bootstrap resampling 
to generate network replicates for each subject,
Section~\ref{subsec:RS}.  We then use a 
beta-binomial model to model the two-level edge probabilities and
employ a beta-binomial two-sample random effects test statistic to
aggregate our edge statistics over the two levels, Section~\ref{subsec:RE}.
Thus, the 
resampling plus random effects portion of our procedure solves the
first popPSI challenge.  The second popPSI challenge of graph
selection errors that bias edge probability estimates is more difficult to
directly solve.  We can dramatically ameliorate the affect of these
errors, however, by using 
a technique introduced by \protect\citenb{Meinshausen:2010ve} - random
regularization penalties,  
discussed in Section~\ref{subsec:RP}.  Random penalties in conjunction
with our 
resampling procedure thus address the second popPSI challenge.  As
subject-level network estimation is integral to the entire $R^{3}$
procedure, we begin by discussing how we estimate each functional
brain network.

\subsection{Preliminaries: Subject Graph Estimation}\label{subsec:graph_est}

Our proposed $R^3$ framework is compatible with any graph selection procedure
for Gaussian graphical models. A popular method for estimating inverse
covariances is the so-called {\em graphical lasso} or penalized maximum
likelihood method, \cite{dAspremont:2006uq,Friedman:2008ys},
which solves the following objective: 
\begin{align} \label{eqn:pmle-objective}
\hat{\Theta}^{\inds}_{\lambda_{i}}(X^{(\inds)}) 
	&= \arg\min_{{\Theta} \succ 0}\ \mathcal{L}(\hat{\Sigma}^{\inds};{\Theta}) + \lambda_{\inds}\|\Theta\|_{1,\text{off}}
	 = \arg\min_{{\Theta} \succ 0}  \Tr(\hat{\Sigma}^{\inds}\Theta) - \log\det(\Theta) + \lambda_{\inds}\|\Theta\|_{1,\text{off}}
\end{align}
where $\hat{\Sigma}^{\inds}$  is the empirical sample covariance,
$ \hat{\Sigma}^{\inds} = \frac{1}{T}(X^{(\inds)\Tp}X^{(\inds)})$ and
$\|\Theta\|_{1,\text{off}} = \sum_{k<l}|\theta_{k,l}|$ is the $\ell_1$
penalty on off-diagonals. 
Other estimation procedures such as the neighborhood
selection of \citenb{meinshausen2006high} or the CLIME estimator
of \citenb{cai2011constrained} could also be employed. In this paper,
we obtain $\hat{\Theta}$ using the QuIC implementation
by \citenb{Hsieh:2011fj}. From hereon, we 
denote the presence of the $(k,l)$ edge, selected by any graph estimation
procedure as: 
\begin{align}\label{eqn:graph_statistic}
Y^{\inds}_{(k,l)}|\hat{\Theta}^{(\inds)}_{\lambda_{\inds},(k,l)}(\design^{(\inds)})
		& = \mathbb{I}(
			\hat{\Theta}^{(\inds)}_{\lambda_{\inds},(k,l)}(\design^{(\inds)}) \ne 0 ) 	
\end{align}

While inverse covariance estimation assumes that $X^{\inds}$ consists
of independent observations from the multivariate normal,
resting-state fMRI data consists of dependent observations. Thus,
neuroimaging data effectively consists of fewer than $T$ independent
observations and is often well-described by an autoregressive
process \cite{worsley2002general}. Hence, we first use an
autoregressive model to whiten the time series 
and the Llung-Box test to verify that whitened observations are
independent before applying graph selection procedures.

Notice, that we also need to estimate the regularization
parameter, $\lambda_{\inds}$, controlling the graph sparsity for each
subject. In the standard approach, sparsity
levels are typically fixed across all subject
networks \cite{Bullmore:2009vn}. As our 
procedure tests for differential sparsity, however, we cannot enforce
identical graph sparsity for each subject.  Hence, we need a good
initial estimate of $\lambda_{\inds}$.  While there are several model
selection procedures proposed for graph
selection, we employ the StARS procedure of \cite{Liu:2010uq}.

\subsection{$R^{3}$: Resampling}\label{subsec:RS}

\renewcommand{\Sel}{\mu}

Recall that, as discussed in Section~\ref{subsec:challenge1}, one of the
challenges with popPSI is accounting for two-levels of network
variability when we typically obtain only one network estimate per
subject.  We address this by using resampling, specifically
bootstrapping \cite{Efron:1993fk}, to obtain both a better estimate of
the network and its variability.  We also note that as we discuss in
the next section, resampling will also be critical in addressing the
second popPSI challenge.

For each subject $i$, we sample $T$ out of $T$
observations with replacement yielding the bootstrapped data,
$\design^{*\indb,(\inds)}$.  We then apply a graph selection
procedure to this bootstrapped data which gives us the bootstrapped
edge selection statistic
$Y^{*b,\inds}_{(k,l)}|\hat{\Theta}^{*b,(\inds)}_{\lambda_{\inds},(k,l)}(\design^{*b,(\inds)})$.
While we could estimate the edge-level probability for each subject by
$\tilde{\mu}^{(i)}_{(k,l)} =  Y^{\inds}_{(k,l)}$, we could also use
bootstrap aggregation \cite{breiman1996heuristics} yielding
$\hat{\mu}^{(i)}_{(k,l)} = 
= \displaystyle \frac{1}{B}\sum_{b=1}^{B} Y^{*b,\inds}_{(k,l)}$.  Many
have recently shown the benefits of using resampling for graph
selection with error
control \cite{Bach:2008pf,Meinshausen:2010ve,Li:2011pu,Liu:2010uq}.
Thus, we prefer the resampled statistic $\hat{\mu}^{\inds}$ to
$\tilde{\Sel}^{(\inds)}$.  Although we cannot expect our estimate to
be unbiased for 
$\mu^{\inds}_{(k,l)}$ in high-dimensional settings or for highly
connected network structures (as discussed in
Section~\ref{subsec:challenge2}), \citenb{Meinshausen:2010ve} and \citenb{Shah:2013lv}
have shown that stability 
based statistics more effectively separate true and false edges.
For our $R^{3}$ procedure, we will use resampling to not only improve
estimation of edge selection probabilities, but also to estimate
variability for two-level random effects models and with random
penalization procedures as discussed subsequently.

\subsection{$R^{3}$: Random Penalization}\label{subsec:RP}

As discussed in Section~\ref{subsec:challenge2}, graph selection errors can bias
estimates of edge selection probabilities which in turn lead to errors
when conducting inference at the group level.  For real fMRI data with
limited samples $T$ and highly connected network structures that
violate irrepresentable-type conditions, we will
likely never be able to fully solve the problems induced by graph
selection errors.  Here, we try to ameliorate their affect by
employing random penalization techniques recently introduced
by \citenb{Meinshausen:2010ve}.  
For each bootstrap sample $\indb = 1,\ldots, B $, we generate a 
$p \times p$ symmetric matrix of regularization parameters that
randomly penalizes 
each edge, denoted $\Lambda^{\indb\inds}$.  We employ random penalization
that modifies the objective, \eqref{eqn:pmle-objective}, through an
element-wise weighted penalty:  
\begin{align}
    \hat{\Theta}^{\inds}_{\Lambda^{\indb\inds}} = \arg\min_{\Theta \succ 0}\left( - 2\mathcal{L}(\hat{\Sigma}^{\indb\inds},\Theta) + \|\Lambda^{\indb\inds}\circ \Theta \|_1 \right)
\end{align}
where  $\circ$ is the element-wise Hadamard product.  Our matrix of
random penalties, $\Lambda^{\indb \inds}$, is obtained by
perturbing the initial pilot estimate of the regularization parameter
for each subject, $\lambda^{\inds}$, as follows:
\begin{align} \label{eq:randreg}
     \Lambda^{\indb\inds}_{\indeK\indeL} &= \lambda^{\inds} +
      c \ \lambda^{i}_{max}W_{k,l}  \ \ \ \forall \ \ k < l
\end{align}
where $\Pr\{W_{k,l} =\pm 1\} = \frac{1}{2}$, 
$c \in (0,.5)$ is fixed as a small fraction, and $\lambda^{i}_{\text{max}}$ is
the regularization parameter for each subject that results in the
fully sparse graph.  Thus, our random perturbation procedure can be
seen to penalize each edge independently as $\lambda \pm
c \lambda_{max}$; for our purposes, we have found that $c = .25$ performs well. Note that our
random penalties are different than the conservative scheme proposed
by \citenb{Meinshausen:2010ve} for the purpose of controlling false
positive edge selection. Other alternatives such as using $c \sim
U(0, .5)$ are also possible, and are closely related to the procedure of \citenb{Li:2011pu} who aggregate edge selection frequencies over a
range of perturbations of $\lambda$.

Intuitively, our randomized regularization scheme decreases the
influence of the inclusion or exclusion of any given edge on the
selection of other edges.  Thus, we expect our approach to improve the
problems associated with graph selection discussed in Section~\ref{subsec:graph_est}.
In fact, several have recently shown that restricted eigenvalue and
irrepresentability-type conditions can be violated for the original
data, but hold when aggregating selection over random
penalizations \cite{Meinshausen:2010ve,buhlmann2011statistics}.
Hence, with random penalization, consistent graph selection can be
achieved while tolerating larger correlations between
variables.  For our popPSI problem, we expect that random penalization
will allow us to tolerate more correlation between 
differential edges and common edges, and differential
edges with non-edges.  We empirically study this intuition through
simulations in Section~\ref{section:sims}.

\subsection{$R^{3}$: Random Effects}\label{subsec:RE}

Recall that in Section~\ref{subsec:challenge1} we introduced a Beta-Binomial
model, \eqref{eqn:beta_binomial}, to 
account for the two sources of network variability at the edge level.
With only one estimated network, however, estimating two levels of
variability and fitting the Beta-Binomial model was not possible.
Now, we can use our bootstrap resampled data to properly fit the
Beta-Binomial model and obtain the corresponding two-sample test
statistics for each edge.  For each subject $i$ and each edge $(k,l)$,
we obtain $B$ resampled edge statistics, $Y^{b,i}_{(k,l)}$.
Suppressing the edge indices for notational simplicity, we then have
that $Z^{*,i} = \frac{1}{B}\sum_{b=1}^{B} Y^{b,i}$ is our statistic
associated with the subject edge probability, $\mu_{i}$.  Hence, we
can re-write the Beta-Binomial model given in \eqref{eqn:beta_binomial}
for our bootstrapped statistics:
\begin{align} \label{eqn:beta_binomial_boot}
Z^{*,i}| \mu^{(i)} & \overset{iid}{\sim} \text{Bin}(\mu^{(i)},B), \quad \ 
\mu^{(i)} \overset{iid}{\sim} \text{Beta}(\pi^{(g)},\rho^{(g)}).
\end{align}
Recall that the Beta-Binomial model is often used for over-dispersed
or group-correlated binary data \cite{crowder1978beta}.  Our bootstrapped edge
statistics over the subjects certainly fit this model as bootstrapping
results in positively correlated statistics within each
subject \cite{bickel2012resampling}.  As previously noted, this model also nicely captures
the variability of edge support both within and between subjects.

We propose to fit our Beta-Binomial model via the widely used moment
estimators for $\pi$ and $\rho$ \cite{Kleinman:1973fk}.  For estimation, assume that
we will always have a balanced number of bootstrap samples per
subject.  Then estimates for $\pi$ and $\rho$ as proposed by \citenb{Kleinman:1973fk,Ridout:1999mg}
are as follows: 
\renewcommand{\indb}{b}
\begin{align}\label{eqn:betabin-est1}
\hat{\pi}_{\indg} = \frac{1}{n_{\indg}}\sum_{i \in \mathcal{G}_{g}} Z^{*,i} \quad \& 
\quad \hat{\rho}^{\indg} = \frac{B}{B-1}\frac{\sum_{i \in \mathcal{G}_{g}}(\hat{\pi}^{\indg}
- Z^{*,i})^2}{\hat{\pi}^{\indg}(1-\hat{\pi}^{\indg})(n_1-1)}
- \frac{1}{B-1}.  
\end{align}
These estimators are consistent for $\pi$ and
$\rho$ \citep{moore1986asymptotic} and are 
asymptotically normal \citep{Kleinman:1973fk}.  For balanced data such
as in our case, these estimators exhibit only mild loss of efficiency
compared to more commonly used likelihood-based
estimators \cite{Kleinman:1973fk}.  We 
choose to employ these estimators, however, as they are very simple to
compute and widely used when conducting inference on $\pi$ as in our
problem.  Specifically for inference on $\pi$, it is well-known in the
teratological literature that failure to account for $\rho$ results in 
inflated Type I error rates \citep{weil1970selection,liang1994use},
but this inference has also been shown to be robust to various
estimators for $\rho$ \citep{moore1986asymptotic}.
Further, many have shown that for balanced data as in our case, the
moment estimators for $\pi$ and $\rho$ give empirical Type I error
control when conducting inference on
$\pi$ \citep{Ridout:1999mg,liang1994use,liang1996asymptotic}.  Given
this wide literature, we thus opt to use the computationally simpler moment 
estimators \eqref{eqn:betabin-est1} to fit our Beta-Binomial model.

With these estimators, we develop a two-sample Wald test statistic
appropriate for our 
hypothesis \eqref{edge_prob_test}.  To this end, we need an estimate of the sampling
variance of $\hat{\pi}^{g}$.  Following
from \eqref{eqn:unconditional_var} and using our 
estimates of $\hat{\pi}^{g}$ and $\hat{\rho}^{g}$, we can easily see
that an estimate of the variance of $\hat{\pi}^{g}$ is given by:
\begin{align*}
s^2_{\indg}
= \frac{\hat{\pi}^{\indg}(1-\hat{\pi}^{\indg})}{m(n_{\indg}-1)}(1+(m-1)\hat{\rho}^{\indg}). 
\end{align*}
Putting everything together, we then arrive at the following
two-sample Wald test-statistic for our problem \eqref{edge_prob_test}:
\begin{align} \label{eqn:bbd_test}
\mathcal{T} = \frac{\hat{\pi}^{A} - \hat{\pi}^{B}}{\hat{\text{se}}(\hat{\pi}^{B}-\hat{\pi}^{B})} = \frac{\hat{\pi}^{A} - \hat{\pi}^{B}}
{\sqrt{\frac{s^2_{A}(n_{A}-1)}{n_{A}} + \frac{s^2_{B}(n_{B}-1)}{n_{B}}}}.
\end{align}
Following from \citet{Kleinman:1973fk}, this test statistic is asymptotically
standard normal as $n_A$ and $n_B \rightarrow \infty$.  In fMRI
studies, however, our sample sizes are typically small.  Thus, we
favor comparing our test statistic $\mathcal{T}$ to a permutation null
distribution to obtain $p$-values \cite{Janssen:1997ve,nichols2002nonparametric}.

\subsection{The $R^{3}$ Procedure}\label{subsec:R3}

\renewcommand{\TS}{\mathcal{T}}
\renewcommand{\indb}{b}
\renewcommand{\inde}{(k,l)}

\begin{alg}[!ht] 
\renewcommand{\arraystretch}{.5}
\normalsize
\caption{ \protect $R^3 := $ Resampling,  Random Penalization and
Random Effects Procedure \protect\label{alg:r3}}      

     \begin{enumerate}

     \item For each subject, $\inds = 1, \ldots n$, obtain pilot estimates of the regularization parameter $\lambda_{\inds}$. (Section~\ref{subsec:graph_est})

     \item RESAMPLING AND RANDOM PENALIZATION: 
     \item[] For $b=1,\ldots B$:
     \begin{enumerate}
     \item Bootstrap data yielding $\X^{*\inds, \indb}$.
     \item Fit weighted graphical lasso using random penalty matrix
     in Eq.~\eqref{eq:randreg} giving $\hat{\Thet}( \X^{*i,b})$.  (Section~\ref{subsec:RP})
     \item Record edge support statistics $Y^{* \inds, \indb}_{k,l} =
     I( \hat{\theta}^{*,i,b}_{(k,l)} \neq 0)$.
     \end{enumerate}
     \item[] End.

     \item EDGE FILTERING: Eliminate edges absent from both groups
     from consideration, giving the set $\mathcal{E}_{F}$ for
     testing. (Section~\ref{subsec:R3}) 

     \item INFERENCE:
     \begin{enumerate}
     \item[] For $(k,l) \in \mathcal{E}_{F}$:
     \begin{enumerate}
     \item Compute test statistics $\TS_{\inde}$ as in Eq.~\eqref{eqn:bbd_test}
     (Section~\ref{subsec:RE}) 
     \item Calculate p-values using a permutation mull distribution
     for $\TS_{\inde}$.  
     \end{enumerate}
       \item[] End.
     \item[iii] Correct for multiplicity via the Benjamini-Yekutieli procedure.
     \end{enumerate}
     \end{enumerate}

\end{alg}

Now, we are ready to put our whole $R^{3}$ procedure together.  We
outline our procedure in Algorithm~\ref{alg:r3} for conducting
 inference on the differential presence of all edges in a
population of graphical models, \eqref{edge_prob_test}.  Note that testing all edges
would result in an ultra-large-scale inference problem as there would be
${p \choose 2}$ hypotheses tested.  This is clearly ill-advised;
especially so since for brain connectivity networks, we expect rather
sparse networks meaning that most edges are absent from both
population groups.  Thus, we limit our consideration to only the edges
that are present in at least one of the population groups:  
$$\mathcal{E}^{c}_{F} \triangleq \{(k,l): Z^{*i,}_{(k,l)} \le  B \tau
, \ \forall \ \inds = 1, \ldots , n\}$$  
We filter out all edges that have edge proportions less than
$\tau$ for all subjects, leaving our filtered edge set,
$\mathcal{E}_{F}$. Notice that filtering is agnostic to group
membership and thus does not affect group-level inference.  We suggest
taking $\tau \in (.2,.5)$ which typically reduces the number of edges
under consideration from thousands to hundreds for real fMRI data.
Additionally, we must correct for multiple testing.  As our test
statistics will be highly dependent, we suggest using the
Benjamini-Yekutieli procedure \cite{Benjamini:2001hl} which controls
the false discovery rate under arbitrary dependencies.  Finally, we
note that instead of testing all edges, our procedure could also be
used to test targeted hypotheses regarding specific edges.

\section{Simulation Studies}\label{section:sims}

\renewcommand{\CSupp}{C}

We study our $R^{3}$ procedure through a series of simulations,
showing that $R^3$ substantially improves statistical power and
error control for our popPSI problem, \eqref{edge_prob_test}.  We will
particularly study how our method and the standard approach 
address each of the challenges outlined in
Section~\ref{section:challenge}. 

\subsection{Simulation Setup}

Henceforth, we will denote the
components of $R^3$ as Resampling (RS), Random Penalization (RP) and
Random Effects (RE). 
To better understand how each challenge outlined in
Section~\ref{section:challenge} as well as the our methodological
solutions to these challenges affect 
inferential procedures, we compare $R^3$ not only to the standard
approach but also variations of our own method: $R^2$ with (RS, RP)
and $R^2$ with (RS,RE). Recall from Section~\ref{section:challenge},
that the standard approach uses two-sample test statistics associated
with the one-level Binomial distribution.  Both the numerator and
denominator of this test statistic are incorrect, with the mean group
level parameters biased by graph selection errors (Challenge II in 
Section~\ref{subsec:challenge2}) and with the
denominator under-estimating the variance components associated with
two levels of network variability (Challenge I in Section~\ref{subsec:challenge1}).
Our first variant,  $R^2 =$ 
(RS, RP), seeks to address only Challenge II by ameliorating the bias
in group-level edge proportions using random penalization. Our second
variant,  $R^2 =$ (RS, RE) seeks to address only Challenge I by using
the correct two-level Beta-Binomial model and test statistics.  
We adopt the same specifications outlined in
Section~\ref{subsec:R3} with $\lambda_{\inds}$ selected using
StARS \citep{Liu:2010uq} for all methods. Methods including the RE
component use random effect statistics 
from Section~\ref{subsec:RE}, while those without RE use the standard
two-sample binomial proportions test as in
Section~\ref{subsec:challenge1}. We control FDR at 10\% for all
methods using the Benjamini-Yekutieli approach \citep{Benjamini:2001hl}. 

We study several simulation scenarios to fully test our methods.
First, we generate multivariate observations, $X^{(\inds)}_{T
  \times p}$, for each subject according to
$\mathcal{N}(0,(\Theta^{(\inds)})^{-1})$.  We simulate the strength
of connections for all edges as $\theta^{\inds}_{(k,l)} \sim
Uniform([-1.25, -1] \cup [\hphantom{-}1,\hphantom{-}1.25])$, and then
add a sufficient amount to the eigenvalues of $\Thet^{i}$ to ensure
positive definiteness.  Each group consists of a
balanced number of subjects, $n_1=20$ and $n_2 = 20$, and we consider
a moderate dimensional case with $p=50, T=400$; we set approximately 150 edges to be
differentially present and evenly divide these between the two
groups.  Second, as functional
connectivity is 
known to exhibit small world and hub-like network structures
\cite{achard2006resilient}, our simulated network models follow a
challenging banded, small world, or hub-like structure.  Third, as 
the location of common and differential edges in the 
population network structure can lead to bias in the group-level edge
estimates (discussed in Section~\ref{subsec:challenge2}), we set the
location of 
differential edges to follow two schemes, referred to as Case I and
Case II.  In Case I, we consider \emph{Clustered Differential Edges}
in which differential edges in one group are highly
correlated with other differential edges as well as common
edges.   In Case II, we consider \emph{Random Differential Edges}
where the differential edges occur at random
throughout the network structure.  Thus, we expect Case I to violate
our conjectured irrepresentable-type 
conditions under which unbiased estimation of the edge probability
and hence overall error control of the inferential procedures is
achievable; Case II should ameliorate these conditions.  
Combining graph types for each of these cases results in a total of
six simulations.  To simplify our investigation into these six
simulation scenarios (results shown in Figure 4 and Table 1), we set
$\pi^{g}_{(k,l)} = 1$ for all edges.

We investigate changing
$\pi^{g}$ for differentially present edges by setting $\pi^{g}_{(k,l)} =
\{1, .5 , .3\}$ in a separate simulation (Figure 5) for the banded and
hub-type graphs for Case II type differential edges.   For this
simulation, we use population
correlations rather than covariances in order to eliminate variations
in scale across subjects. Additionally for fair
comparisons here, we limit the number of differential edges to 25 per group and
fix the degree of the common support to be $0.12p$.

Further simulations studies as well as additional supporting material
for our simulations can be found in the supplementary
materials \citep{Narayan:2014siSupp}.  These include confusion
adjacency matrices showing the location of false positives and false
negatives in the network structure for all methods, an analogous
simulation study in a high-dimensional setting $p>T$, and a study of
the impact of graph sparsity for both common and differential edges on
our methods.

\begin{figure}[!tb]
    \centering{
      \subfloat[Small World Graph, I]{\label{fig:2:smallw}
            \includegraphics[width=.33\textwidth,height=.3\textwidth]{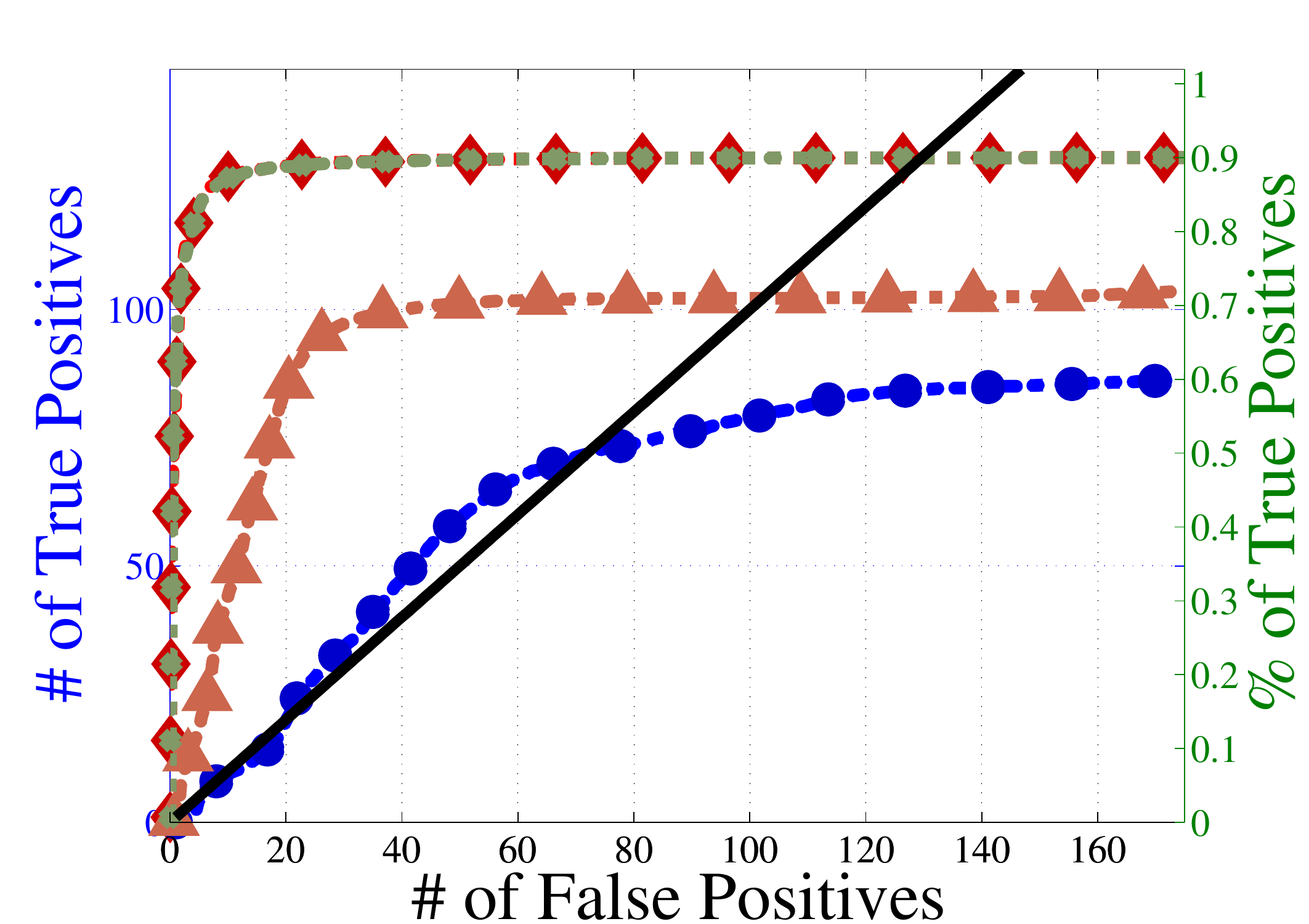}}
            \subfloat[Banded Graph, I]{\label{fig:2:banded}
            \includegraphics[width=.33\textwidth,height=.3\textwidth]{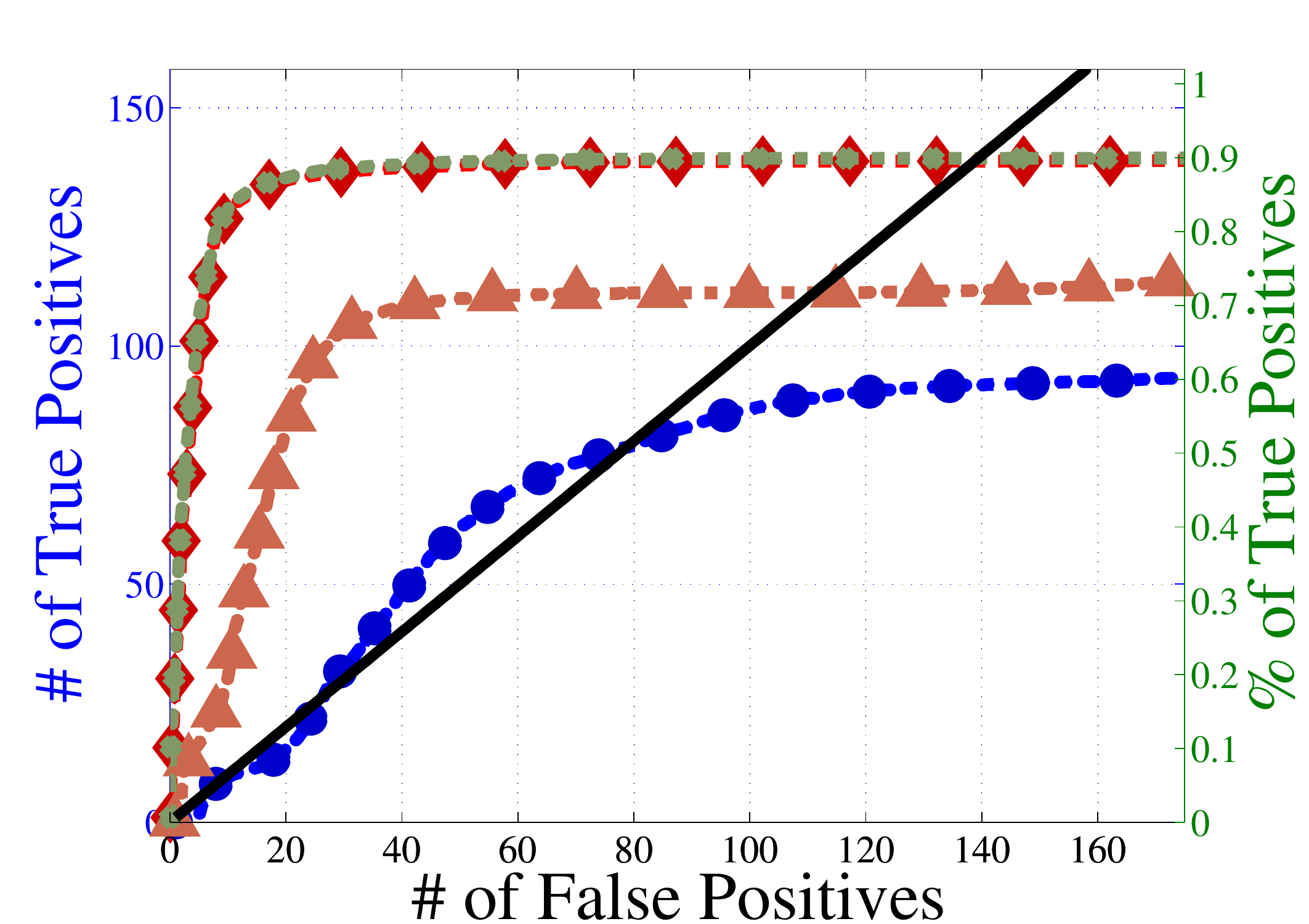}}
            \subfloat[Hub Graph, I]{\label{fig:2:hub}
            \includegraphics[width=.33\textwidth,height=.3\textwidth]{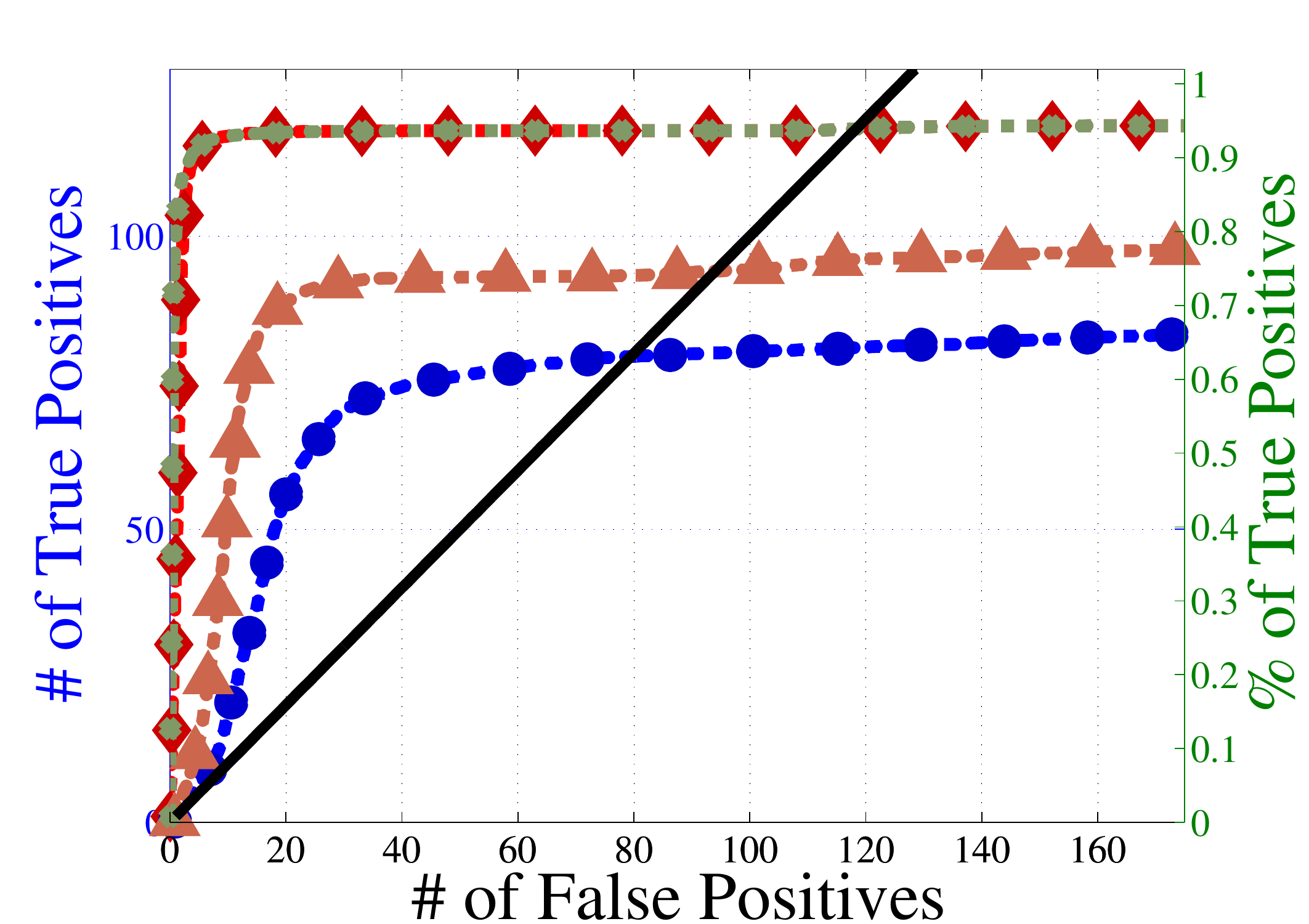}}\\
           \subfloat[Small World Graph, II]{\label{fig:3:smallw}
            \includegraphics[width=.33\textwidth,height=.3\textwidth]{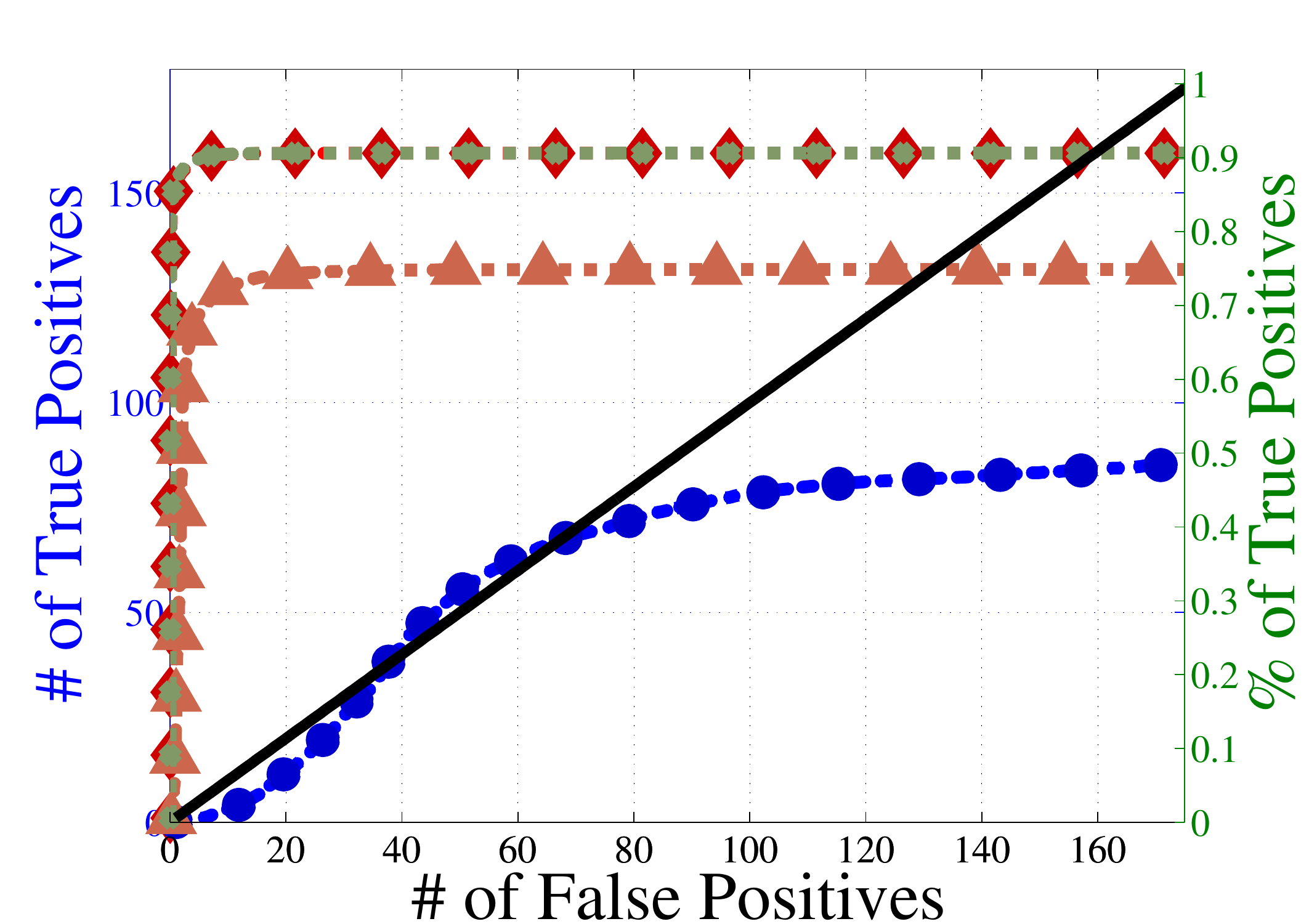}}
            \subfloat[Banded Graph, II]{\label{fig:3:banded}
            \includegraphics[width=.33\textwidth,height=.3\textwidth]{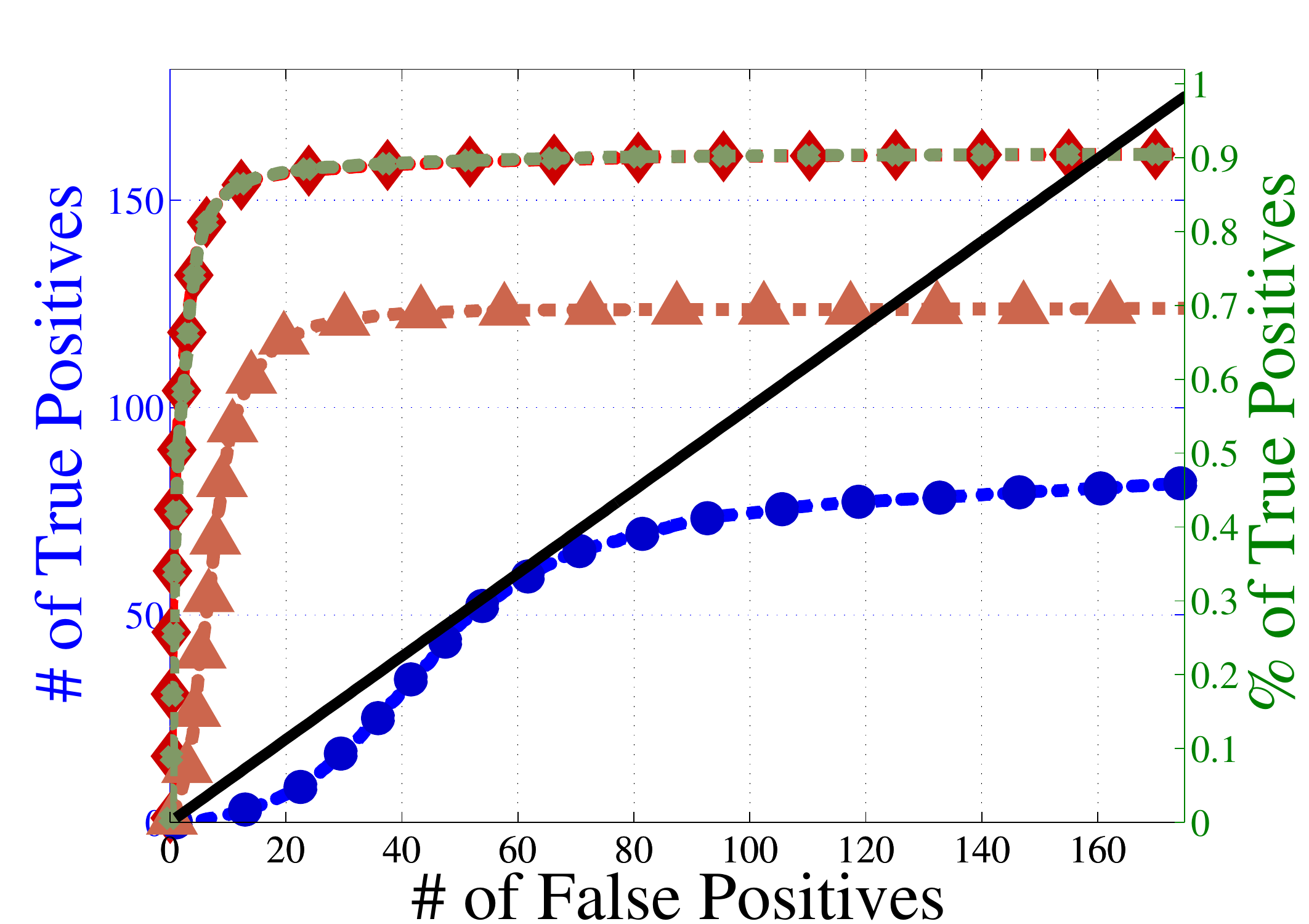}}
            \subfloat[Hub Graph, II]{\label{fig:3:hub}
            \includegraphics[width=.33\textwidth,height=.3\textwidth]{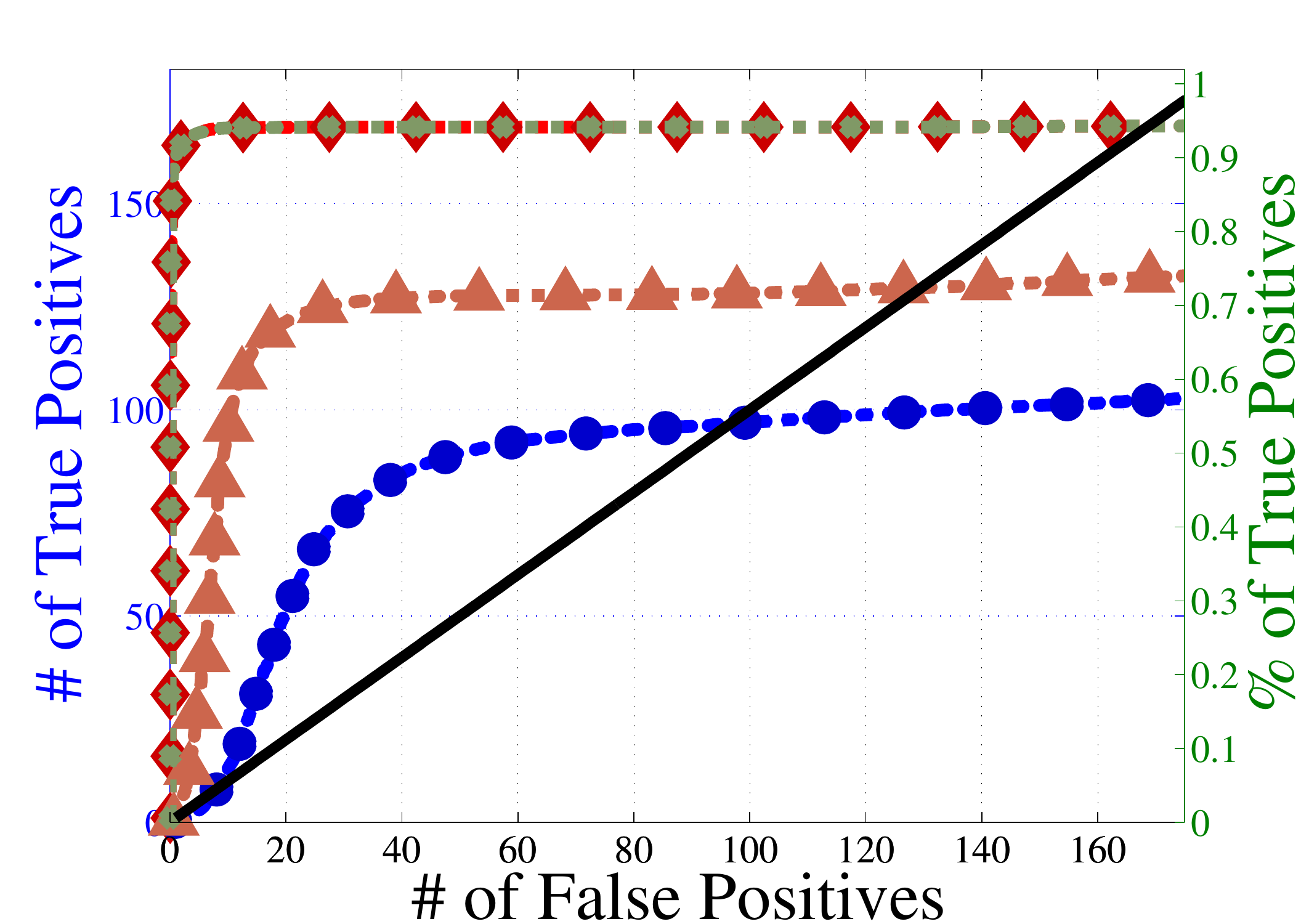}} \\
            \includegraphics[width=\textwidth]{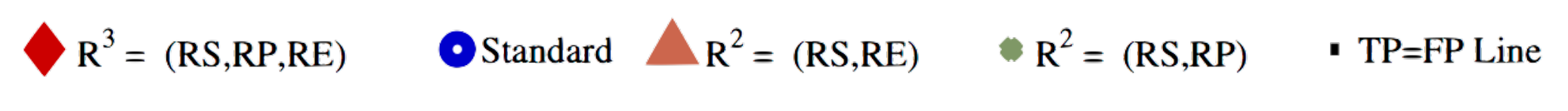}  
  \caption{Average ROC curves for sequentially
    rejected tests comparing our 
    method to the standard approach, $R^{2} = (RS,RE)$, and $R^{2} =
    (RS,RP)$ for each network structure type and Case I and II type
    differential edges.  Methods employing random penalization (RP)
    improve statistical power as they ameliorate graph selection
    errors that bias group-level estimates.  \label{fig3a:CountROCs-lowp}}}
\end{figure}  

\begin{table}[hbt]
  \begin{tabular} {|c|l|c|c >{(}r<{)}|c>{(}r<{)}|c>{(}r<{)}|c>{(}r<{)}|} 
    \hline
      \centering{Case} & \multicolumn{1}{|p{1.3cm}|}{Sim Type} & Metric 
      & \multicolumn{2}{p{2cm}|}{\centering $R^{3}$} & \multicolumn{2}{p{2cm}|}{\centering Standard Test} 
      & \multicolumn{2}{p{2cm}|}{\centering $\Rs$} & \multicolumn{2}{p{1.5cm}|}{\centering $\Rp$} \\
    \hline \hline
    \multirow{2}{*}{\hfil I}
          & \multirow{2}{*}{\hfil SmallW} 
            & \text{TPR}  & 0.934 & 0.036 & 0.542 & 0.096 & 0.739 & 0.082 & 0.937 & 0.037 \\
                          & & \text{$\text{FDP}$}  &  \text{0.245} & 0.079 & 0.661 & 0.028 & 0.476 & 0.040 &\textcolor{black}{ 0.643} & 0.016 \\ \hline 
    \multirow{2}{*}{\hfil I}     
         & \multirow{2}{*}{\hfil Banded}
          & \text{TPR} & 0.921 & 0.069 & 0.524 & 0.105 & 0.735 & 0.091 & 0.933 & 0.038\\
                      & & \text{$\text{FDP}$}   & \text{0.261} & 0.122 & 0.645 & 0.039 & 0.446 & 0.053 & \textcolor{black}{ 0.624} & 0.041 \\ \hline
    \multirow{2}{*}{ \hfil I}     
          & \multirow{2}{*}{\hfil Hub}
                        & \text{TPR} & 0.967 & 0.021 & 0.616 & 0.131 & 0.763 & 0.099 & 0.968 & 0.021 \\
          & & \text{$\text{FDP}$}     & 0.107 & 0.056 & 0.497 & 0.040 & 0.306 & 0.048 & 0.474 & 0.019 \\ \hline
       \centering{Case} & \multicolumn{1}{|p{1.3cm}|}{Sim Type} & Metric 
      & \multicolumn{2}{p{2cm}|}{\centering $R^{3}$} & \multicolumn{2}{p{2cm}|}{\centering Standard Test} 
      & \multicolumn{2}{p{2cm}|}{\centering $\Rs$} & \multicolumn{2}{p{2cm}|}{\centering $\Rp$} \\
    \hline
    \multirow{2}{*}{ \hfil II}
          & \multirow{2}{*}{ \hfil SmallW} 
          & \text{TPR} & 0.959 & 0.026 & 0.483 & 0.097 & 0.792 & 0.071 & 0.959 & 0.026 \\
                              & & \text{$\text{FDP}$}& \text{ 0.112} & 0.079 & 0.659 & 0.042 & 0.370 & 0.037 &\textcolor{black}{ 0.615} & 0.012 \\ \hline
    \multirow{2}{*}{ \hfil II}      
         & \multirow{2}{*}{  \hfil Banded}
            & \text{TPR} & 0.941 & 0.044 & 0.450 & 0.105 & 0.724 & 0.105 & 0.946 & 0.043 \\
                               & & \text{$\text{FDP}$} & \text{ 0.199} & 0.095 & 0.667 & 0.042 & 0.425 & 0.058 & \textcolor{black}{ 0.623} & 0.018 \\ \hline
    \multirow{2}{*}{ \hfil II}      
          & \multirow{2}{*}{ \hfil Hub}
          & \text{TPR}         & 0.971 & 0.023 & 0.533 & 0.133 & 0.735 & 0.106 & 0.972 & 0.023 \\
                  & & \text{$\text{FDP}$}   & 0.051 & 0.029 & 0.463 & 0.046 & 0.262 & 0.042 & 0.406 & 0.009 \\ \hline
    \hline
  \end{tabular}
  \caption{ \label{tab:fdp0} Average true positive
    rate (TPR) and 
    false discovery proportion (FDP) for tests rejected by each method
    when controlling the FDR at 10\% via the Benjamini-Yekutieli
    method; standard errors are given in parentheses.  Methods
    employing random effects models and test statistics yield improved
    Type I error rates.  }
\end{table}

\begin{figure}[!htb]
	\centering{
          \subfloat[Banded Graph, II]{\label{fig:3:banded}
          \includegraphics[width=.45\textwidth]{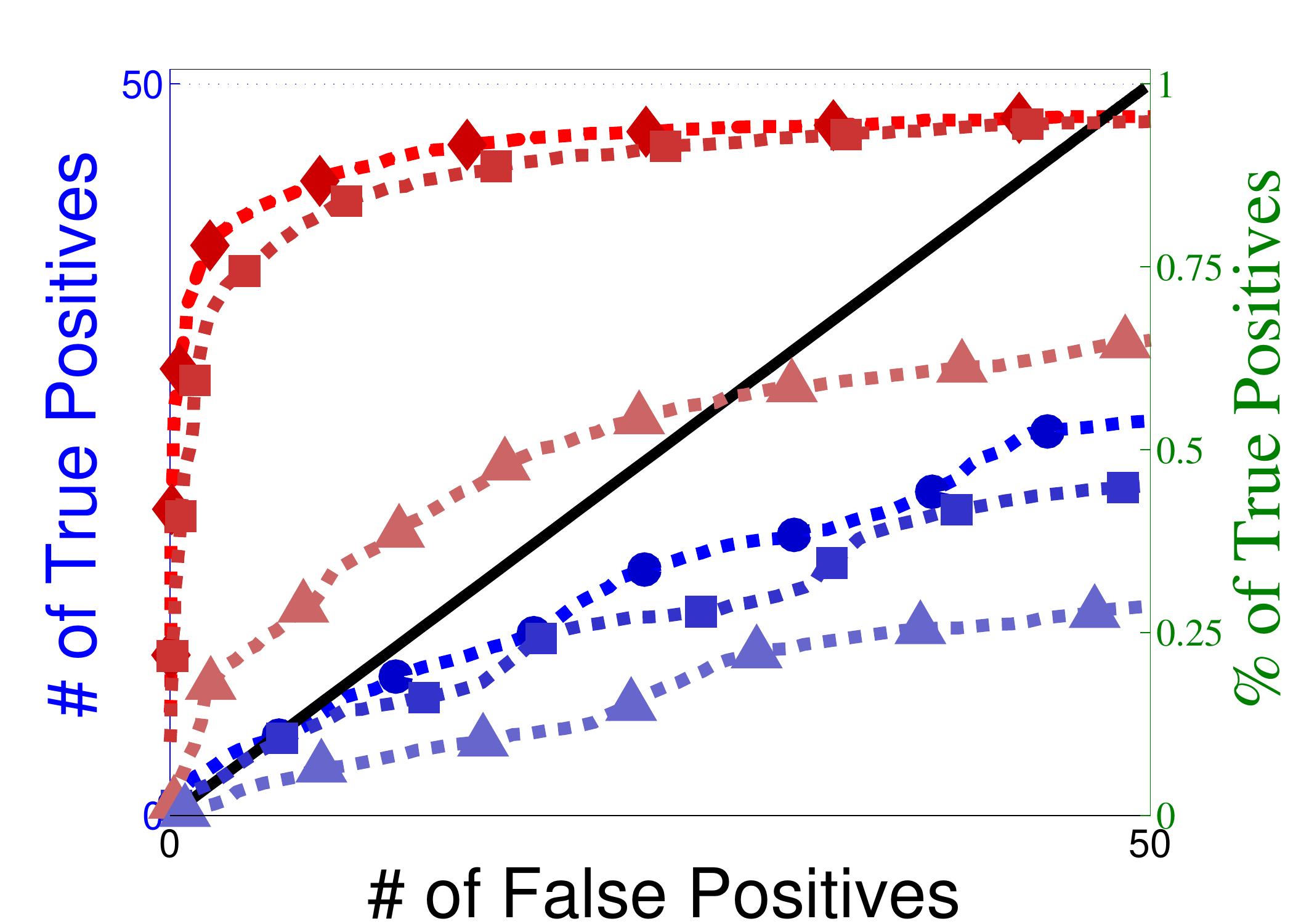}} 
          \subfloat[Hub Graph, II]{\label{fig:3:hub}
          \includegraphics[width=.47\textwidth]{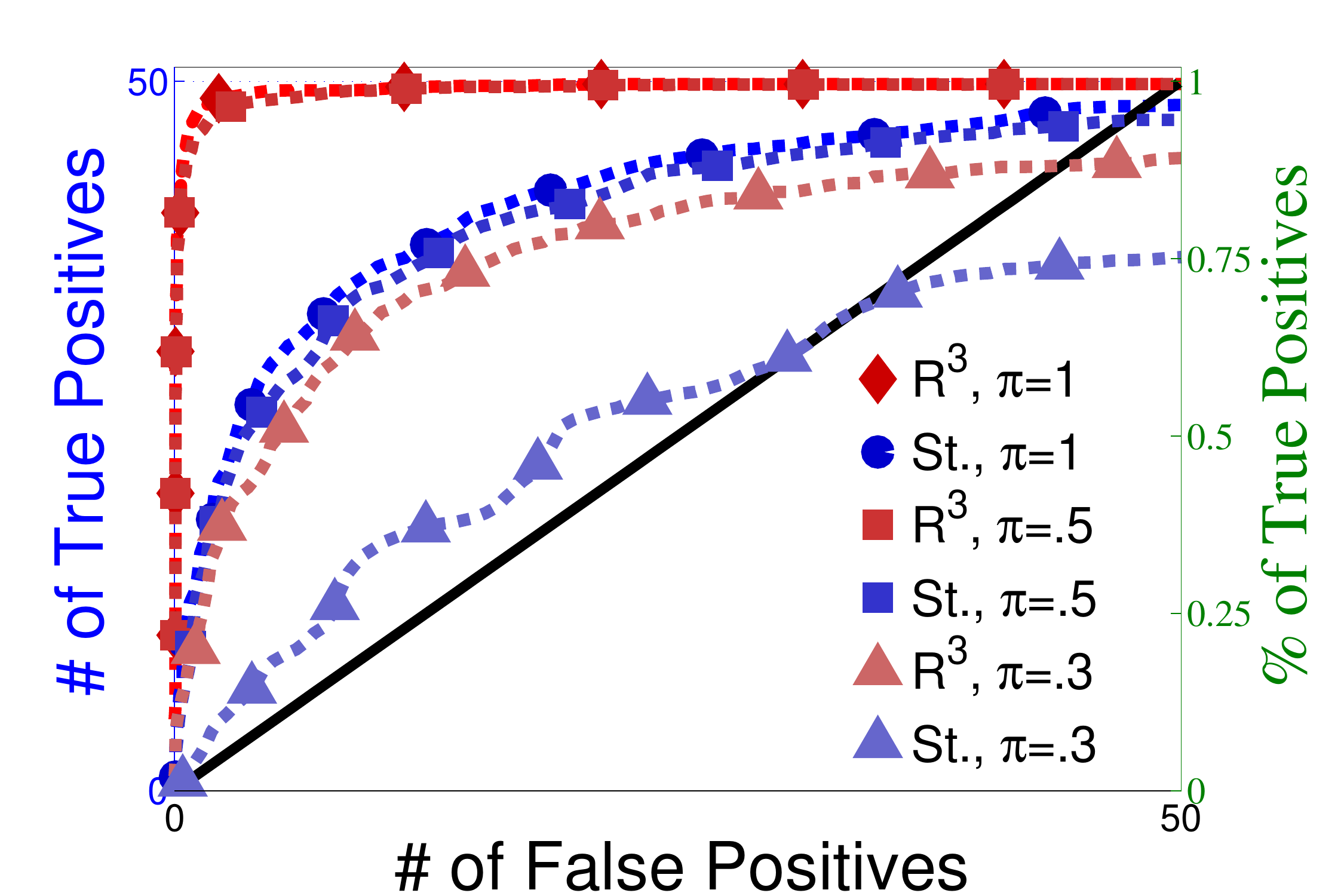}} 
          \caption{Impact of network variability between subjects.
            While decreasing $\pi^{g}$ for the differentially
            present edges, we report average ROC
            curves for sequentially 
            rejected tests for our $R^{3}$ method compared to the standard
            approach for banded and hub-type graphs in Case II
            scenarios.  As the proportion of subjects with
            differential edges decreases, performance degrades but
            $R^{3}$ continues to well outperform the standard approach.
          }\label{fig5:ROC_VarySub_lowp}}
 \end{figure}

\subsection{Results}

In Figure~\ref{fig3a:CountROCs-lowp} and Table~\ref{tab:fdp0}, we
present our main simulation results comparing $R^{3}$ to the two
variations of our $R^{2}$ method and the standard approach for three
network structures and Case I and II type differential edges.  First
for Figure~\ref{fig3a:CountROCs-lowp}, we report results in terms of
operating characteristics 
averaged across 50 replicates with the number of true positives
(y-axis) plotted against the number of false positives (x-axis) for
each test statistic, rejected sequentially from largest to smallest in
absolute magnitude. Overall, all of our methods and particularly
$R^{3}$ yield substantial improvements over the standard approach in
all scenarios.

Notice that both $R^3$ and $R^2$=(RS,RP) share similar orderings of test
statistics, and consequently similar ROC curves.   Overall, methods
that include random penalization yield major improvements in
statistical power over those that do not.  This indicates that the
second popPSI challenge outlined in Section~\ref{subsec:challenge2} is a significant
contributor to the poor performance of the standard method.  Recall our
discussion of how graph selection errors at the subject stage occur
non-randomly and hence bias our group-level estimates of
$\hat{\pi}^{g}$.  Our results empirically demonstrate that random
penalization dramatically improves these biases, leading to less bias
in our test statistics and hence improvements in both Type I and Type II
error rates.  Furthermore, in Case II scenarios where
selection errors are moderate, the performance gap between any method
containing $RP$ over $R^2 = (RS,RE)$ reduces compared to Case I
scenarios where selection errors are more severe. Thus, the benefits of
random penalization are greater when selection errors are more
abundant.  Confusion adjacency matrices illustrating the location of
inferential errors for our methods shown in the supplemental
materials also indicate that random penalization improves graph
selection in cases where there are larger correlations between
differential edges and common edges.  Similar results hold for our
high-dimensional study presented in the supplemental material.

Table~\ref{tab:fdp0}, which accompanies
Figure~\ref{fig3a:CountROCs-lowp}, gives the empirical true positive
and false discovery rates (FDR) averaged over 50 simulation replicates when
the Benjamini-Yekutieli \citep{Benjamini:2001hl} procedure controlling the FDR at 10\%
is used to determine the number of tests to reject.  First, notice
that the observed false discovery proportion (FDP) of our $R^{3}$
procedure is not 10\% on average, indicating that our method does not
fully control the FDR.  This occurs because we specifically simulate
difficult and realistic fMRI scenarios with graph structures that
severely violate irrepresentable-type conditions.  In situations (not
shown) where irrepresentable-type conditions are met that ensure graph
selection consistency, our procedure as well as the standard method
correctly control the FDR.  As discussed in Section~\ref{subsec:challenge2}, in
situations where graph selection errors occur with high probability,
it is likely impossible to provably control the FDR, consistent with
our empirical results.  Yet even though $R^{3}$ does not fully control
the FDR, our error rates are dramatically improved over the standard
approach and other variations of our procedure.

Also in Table~\ref{tab:fdp0}, observe that $R^{2}=$(RS,RP), which had
similarly ordered test statistics to $R^{3}$, has dramatically worse
Type I error rates that do not come close to controlling the FDR.
While $R^{2}=$(RS,RE) also does not control the FDR, the error rates
are much improved over $R^{2}=$(RS,RP).  These results demonstrate
that using two-level models with the correct random effects test
statistics are crucial to Type I error control.  Recall from
Section~\ref{subsec:challenge1}, that using the one-level Binomial model leads to an
under-estimation of the variance term which in turn inflates test
statistics and leads to an increase in false positives.  Note also
that the estimated FDP of $R^{3}$ is still a major improvement over
that of $R^{2}=$(RS,RE).  This occurs as the problem of graph selection
errors induces both Type I and Type II errors.  Hence, these results
demonstrate the necessity of all three of our $R^{3}$ ingredients.
Finally, observe that our error rates in Case II scenarios are better
than those for Case I scenarios, again indicating that differential
edges that are highly correlated with non-edges and common edges pose
particular challenges for our popPSI problem.  These results are also
corroborated in our high-dimensional study presented in the
supplemental materials.

Lastly, in Figure~\ref{fig5:ROC_VarySub_lowp}, we study the effect of
letting the network structure vary across subjects by decreasing the
differential group edge probability, $\pi^{g}$.  Our method continues
to perform well for $\pi^{g} \in [.5 \  1]$. However, when the
differential edge probability drops further to $\pi^{g} = .3$,
we see that both $R^{3}$ and the standard approach have greatly
reduced statistical power, as one would expect. Despite this, $R^{3}$
continues to outperform the standard approach.

Overall, our results demonstrate the difficulty of solving the
challenges associated with our popPSI problem.  In particular, using
the correct two-level models are critical to Type I 
error control while solving or ameliorating the problem of graph
selection errors at the subject level are critical for both Type I and
Type II error control.  Our results also demonstrate the 
substantial outperformance of our new $R^{3}$ method over existing
state-of-the-art methods in neuroimaging.

\section{Case Study: Differential Functional Connections in
  Autism}\label{section:abide}

We apply our $R^{3}$ method to identify differential functional
connections associated with Autism Spectrum Disorders (ASD) in a
publicly available fMRI study from the Autism Brain Imaging Data Exchange
(ABIDE) \cite{ABIDE:2013fk} consortium. 

\subsection{fMRI data acquisition and preprocessing}

We use resting state fMRI data from the UCLA Sample
1, \cite{Rudie:2012ul,ABIDE:2013fk} which consists of 73 subjects,
with 32 controls and 41 subjects diagnosed with Autism Spectrum
Disorder (ASD) based on the ADOS or ADI-R criteria.  In addition to no
history of illness, the controls did not have a first 
degree relative with autism. The fMRI scans from the UCLA site were
acquired when the subjects were passively at rest for 6 minutes using
a Siemens 3 T Trio.(T2$^*$-weighted functional images: TR = 3000 ms, TE =
28 ms, matrix size $64 \times 64$, 19.2 cm FoV, and 34 4-mm thick slices (no
gap), interleaved acquisition, with an in-plane voxel dimension of $3.0 \times 3.0$ mm). 
This results in a total of 120 images per subject.   
We preprocess the data minimally using FMRIB's Software Library
(\url{www.fmrib.ox.ac.uk/fsl}). These steps include brain extraction,
spatial smoothing of the images using a Gaussian kernel with FWHM =
5mm, band-pass filtering (.01Hz < f < 0.1Hz) to eliminate respiratory
and cardiovascular signals, and registering images to standard MNI
space. We parcellate the images using the anatomical Harvard-Oxford
Atlas (Desikan et al., 2006) to obtain 113 regions of interest. We
average all voxel time-series within a region to obtain a single
time-series per region.  Thus, our final processed data matrix
consists of 113 regions $\times$ 120 time points $\times$ 73
subjects. 

\begin{figure}[!ht]
\begin{minipage}{.9\textwidth}
	\centering{
	\includegraphics[width=.9\textwidth]{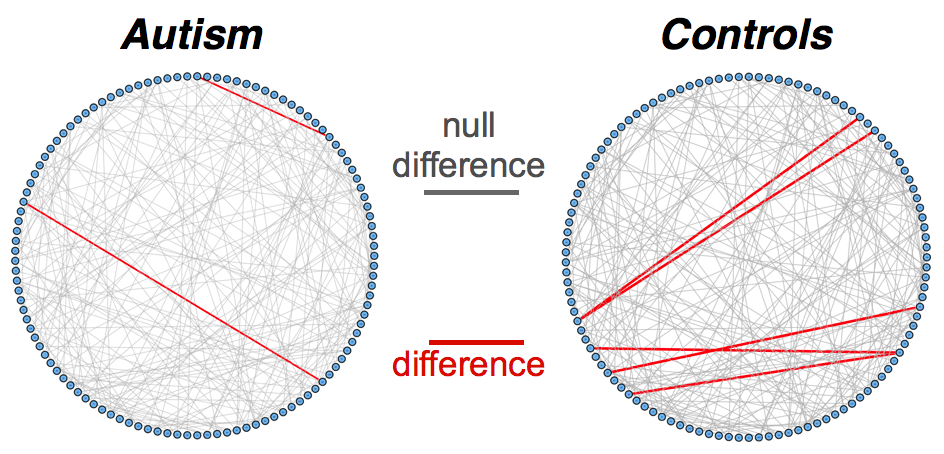}
	\caption{Population Level Differences between Austism and Control Groups.}
\label{fig:AutismGrphs}}
\end{minipage}
\end{figure}

\begin{table}[t]
	\begin{tabular}{ccp{3cm}p{3cm}cc}
	\hline
	& \textbf{Group} & \textbf{ROI} & \textbf{ROI} & \textbf{Raw P-value}  & $\widehat{\textbf{FDR}}$ \\ \hline
	1. & Control &  {Right IFG, po.(27)} & {Right po. supramarginal(55)} & .0012 & .23 \\
	2. & Autism & Left frontal pole (16) &  Left insula (18) & .0016  & .29 \\
	3. & Control &  Left sup. parietal (50) &  Right sup. parietal (51) & .0028 & .27 \\
	4. & Control & Right IFG, pt(25) & Right posterior supramarginal (55) & .0067 & .29 \\
	5. & Control & Right post. central (49) & Left superior parietal (50) &  .0076 & .29 \\ 
	6. & Control & Left lateral occipital cortex(58) & Superior Right fusiform (95) & .0098 & .29 \\
	7. & Autism & Left Caudate(2) &  Left subcallosal cortex (68) &  .0102 & .29 \\ 
	\hline 	
	\end{tabular}
	\vskip\baselineskip
	\caption{We provide a list of top 7 differential edges and
	report corresponding false discovery rates estimated using Storey's method.}\label{tabl:autism}
\end{table}	

\subsection{$R^3$ Data Analysis}

We use $R^3$ to find differential edges between ASD and control
groups.  
In order to obtain initial estimates of the regularization parameters
$\lambda_{\inds}$, we use StARS with instability parameter $\beta =
.1$ following the procedure in Section~\ref{subsec:graph_est}.  We
then apply $R^3$ procedure as outlined in Section \ref{subsec:R3}.  In
Figure \ref{fig:AutismGrphs} and Table \ref{tabl:autism}, we report the 7
most differential edges detected by our method.  We apply Storey's
direct method \cite{Storey:2002yq} to estimate the FDR associated with 
these discoveries. Notice that the estimated FDR is perhaps larger
than expected.  This likely occurs because of the limited sample size
and large subject heterogeneity. For instance, of the 7
edges we identify, none of them were found to be present in more than
20 individual subject networks in any particular group. Such
hetergeneity is to be expected in clinical
populations \cite{lenroot2013heterogeneity}. Moreover differences in
anatomy and brain parcellation also contribute to this
heterogeneity. Nonetheless, $R^{3}$ was able to detect marginally
significant and biologically relevant edges which could not have been
found by mere qualitative inspection of subject networks.  

The differential edges we identify align
with trends observed in the wider ASD literature 
\cite{dichter2012functional,just2012autism,rudie2012reduced,vissers2012brain}.
Overall, our results support three general patterns that consistently
describe 
the Autistic brain: increased local connectivity, decreased connectivity
between the two hemispheres, and decreased activity in inferior frontal
gyri (IFG) and fusiform cortex.  We summarize specific differential
edges and their relevance for ASD below:
\begin{itemize}
\item \textit{Differential edges 2,7.}
In connectivity studies, ASD networks are characterized by increased
short-range connections that connect proximate brain
areas \cite{dichter2012functional,Rudie:2012ul}, and a noticeable
absence of anterior-posterior
connections \cite{just2012autism,vissers2012brain}, while controls
have more frequent long-range connections.  The two edges that are
more present in our ASD
cohort connect regions that are physically close to one
another and located in the same hemisphere.  
\item \textit{Differential edges 4, 6, 7.}
In contrast, 3 of the 5 edges that are more present in controls,
connect distant brain regions that are located in opposite
hemispheres. This supports previous reports of enhanced local
connectivity and laterality in ASD subjects, a feature so common in
ASD networks that it is anecdotally used as a benchmark for confirming
diagnoses.  
\item \textit{Differential edges 3,5.}
Another hallmark of ASD neuroimaging studies is a noticeable absence
of activity in the inferior frontal gyrus (IFG).  The IFG is a primary
component of the mirror neuron network, a network that is often less
active in ASDs compared to controls when observing or imitating human
activity \cite{leslie2004functional,iacoboni2009imitation}.  Our data
support this finding by showing two of the five edges more present in
controls and absent from ASD subjects connecting the IFG to areas in
the parietal lobe.  
\item \textit{Differential edge 6.}
Hypoactivity in the fusiform gyrus is another common finding in ASD 
subjects compared to controls \cite{corbett2009functional,pierce2008fusiform}.  Our results suggest an edge between the right fusiform
gyrus and the left lateral occipital cortex that does not occur in 
ASD.   
\end{itemize}

Our results corroborate a number of common trends in the ASD
neuroimaging literature. However, since this is the first study to
specifically investigate differential functional connections in ASD,
we cannot validate the biological significance of edges we identified
using existing literature. We plan to verify these findings using
ABIDE data from alternative sites as well as other independent ASD
datasets.

\section{Discussion}\label{section:disc}

In this paper, we have studied a new statistical problem that arises
when conducting inference for multi-subject functional connectivity.
Our problem assumes a two-level model where subject data arises from a
Gaussian graphical model and the edge support is governed by a group
level probability, with inference conducted on these group level
parameters.  This leads to a completely new class of statistical
problems that we term Population Post Selection Inference (popPSI).
In this paper, we have discussed some of the challenges of our popPSI
problem and proposed a new procedure that partially solves these
challenges.  As we work with a new class of inference problems,
however, there are many remaining questions and open areas of related
research.

Our model and inference problem is similar in spirit to testing for
differences in the elements of two covariance or inverse covariance
matrices \cite{cai2013two}.  If we let $\pi^{A}$ and $\pi^{B}$ only take values
in $\{0,1\}$, then in fact our problem perfectly coincides with that
of \citenb{zhao2014direct}.  Given this, some may argue that we needlessly complicate the problem and potentially lose statistical power by using separate estimators for subject-level networks instead of a joint group
estimator.  While this would certainly be true in the 
case where subjects within a group all follow the same distribution,
there are a number of reasons why our approach is advantageous for
real fMRI data: (1) Assuming that each subject follows a potentially
different brain network model mimics assumptions for functional
connectivity where neuroscientists expect each subject to have a
slightly different brain network; this is especially true for
resting-state data in which passive subjects may be thinking about
different items.  (2) Our models and methods are more closely aligned
with the goal of neuroscientists who along with inference wish to
examine each subject's brain network and understand the network
differences between subjects both within a group and between groups.
Estimating group networks does not permit this analysis across
subjects. Most importantly, since group networks do not model between-subject variability, corresponding inferential results do not generalize to larger populations. (3) Our approach is also more robust to potential outliers
and artifacts common in neuroimaging \cite{Smith:2011uq,power2012spurious}.  If a few subjects have
gross artifacts or head motion, these can easily corrupt group-level
graph estimation and hence group-level inference.  By estimating
separate graphs for each subject and testing the stability proportion
of each edge, our procedure will be more robust to subject-level
artifacts.  (4) Our framework can easily be extended beyond the
two-group model to testing parameters for multiple groups or even for
continuous clinical outcomes by using linear and generalized linear
mixed effects models and test statistics \cite{Searle:2009jh}.  This is possible as we directly estimate separate subject networks and the between and within
network variability; testing group-level parameters directly cannot
easily be extended in this manner. 
Finally, as our approach can be seen as building upon
the standard approach in neuroimaging by incorporating
resampling, random effects, and random penalization, it is more likely
to be adopted by neuroimagers than a new and unfamiliar paradigm.

This paper has focused on characterizing our particular popPSI
problem, providing intuition as to why standard approaches in
neuroimaging fail, and suggesting a methodological solution.  Studying
this problem from a theoretical perspective is beyond the scope of
this paper and hence, many open theoretical questions remain.  We seek
to control the overall error rate or FDR of the 
procedure; studying if this is achievable and under what theoretical
conditions is an important open problem.  Based on our empirical
investigations in Section~\ref{section:sims} and our discussion in Section~\ref{subsec:challenge2},
we conjecture that this is possible under extensions of an
irrepresentable or incoherence condition.  More specifically, these
conditions limit the correlation between edges that
are common across subjects and those that have differential edge
probabilities, especially when such common and differential edges belong to the same connected component, as well as limit correlations between those that have differential edge probabilities and the non-edges.
Another interesting line of theoretical investigation is relating the
overall error rate of the inferential procedure to that of the error
rate for graph selection at the subject level.  Additionally, there
are likely many interesting theoretical questions that arise from
characterizing and studying popPSI problems more generally beyond our
specific model.

Our model and methods can be extended well beyond testing for the
differential presence of an edge in a population of Gaussian graphical
models.  In particular, our basic approach is applicable to inference
for any support related metric such as overall network sparsity or the
degree of each node.  Here, one could assume the group level 
parameters follow a Poisson or normal distribution and adjust our
random effects Beta-Binomial estimators and test statistics
accordingly.  Also as mentioned above, we can extend our
framework to test multiple groups or continuous clinical outcomes by
using linear and generalized linear mixed effects models. 
Further, other network models such as a sparse covariance model could be used for the subject networks.

For application to multi-subject fMRI data, there are also many other
considerations and items to investigate.  In Section~\ref{section:intro}, we
mentioned that data for each subject must be parcellated into a matrix
of brain regions by time series.  More investigations are needed to
determine which parcellation method to use and how this affects
network estimation and group-level inference.  Additionally, many have
noted that resting state fMRI data is particularly sensitive to
physiological artifacts such as head motion \cite{power2012spurious}; further studies are also needed to determine how head motion affects group-level
network inference.  Finally we note, that our inference paradigm can additionally
be employed for task related fMRI experiments as well as in \citenb{Tomson:2013uq}.

While we introduce a new class of statistical problems, Population PSI,
these problems actually arise often in neuroimaging.  First, the
problems and challenges we outline for the standard approach used for
functional connectivity with fMRI data extend to many other forms of
connectivity in neuroimaging.  Consider structural connectivity which
estimates networks based on the number of tracts connecting different
brain regions.  These tracts, however, are estimated by complicated
probabilistic tractography algorithms \cite{ciccarelli2003diffusion} from diffusion tensor
imaging.  Hence, estimates of subject-level networks are imperfect,
resulting in two-levels of network variability and then a popPSI-type
group level inference problem.  Similar problems arise for vector autoregressive graphical models or effective connectivity estimated from EEG, MEG, fMRI, or ECoG data. Additionally, popPSI problems arise in neuroimaging applications
beyond network inference where a selection based statistical learning
procedure is applied at the subject level but inference is to be
conducted across subjects.  Finally, our work provides a cautionary
tale about conducting inference on estimated parameters without
properly accounting for multiple levels of variability.

In conclusion, we have studied a new problem arising in inference for
multi-subject connectivity.  As with any new framework many open
questions and directions for future research remain.  Software for our
$R^{3}$ procedure will be available as part of the Markov Network Matlab Toolbox
\url{https://bitbucket.org/gastats/monet}.

\section*{Acknowledgments}
M.N. and G.A. are supported by NSF DMS 1209017 and 1264058; S.T. is
supported by an NIH training grant 5T32NS048004-08; M.N. is supported by Amazon Web Services (AWS)
research grant for computational resources.


\begin{supplement}
\setattribute{journal}{name}{}
\sname{Supplement A}
\stitle{Supplementary Material to \textit{Two Sample Inference for Populations of Graphical Models, with Applications to Functional Brain Connectivity}}{}
\slink[doi]{arxiv.org/abs/0000.0000}
\sdescription{Additional simulations demonstrate that $R^{3}$ outperforms the standard approach in high dimensional regimes and when increasing differential or common graph density.}
\end{supplement}
\bibliographystyle{imsart-nameyear}
\bibliography{sarxiv_r3}

\appendix

\renewcommand\thefigure{\thesection.\arabic{figure}}    
\renewcommand\thetable{\thesection.\arabic{table}}

\section{Additional Figures and Simulation Studies}

\begin{figure}[!htb]
\renewcommand{\arraystretch}{.01}
  \centering{
      \subfloat[$R^3$, Small world~I]{\label{fig:sim:smallwedge}
          \includegraphics[width=.25\textwidth]{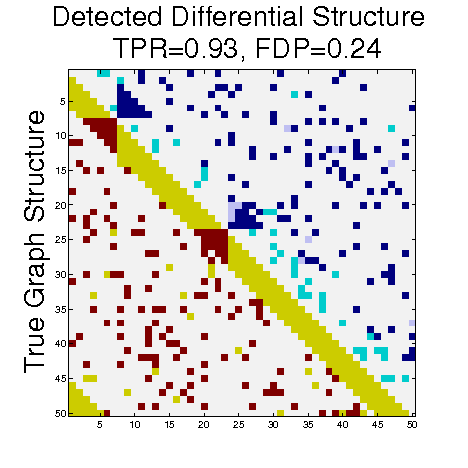}}
      \subfloat[$R^3$, Banded~I]{\label{fig:sim:bandededge}
          \includegraphics[width=.25\textwidth]{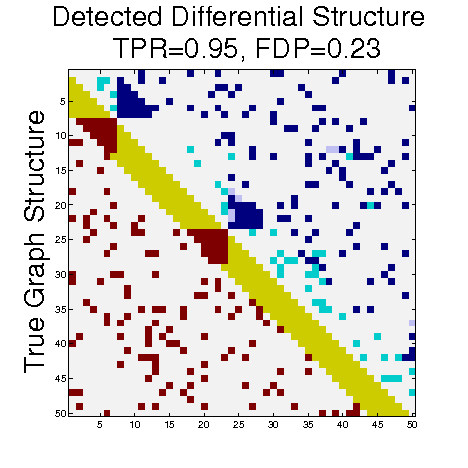}}
      \subfloat[$R^3$, Clusters~I]{\label{fig:sim:hubedge}
          \includegraphics[width=.25\textwidth]{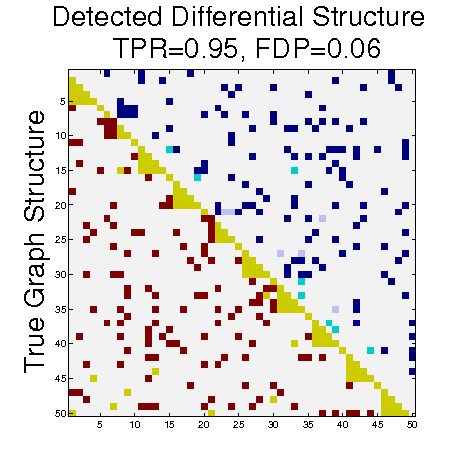}} \\
      \vspace{-2mm}
      \subfloat[Standard, Small world~I]{\label{fig:sim:smallwedge2}
          \includegraphics[width=.25\textwidth]{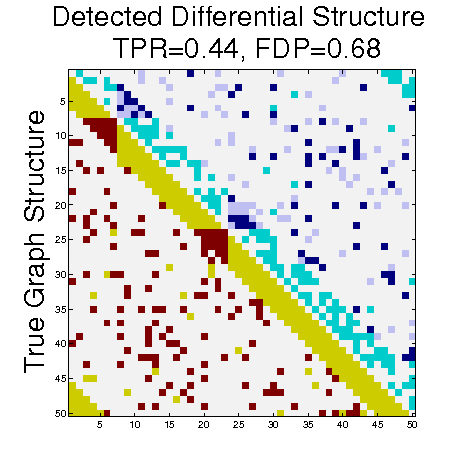}}
      \subfloat[Standard, Banded~I]{\label{fig:sim:bandededge2}
          \includegraphics[width=.25\textwidth]{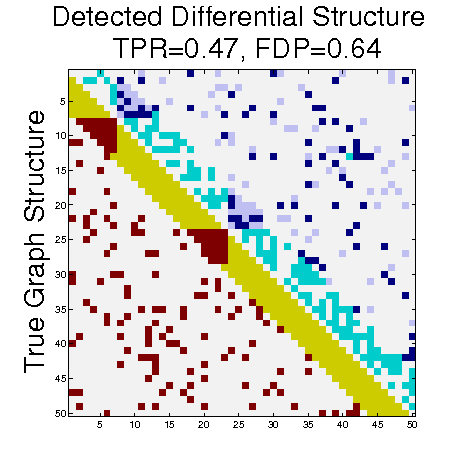}}
      \subfloat[Standard, Clusters~I ]{\label{fig:sim:hubedge2}
          \includegraphics[width=.25\textwidth]{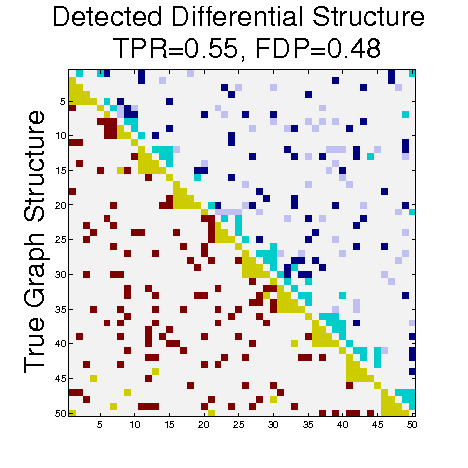}} \\
  \centering{
     \subfloat[$R^3$, Small world~II]{\label{fig:sim:smallwedge3}
          \includegraphics[width=.25\textwidth]{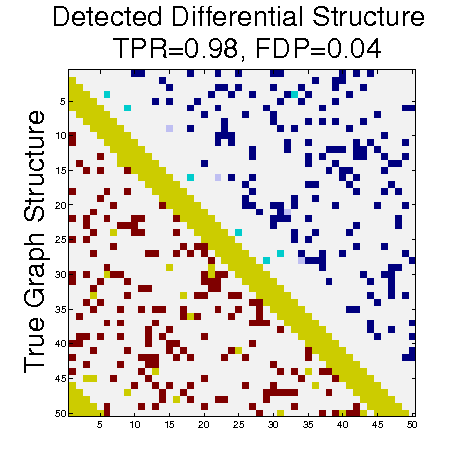}}
      \subfloat[$R^3$, Banded~II]{\label{fig:sim:bandededge3}
          \includegraphics[width=.25\textwidth]{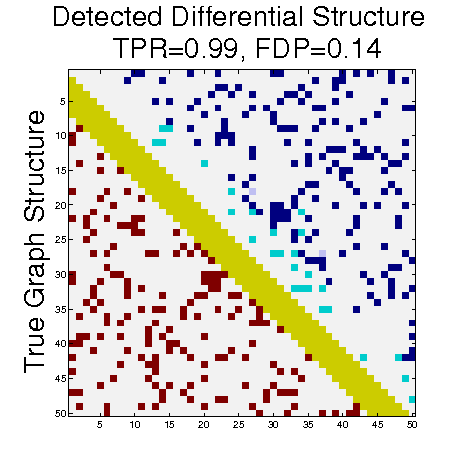}}
      \subfloat[$R^3$, Clusters~II]{\label{fig:sim:hubedge3}
          \includegraphics[width=.25\textwidth]{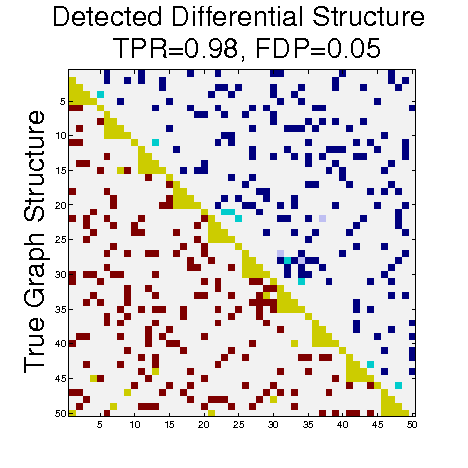}} \\
      \vspace{-2mm}
      \subfloat[Standard, Small world~II]{\label{fig:sim:smallwedge4}
          \includegraphics[width=.25\textwidth]{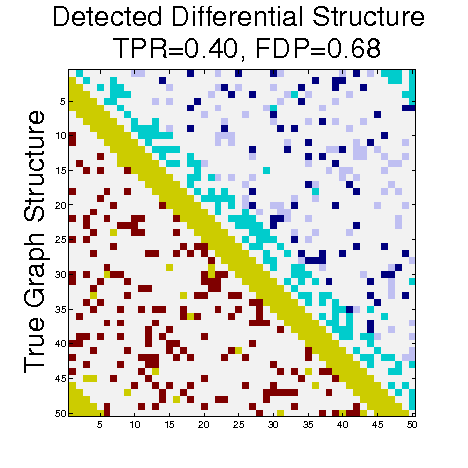}}
      \subfloat[Standard, Banded~II]{\label{fig:sim:bandededge4}
          \includegraphics[width=.25\textwidth]{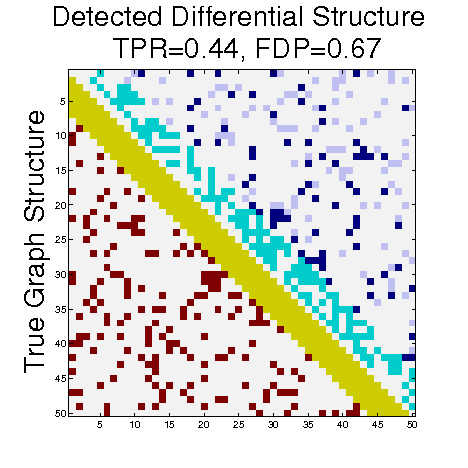}}
      \subfloat[Standard, Clusters~II ]{\label{fig:sim:hubedge4}
          \includegraphics[width=.25\textwidth]{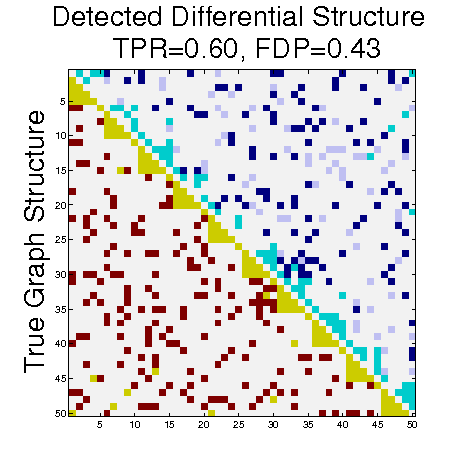}} \\
  \includegraphics[width=.6\textwidth,height=.14\textwidth]{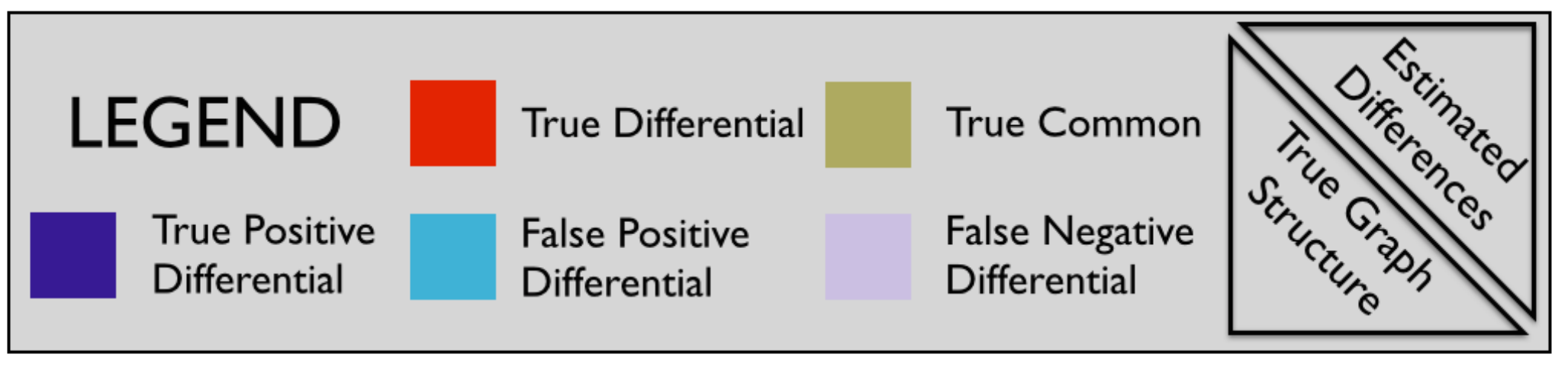}} 
  \caption{Confusion adjacency matrices illustrating the location of
    false positives and false negatives for group-level inference
    detected by both our $R^{3}$ and the standard approach.  Each
    example accompanies one of the six simulation scenarios described
    in \citet{Narayan:2014si}.  
\label{fig4b:location}}}
\end{figure}

We present additional figures and simulation studies that complement those in
\citet{Narayan:2014si}.  First in Figure~\ref{fig4b:location}, we
present confusion adjacency matrices for specific examples from our Simulation
Study described in \citet{Narayan:2014si} and corresponding to Figure~4
and Table~1.  These illustrate the location of false positive and
false negative edges for group inference with the lower-triangular portion
showing the true common and differential edges and the upper-triangular
portion showing the rejected edges that are declared differential.
Overall, we can see that $R^{3}$ offers substantial improvements over
the standard method which has both high Type I and Type II error
rates.  Notice also, that false positives where common edges are
mistaken as differential edges are far more likely to occur when graph
selection errors are more severe.  These are either (i) when
differential edges are highly correlated as in Case I scenarios where
the differential edges are clustered, or (ii) when the differential
and common edges are highly connected as in small world or hub graphs.

\begin{figure}[!htb]
\vspace{-.75cm}
\hspace{-.5cm}
	\centering{
         \subfloat[Small World Graph, II]{\label{fig:5:smallw}
          \includegraphics[width=.33\textwidth]{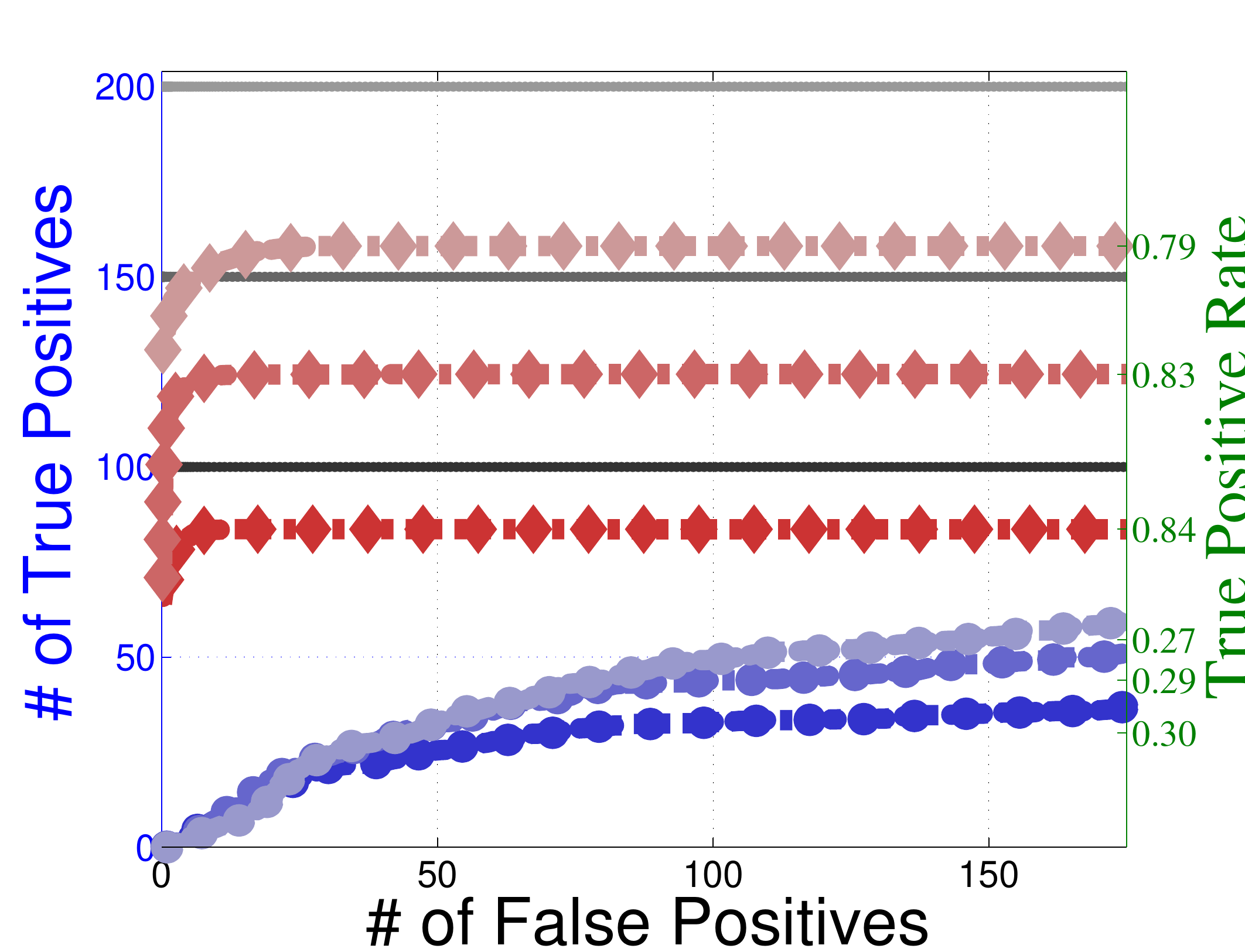}}
          \subfloat[Banded Graph, II]{\label{fig:3:banded}
          \includegraphics[width=.33\textwidth]{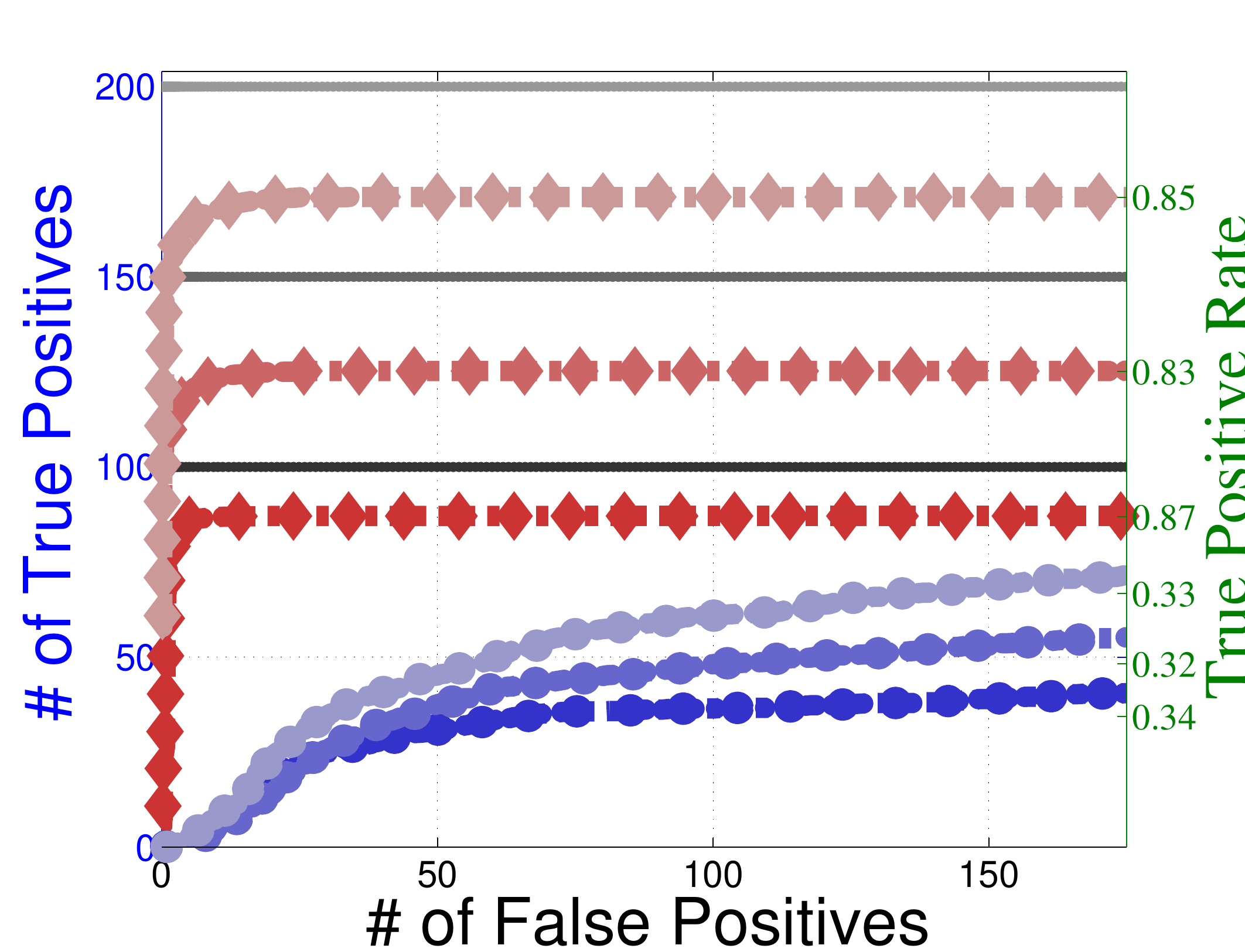}}
          \subfloat[Hub Graph, II]{\label{fig:3:hub}
          \includegraphics[width=.33\textwidth]{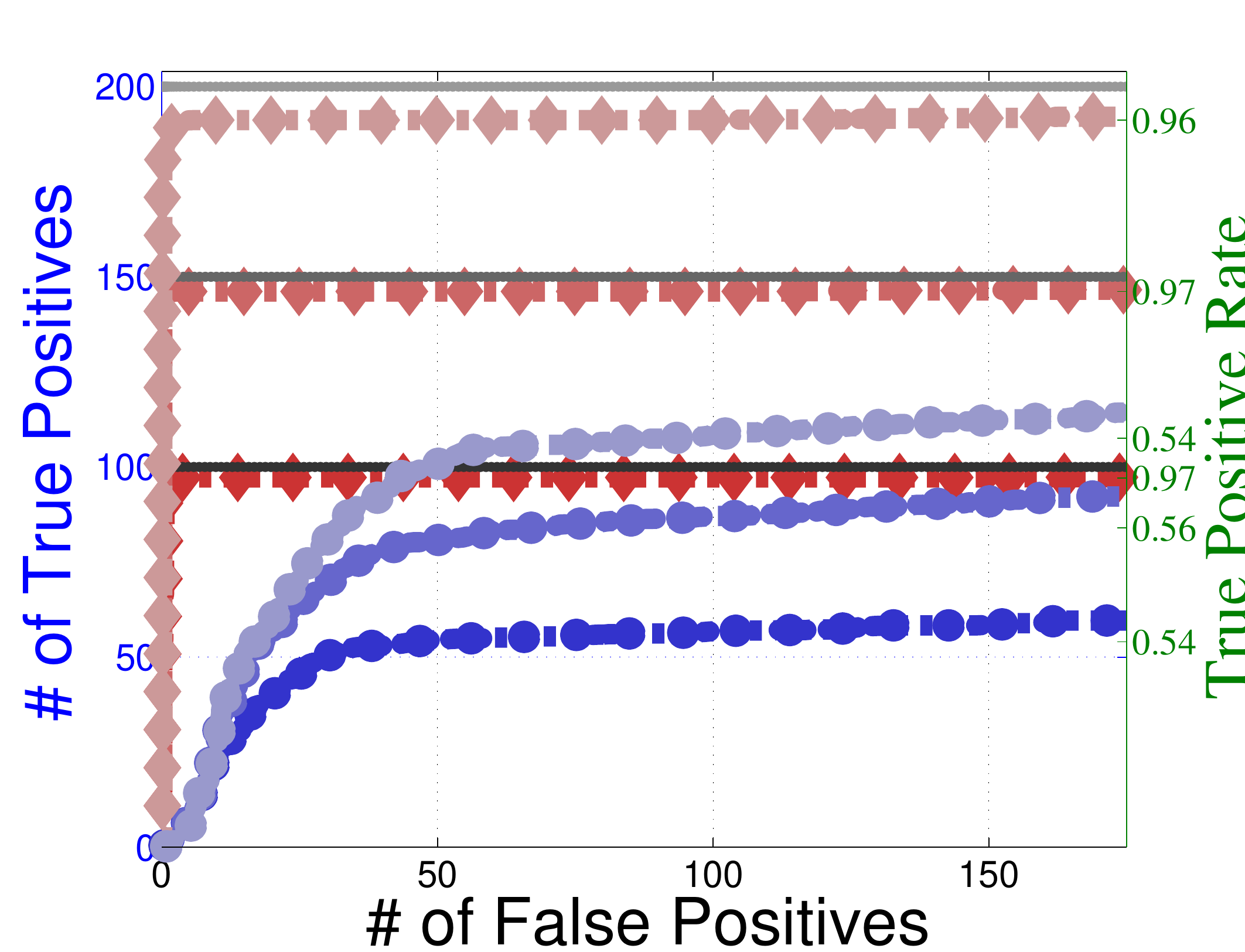}} \\
          \includegraphics[width=.75\textwidth,height=.09\textwidth]{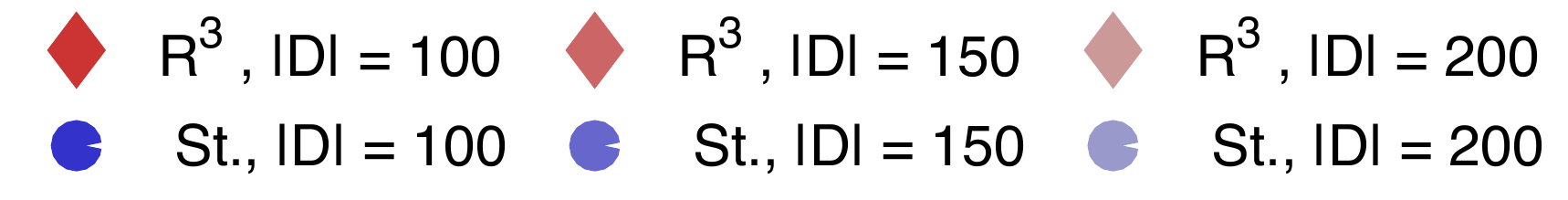}
          \caption{Increasing the Number of Differential Edges.  ROC
            curves for sequentially rejected tests comparing $R^{3}$
            to the standard method when the number of differential
            edges are increased:  $|D| = \{100,150,200\}$; otherwise 
            the simulation is as described in \citet{Narayan:2014si}.
            As the number 
            of differential edges increases, the number of alternative
            hypotheses also increases; hence, true positive rates for
            each curve are denoted on the right $y-axis$.  Here,
            increases in the number of differential edges result in a
            mild loss of statistical power.  
          \label{fig6a:ROC_VaryDiff_lowp}}}        
\hspace{-.5cm} 
	\centering{
         \subfloat[Small World Graph, II]{\label{fig:5:smallw}
          \includegraphics[width=.33\textwidth]{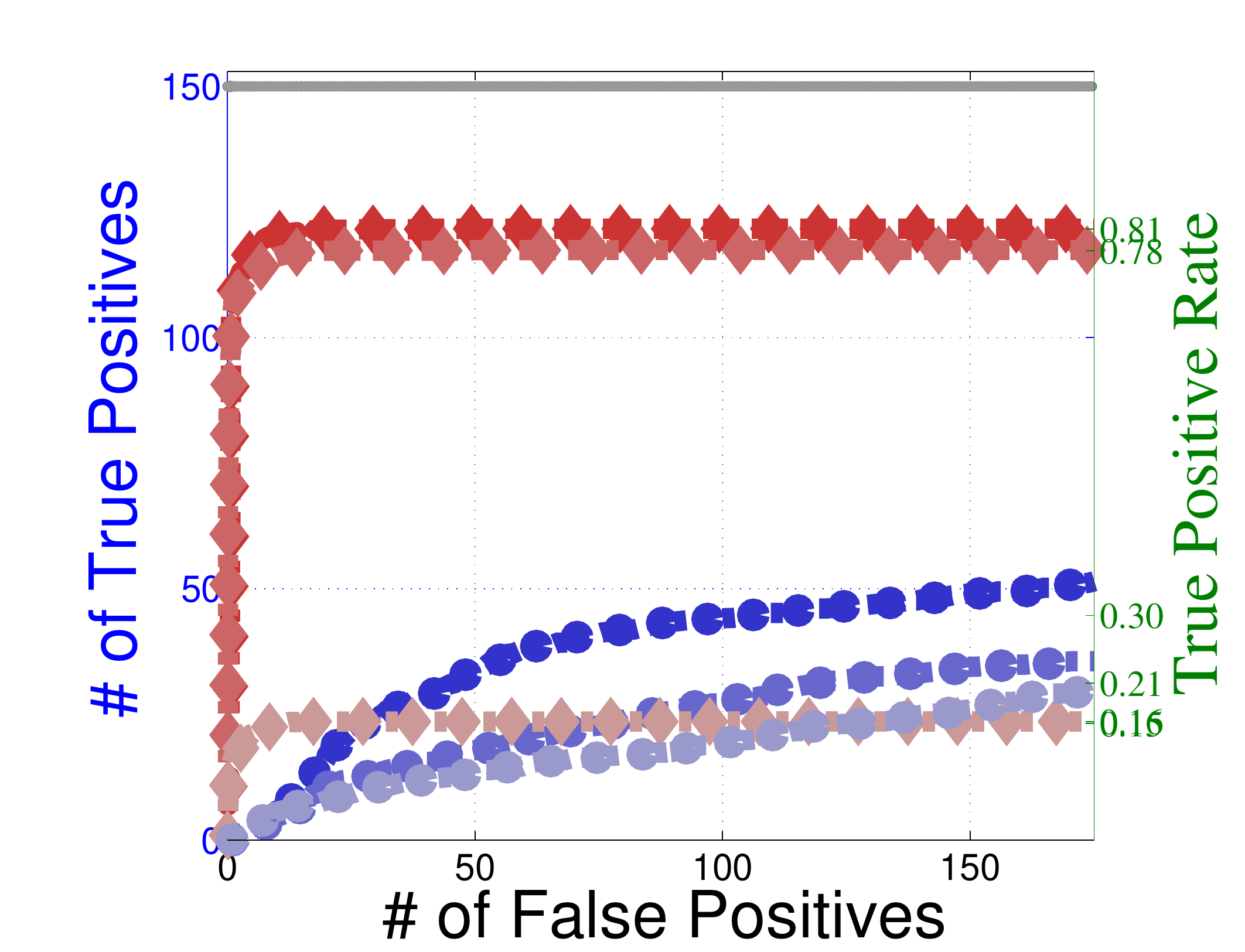}}
          \subfloat[Banded Graph, II]{\label{fig:5:banded}
          \includegraphics[width=.33\textwidth]{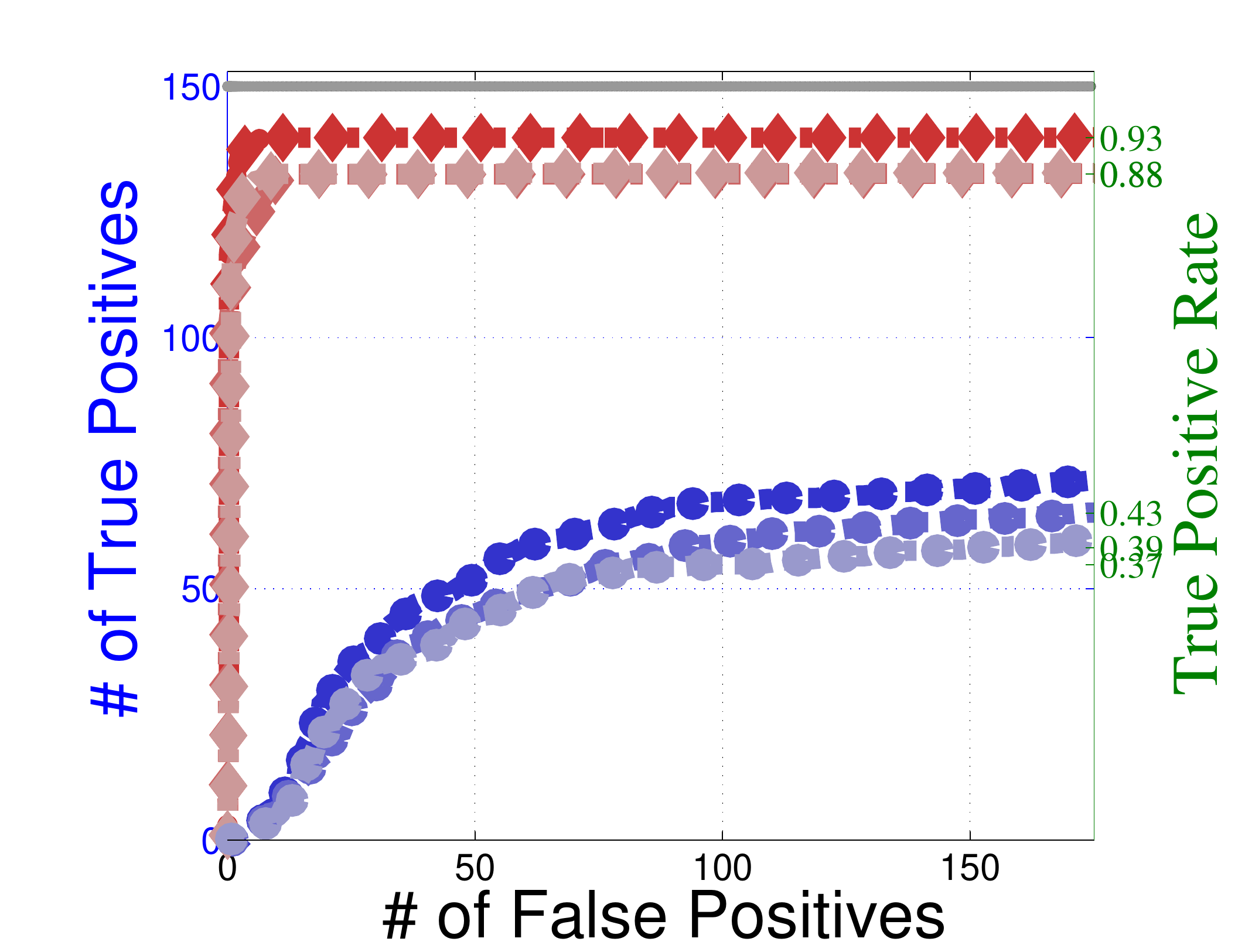}}
          \subfloat[Hub Graph, II]{\label{fig:5:hub}
          \includegraphics[width=.33\textwidth]{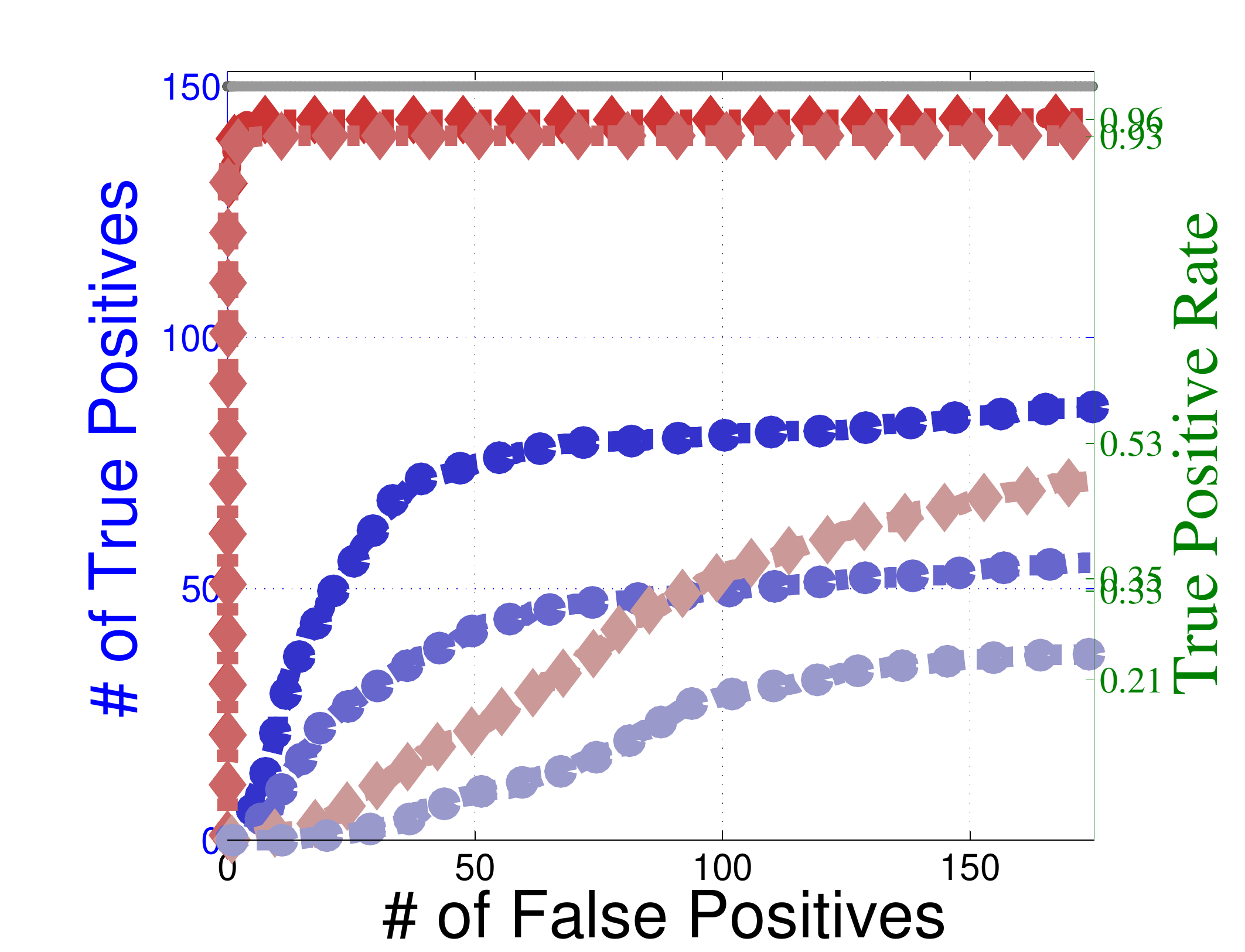}}\\
         \includegraphics[width=.75\textwidth,height=.09\textwidth]{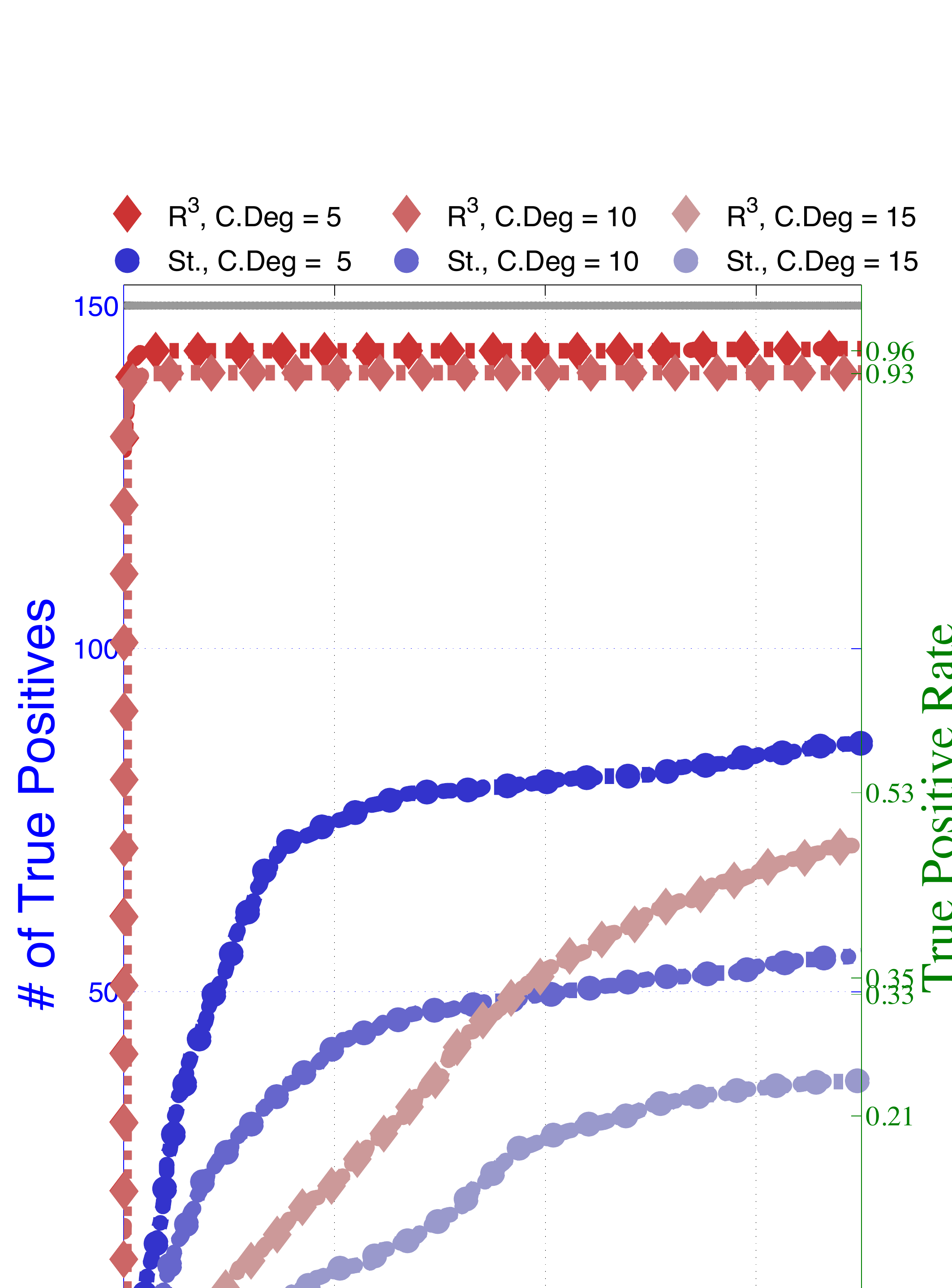}
          \caption{Increasing the Number of Common Edges and Common
            Network Degree.  ROC
            curves for sequentially rejected tests comparing $R^{3}$
            to the standard method when the number of common
            edges are increased such that the average network degree is
            equal to 5, 10, or 15; otherwise 
            the simulation is as described in \citet{Narayan:2014si}.
            As the common network degree increases, the number of alternative
            hypotheses also increases accordingly; hence, true
            positive rates for 
            each curve are denoted on the right $y-axis$.  Here,
            increases network degree are known to increase graph
            selection errors.  As a result, statistical power
            dramatically decreases for networks with high degree. 
\label{fig7a:ROC_VaryCom_lowp}}}
 \end{figure}

Next, we study how changing the number of differential and common
edges in the network structure affects our method and the standard
approach.  In
Figures~\ref{fig6a:ROC_VaryDiff_lowp} and
\ref{fig7a:ROC_VaryCom_lowp}, we present ROC curves as the number of
differential edges is increased and the number and degree of common
edges are increased respectively.  Other that these described changes,
simulation scenarios are as described in
\citet{Narayan:2014si}. Increasing the number of differential edges
results in a slight loss of statistical power.   Increasing the number
and average degree of common edge support in the network structure, on
the other hand, can result in severe loss of statistical power.  Here,
dense graph structures for highly connected graph types are known to
lead to high graph selection error rates which in turn severely bias
our test statistics, resulting in increased Type I and Type II
errors.  These results are also born out in the accompanying
Tables~\ref{tab:fdp2} and \ref{tab:fdp3} which give the average true
positive rate and false discovery portion for the rejected tests.

\begin{figure}[!htb]
  \centering{
    \subfloat[Small World Graph, I]{\label{fig:4:smallw}
         \includegraphics[width=.33\textwidth,height=.3\textwidth]{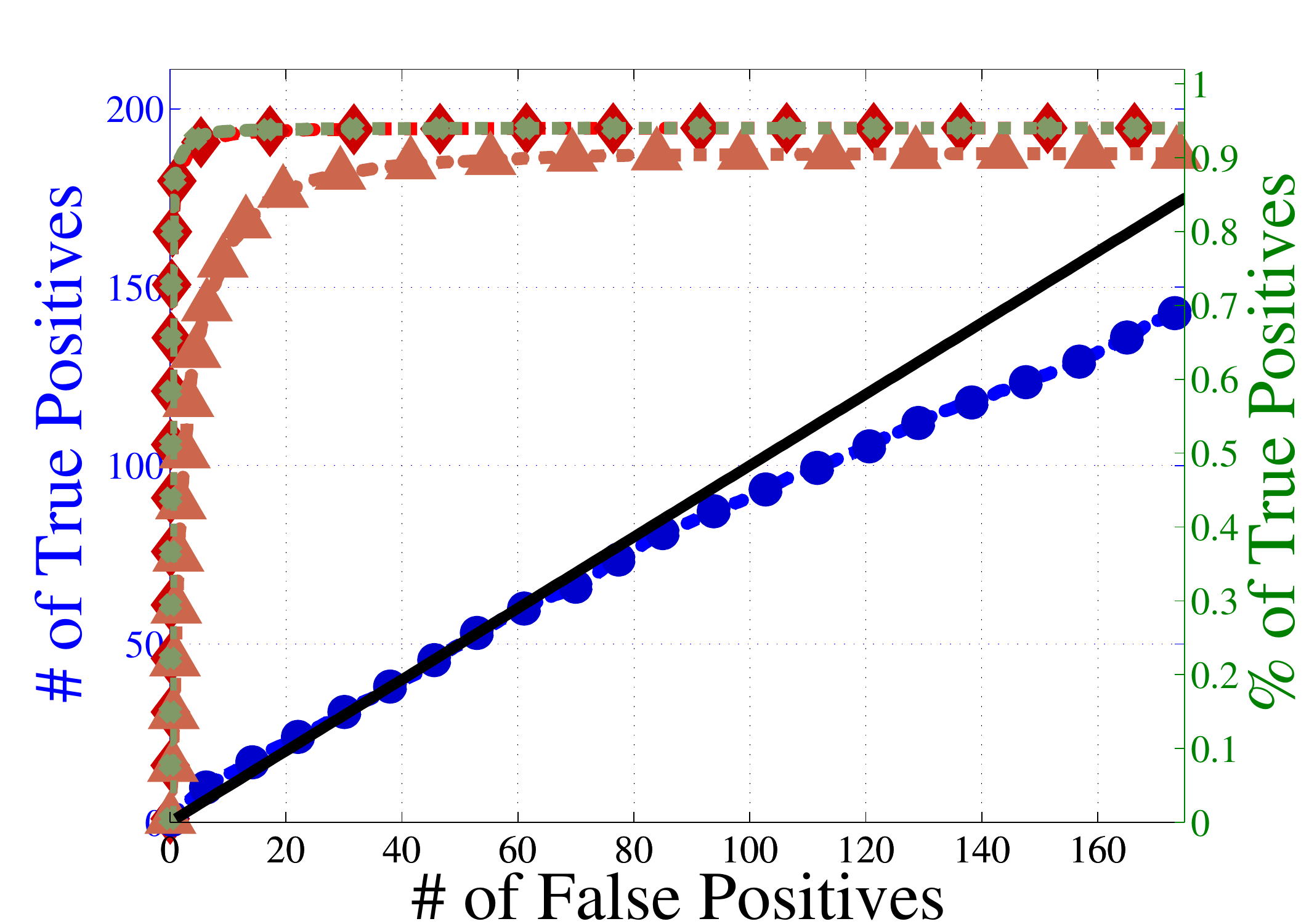}}
         \subfloat[Banded Graph, I]{\label{fig:4:banded}
         \includegraphics[width=.33\textwidth,height=.3\textwidth]{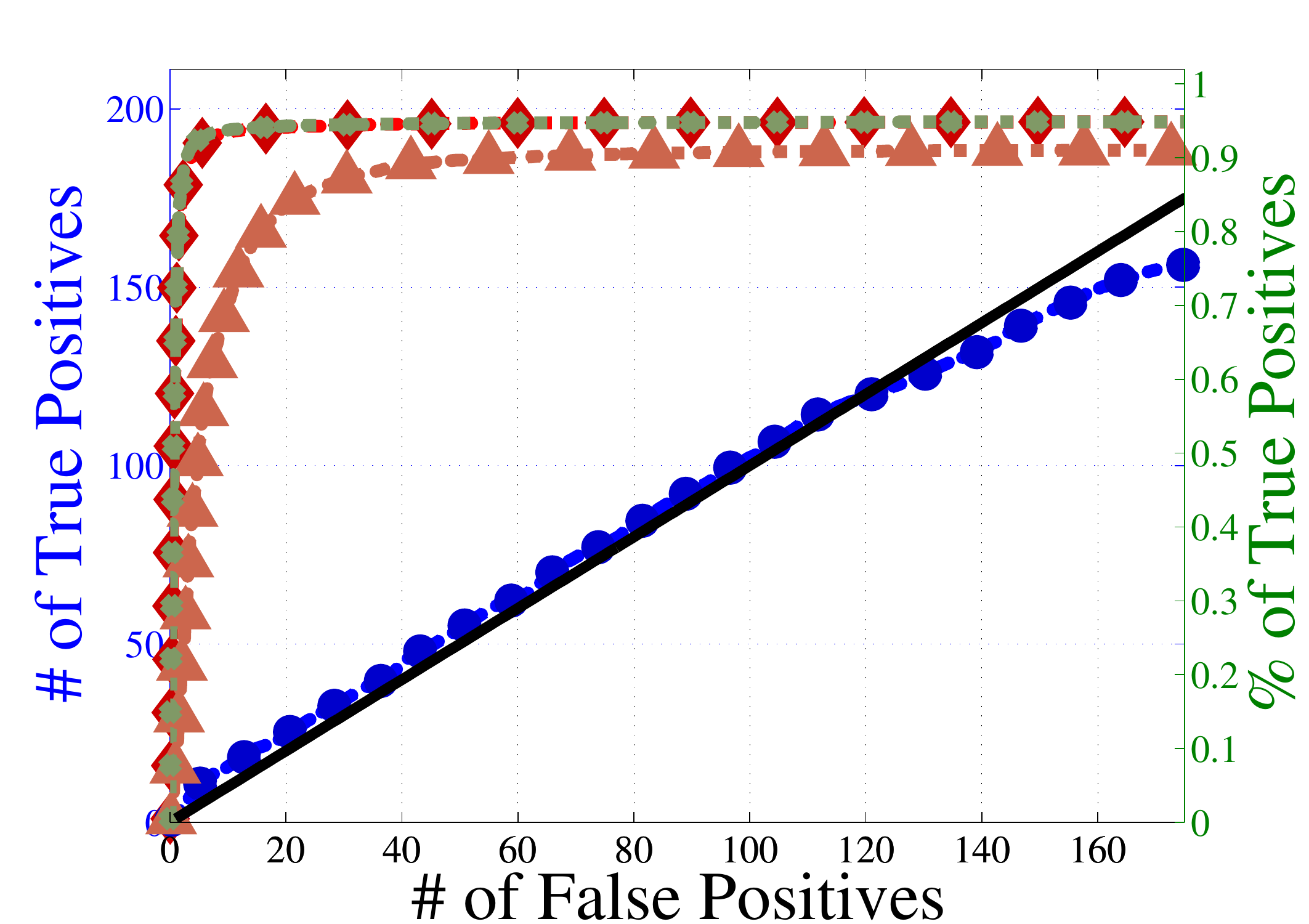}} 
         \subfloat[Hub Graph, I]{\label{fig:4:hub}
         \includegraphics[width=.33\textwidth,height=.3\textwidth]{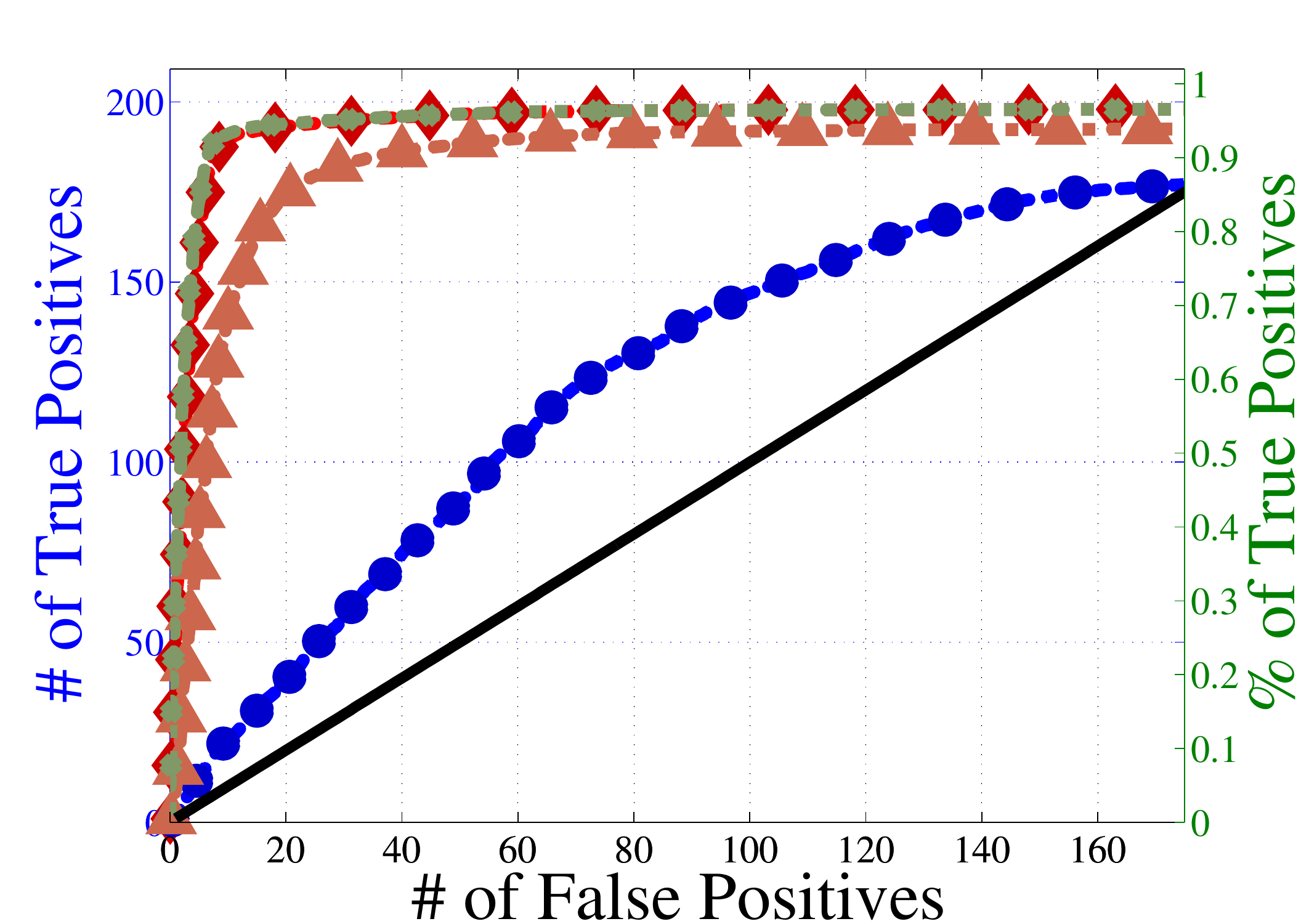}}\\
        \subfloat[Small World Graph, II]{\label{fig:5:smallw}
         \includegraphics[width=.33\textwidth,height=.3\textwidth]{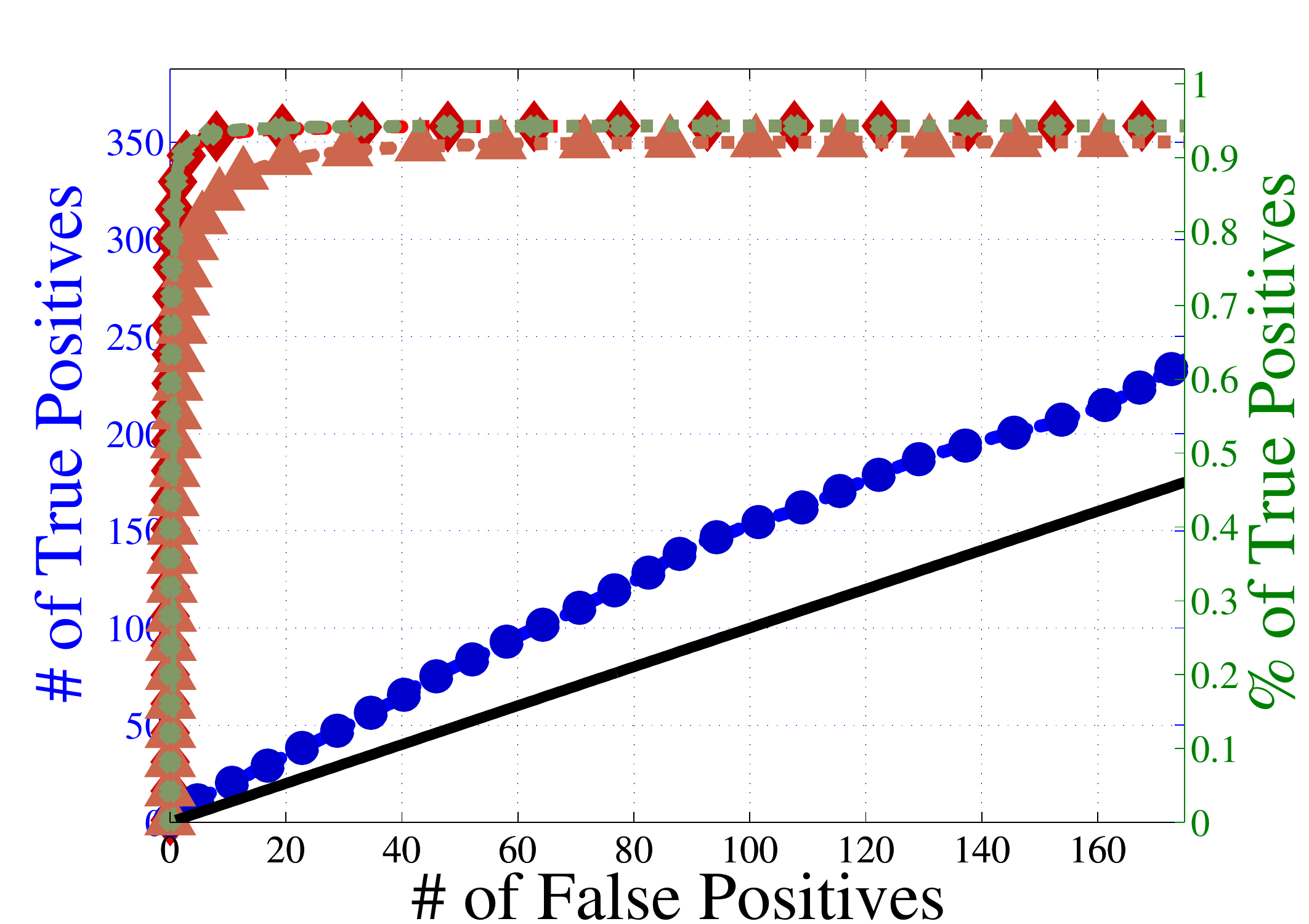}}
         \subfloat[Banded Graph, II]{\label{fig:5:banded}
         \includegraphics[width=.33\textwidth,height=.3\textwidth]{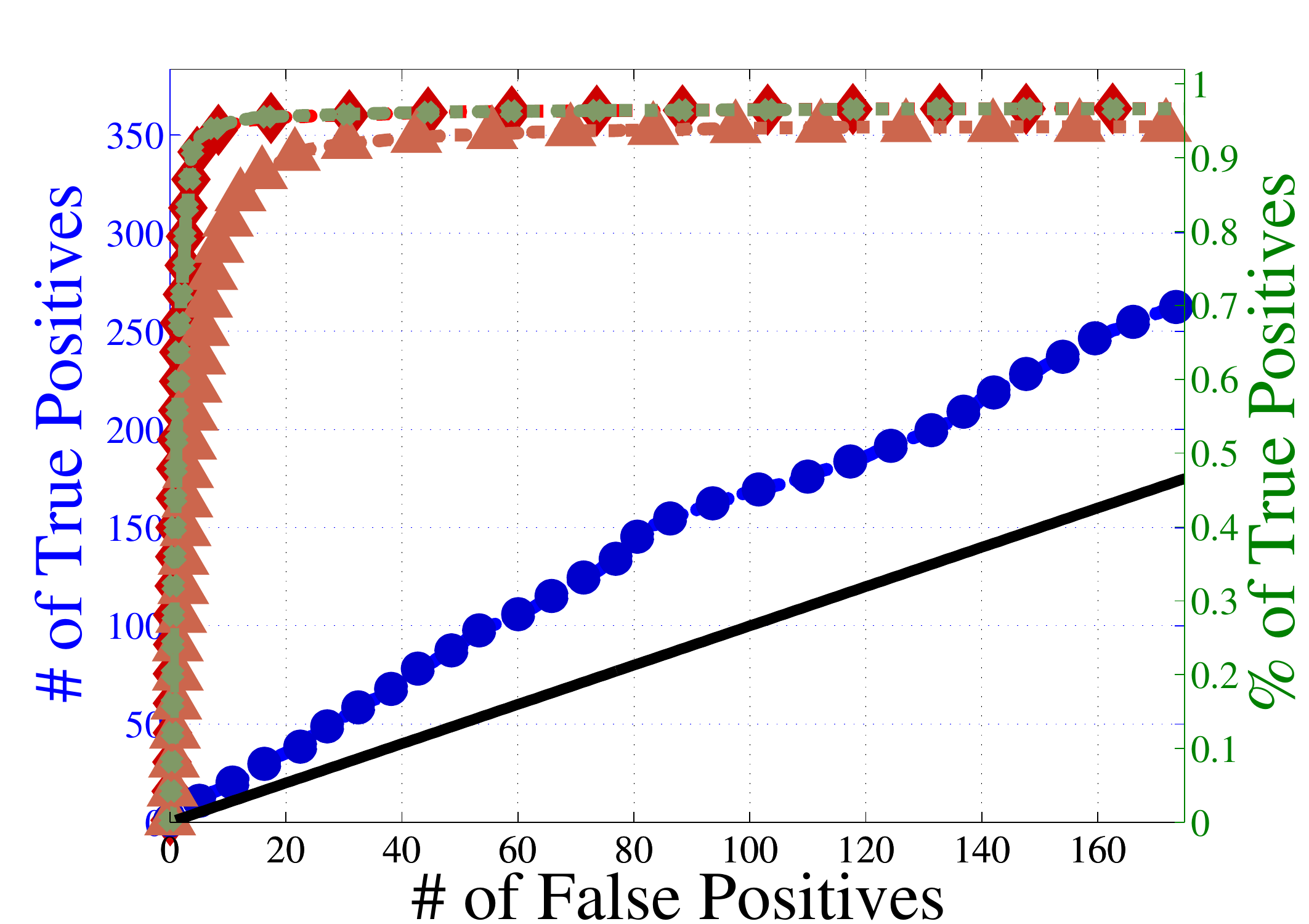}}
        \subfloat[Hub Graph, II]{\label{fig:5:hub}
        \includegraphics[width=.33\textwidth,height=.3\textwidth]{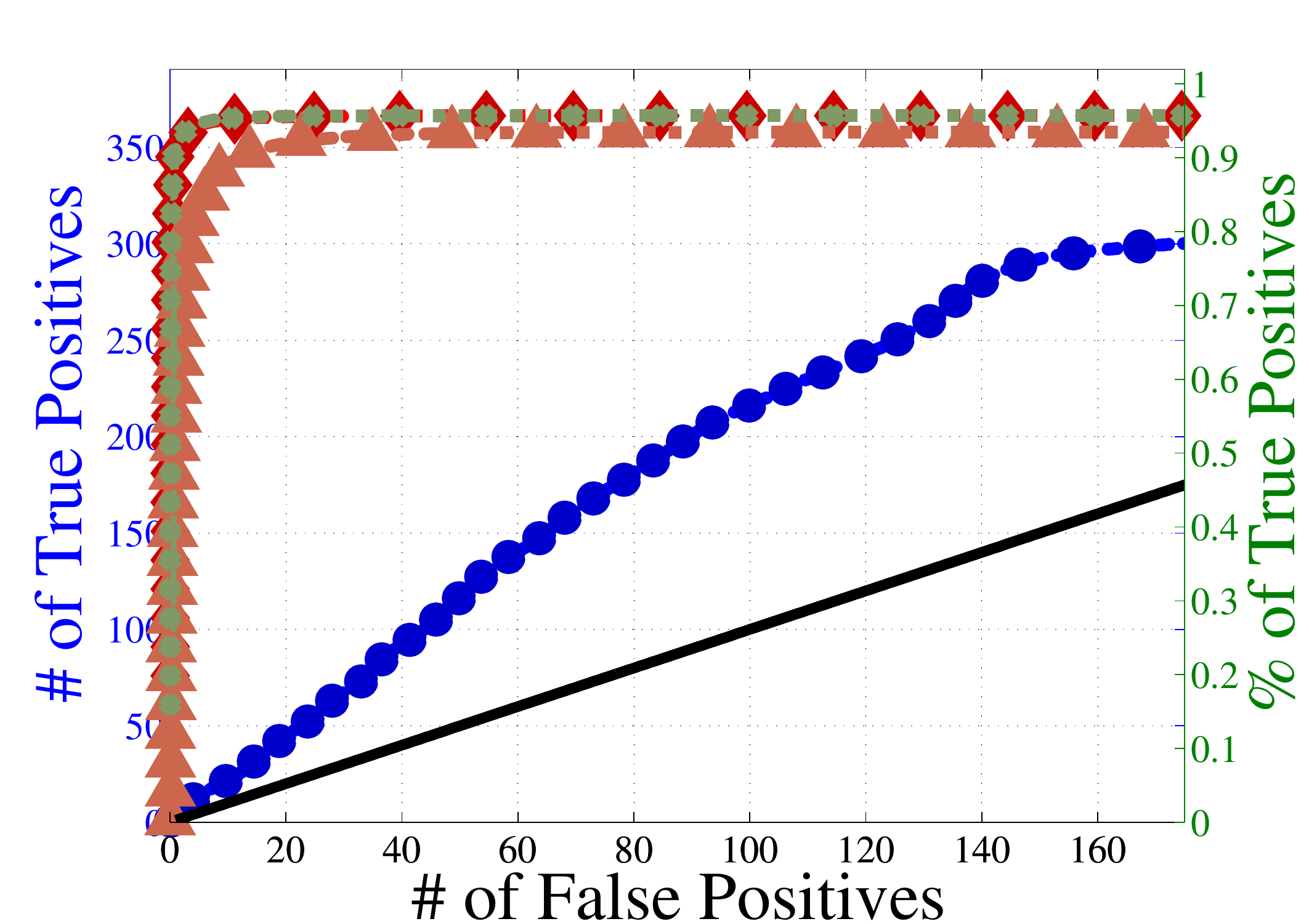}} \\
        \includegraphics[width=\textwidth]{figure_3/ROCLegend.pdf}  
    \caption{\label{fig3b:CountROCs-highp} High-dimensional results in
      terms of average ROC curves for sequentially
    rejected tests comparing our 
    method to the standard approach, $R^{2} = (RS,RE)$, and $R^{2} =
    (RS,RP)$ for each network structure type and Case I and II type
    differential edges. Analogously to the lower dimensional case,
    methods employing 
    random penalization (RP) 
    improve statistical power.}}
    \end{figure}

Finally, we present a high-dimensional simulation study in
Figure~\ref{fig3b:CountROCs-highp} 
and Table~\ref{tab:fdp1} that complements Figure 4 and Table 1 in
\citet{Narayan:2014si}. 
For this simulation, we let $p = 100 >  T = 80$ and all other
parameters are as described in \citet{Narayan:2014si}.  Trends
observed in the lower-dimensional case are also observed here.
Interestingly, this simulation comparatively shows a slight increase
in statistical power with lower estimated false positive rates.  This
likely occurs as the number of differential edges is fewer relative to
$p$ and hence networks are less dense resulting in fewer graph
selection errors.

Overall, these simulations reveal that $R^{3}$ continues to perform
well in a variety of settings.  Combining these findings with
investigations in \citet{Narayan:2014si}, we see that $R^{3}$ is
most affected by graph selection errors that occur for highly dense
and correlated network structures.

 \begin{table}[!hbt]
	\begin{tabular} {|c|l|c|c >{(}r<{)}|c>{(}r<{)}|c>{(}r<{)}|c>{(}r<{)}|} 
    \hline
\centering{Case} & \multicolumn{1}{|p{1.3cm}|}{Sim Type} & Metric 
      & \multicolumn{2}{p{2cm}|}{\centering $R^{3}, |D|=100$} & \multicolumn{2}{p{2cm}|}{\centering St.,~$|D|=100$} 
      & \multicolumn{2}{p{2cm}|}{\centering $R^{3}, |D|=200$} & \multicolumn{2}{p{2cm}|}{\centering St.,~$|D|=200$} \\
		\hline
	\multirow{2}{*}{\hfil II}    
	      & \multirow{2}{*}{\hfil  SmallW}
          & \text{TPR} & 0.836 & 0.092 & 0.323 & 0.125 & 0.790 & 0.088 & 0.250 & 0.065 \\
          & & \text{FDP} & 0.264 & 0.059 & 0.798 & 0.042 & 0.211 & 0.047 & 0.711 & 0.028   \\ \hline 
	\multirow{2}{*}{\hfil II}    
          & \multirow{2}{*}{\hfil  Banded}
          & \text{TPR}   & 0.871 & 0.062 & 0.362 & 0.104 & 0.855 & 0.068 & 0.311 & 0.100  \\
          & & \text{FDP} &  0.273 & 0.068 & 0.767 & 0.029 & 0.166 & 0.046 & 0.654 & 0.045   \\ \hline
	\multirow{2}{*}{\hfil II}    
          & \multirow{2}{*}{\hfil Hub}
          & \text{TPR} & 0.972 & 0.028 & 0.540 & 0.166 & 0.956 & 0.026 & 0.518 & 0.094 \\
		& & \text{FDP} & 0.038 & 0.026 & 0.613 & 0.066 & 0.040 & 0.013 & 0.446 & 0.033  \\ \hline
		\hline
	\end{tabular}
	\caption{\label{tab:fdp2} Increasing the Number of
          Differential Edges.  Average true positive rate (TPR) and 
    false discovery proportion (FDP) for tests rejected by $R^{3}$ and
    the standard approach
    when controlling the FDR at 10\% via the Benjamini-Yekutieli
    method; standard errors are given in parentheses.  Both approaches
    show a mild loss of statistical power as the 
    number of differential edges increases.} 
\begin{tabular} {|c|l|c|c >{(}r<{)}|c>{(}r<{)}|c>{(}r<{)}|c>{(}r<{)}|} 
    \hline
\centering{Case} & \multicolumn{1}{|p{1.3cm}|}{Sim Type} & Metric 
      & \multicolumn{2}{p{2cm}|}{\centering $R^{3}, \textrm{Deg}=5$} & \multicolumn{2}{p{2cm}|}{\centering St.,~$\textrm{Deg}=5$} 
      & \multicolumn{2}{p{2cm}|}{\centering $R^{3}, \textrm{Deg} = 15$} & \multicolumn{2}{p{2cm}|}{\centering St.,~$\textrm{Deg}=15$} \\
		\hline
	\multirow{2}{*}{\hfil  II}    
	      & \multirow{2}{*}{\hfil  SmallW}
          & \text{TPR} & 0.811 & 0.071 & 0.295 & 0.115 & 0.157 & 0.055 & 0.235 & 0.038 \\
          & & \text{FDP} & 0.230 & 0.095 & 0.724 & 0.061 & 0.613 & 0.044 & 0.866 & 0.022   \\ \hline 
	\multirow{2}{*}{\hfil  II}    
          & \multirow{2}{*}{\hfil  Banded}
          & \text{TPR}   & 0.932 & 0.055 & 0.443 & 0.105 & 0.884 & 0.057 & 0.361 & 0.066 \\
          & & \text{FDP} & 0.158 & 0.059 & 0.653 & 0.035 & 0.193 & 0.081 & 0.668 & 0.040    \\ \hline
	\multirow{2}{*}{\hfil  II}    
          & \multirow{2}{*}{\hfil  Hub}
          & \text{TPR} & 0.956 & 0.039 & 0.516 & 0.151 & 0.572 & 0.155 & 0.251 & 0.091 \\
		& & \text{FDP} & 0.054 & 0.038 & 0.513 & 0.074 & 0.744 & 0.049 & 0.809 & 0.030  \\ \hline
		\hline
	\end{tabular}
	\caption{\label{tab:fdp3} Increasing the Number of Common
          Edges and Common Network Degree. Average true positive rate (TPR) and 
    false discovery proportion (FDP) for tests rejected by $R^{3}$ and
    the standard approach
    when controlling the FDR at 10\% via the Benjamini-Yekutieli
    method; standard errors are given in parentheses.  As expected,
    both methods have a severe loss of statistical power as the graph density
    increases and hence graph selection errors increase for
    highly-connected small world and hub graphs.
        }
\end{table}

\begin{table}[hbt]
  \begin{tabular} {|c|l|c|c >{(}r<{)}|c>{(}r<{)}|c>{(}r<{)}|c>{(}r<{)}|} 
    \hline
\centering{Case} & \multicolumn{1}{|p{1.3cm}|}{Sim Type} & Metric 
      & \multicolumn{2}{p{2cm}|}{\centering $R^{3}$} & \multicolumn{2}{p{2cm}|}{\centering Standard~Test} 
      & \multicolumn{2}{p{2cm}|}{\centering $\Rs$} & \multicolumn{2}{p{2cm}|}{\centering $\Rp$} \\
    \hline \hline
    \multirow{2}{*}{\hfil  I}
          & \multirow{2}{*}{\hfil  SmallW} 
           & \text{TPR}  & 0.959 & 0.019 & 0.770 & 0.068 & 0.915 & 0.037 & 0.961 & 0.019 \\
                              & & \text{FDP} & 0.131 & 0.048 & 0.606 & 0.019 & 0.200 & 0.053 & 0.612 & 0.012 
          \\ \hline
    \multirow{2}{*}{\hfil  I}     
         & \multirow{2}{*}{\hfil  Banded}
          & \text{TPR}  & 0.968 & 0.017 & 0.786 & 0.073 & 0.922 & 0.038 & 0.970 & 0.016 \\
                              & & \text{FDP} & 0.151 & 0.070 & 0.565 & 0.025 & 0.222 & 0.063 & 0.585 & 0.023 \\ \hline
    \multirow{2}{*}{\hfil  I}     
          & \multirow{2}{*}{\hfil  Hub}
                  & \text{TPR}   & 0.979 & 0.014 & 0.880 & 0.065 & 0.946 & 0.032 & 0.981 & 0.014 \\
                              & & \text{FDP}  & 0.185 & 0.146 & 0.522 & 0.039 & 0.230 & 0.086 & 0.567 & 0.078 \\ \hline
    \centering{Case} & \multicolumn{1}{|p{1.3cm}|}{Sim Type} & Metric 
      & \multicolumn{2}{p{2cm}|}{\centering $R^{3}$} & \multicolumn{2}{p{2cm}|}{\centering Standard~Test} 
      & \multicolumn{2}{p{2cm}|}{\centering $\Rs$} & \multicolumn{2}{p{2cm}|}{\centering $\Rp$} \\
    \hline
    
    \multirow{2}{*}{\hfil  II}
          & \multirow{2}{*}{\hfil  SmallW} 
          & \text{TPR} & 0.960 & 0.016 & 0.761 & 0.056 & 0.930 & 0.023 & 0.964 & 0.016 \\
                              & & \text{FDP} & 0.046 & 0.019 & 0.492 & 0.017 & 0.078 & 0.021 & 0.506 & 0.011\\ \hline
    \multirow{2}{*}{\hfil  II}      
         & \multirow{2}{*}{\hfil  Banded}
          & \text{TPR}     & 0.962 & 0.012 & 0.757 & 0.081 & 0.930 & 0.033 & 0.967 & 0.014 \\
          & & \text{FDP}    & 0.056 & 0.044 & 0.462 & 0.025 & 0.087 & 0.055 & 0.489 & 0.033 \\ \hline
    \multirow{2}{*}{\hfil  II}      
          & \multirow{2}{*}{\hfil  Hub}
         & \text{TPR} & 0.974 & 0.014 & 0.803 & 0.057 & 0.946 & 0.023 & 0.976 & 0.014 \\
          & & \text{FDP} & 0.038 & 0.016 & 0.396 & 0.015 & 0.070 & 0.021 & 0.410 & 0.011 \\ \hline
    \hline
  \end{tabular}
  \caption{  \label{tab:fdp1}High-dimensional results in terms of average true positive
    rate (TPR) and 
    false discovery proportion (FDP) for tests rejected by each method
    when controlling the FDR at 10\% via the Benjamini-Yekutieli
    method; standard errors are given in parentheses.}  
\end{table}

\end{document}